\documentclass{aa}

\usepackage{fabien} 
\usepackage{hyperref}
\hypersetup{breaklinks,colorlinks,citecolor=blue}

\usepackage{natbib}
\bibpunct{(}{)}{;}{a}{}{,}

\begin{document}

\title{Covariance of the galaxy angular power spectrum with the halo model}
\titlerunning{Covariance of galaxy $C_\ell$}

\author{Fabien Lacasa\thanks{fabien.lacasa@unige.ch}}
\institute{
D\'{e}partement de Physique Th\'{e}orique and Center for Astroparticle Physics, Universit\'{e} de Gen\`{e}ve, 24 quai Ernest Ansermet, CH-1211 Geneva, Switzerland\label{inst1}
}

\date{\today}

\abstract
{
As the determination of density fluctuations becomes more precise with larger surveys, it becomes more important to account for the increased covariance due to the non-linearity of the field. Here I have focussed on the galaxy density, with analytical prediction of the non-Gaussianity using the halo model coupled with standard perturbation theory in real space. I carried out an exact and exhaustive derivation of all tree-level terms of the non-Gaussian covariance of the galaxy $C_\ell$, with the computation developed up to the third order in perturbation theory and local halo bias, including the non-local tidal tensor effect.\\
A diagrammatic method was used to derive the involved galaxy 3D trispectra, including shot-noise contributions. The projection to the angular covariance was derived in all trispectra cases with and without Limber's approximation, with the formulae being of potential interest for other observables than galaxies. The effect of subtracting shot-noise from the measured spectrum is also discussed, and does simplify the covariance, though some non-Gaussian shot-noise terms still remain.\\
I make the link between this complete derivation and partial terms which have been used previously in the literature, including super-sample covariance (SSC). I uncover a wealth of additional terms which were not previously considered, including a whole new class which I dub braiding terms as it contains multipole-mixing kernels. The importance of all these new terms is discussed with analytical arguments. I find that they become comparable to, if not bigger than, SSC if the survey is large or deep enough to probe scales comparable with the matter-radiation equality $k_\mathrm{eq}$.\\
A short self-contained summary of the equations is provided in Section \ref{Sect:summary} for the busy reader, ready to be implemented numerically for analysis of current and future galaxy surveys.
}
\keywords{methods: analytical - galaxies:statistics - large-scale structure of the universe}

\maketitle



\section{Introduction}\label{Sect:intro}

Tracers of the large scale matter distribution in the Universe are one of the main probes of cosmology, with current galaxy surveys such as KiDS \citep{Hildebrandt2017} and Dark Energy Survey (DES) already providing constraints competitive with Planck on some cosmological parameters \citep{DES2017}, and future surveys such as Euclid \citep{Laureijs:2011gra} and the Large Synoptic Sky Telescope \citep[LSST,][]{Abell:2009aa} as scheduled to greatly improve our understanding of dark energy and the structuration of the universe.

Late time tracers of the large scale structure (LSS) have, however, undergone significant non-linear evolution, which makes their probability distribution function deviate from the Gaussianity of the initial density field. As such, two-point statistics no longer retain all the information, and have their covariance increased by the presence of a non-vanishing trispectrum. The focus of the present article is on this increased covariance, in the case of galaxy clustering analysed with the angular power spectrum $C_\ell$.

Covariances are central to the statistical inference process, being the only element besides signal prediction when using the popular Gaussian likelihood, though see \cite{Sellentin2018} for indications that current data may need a non-Gaussian likelihood. In the past, covariances have been estimated through a variety of techniques. Jackknife or bootstrap methods allow estimates from the data itself, however it is known that the results from these methods are very noisy, and \cite{Lacasa2017} showed that they give biased estimates of the effect of super-sample covariance, which I will introduce later. Numerical covariance estimation through simulations of the large scale structure remains costly, especially if one wants cosmology-dependent covariance matrices, although newer techniques allowing fast mock creations \citep{Klypin2017} or data compression \citep{Heavens2017} may help cut down that cost. Furthermore the induced numerical noise in the covariance must be propagated in the likelihood \citep{Sellentin2016,Sellentin2017}, enlarging our uncertainties. Finally \cite{Lacasa2017} showed that simulations also give biased estimates of super-sample covariance, unless the simulation is orders of magnitude larger than the survey, a difficult task for future surveys covering a large part of the sky up to high redshifts \citep{Schneider2016}.

Analytical modelling of the covariances is an approach which has the logical advantage of yielding a self-consistent data analysis with complete analytical understanding of the physics. Conversely, not being able to predict the covariance analytically would question the confidence with which analytical prediction of the signal itself should be trusted.\\
Analytical covariances have been developed recently for LSS tracers including some non-Gaussian effects \citep[e.g.][]{Lacasa2016,Krause2017}. They can provide both fast and noiseless covariance matrices, which could possibly be varied with model parameters. The current approaches are based on the halo model \citep{Cooray2002} coupled with perturbation theory, and are state of the art applied to recent galaxy surveys \citep{Hildebrandt2017,vanUitert2017,Krause2017b,DES2017}.

The Gaussian contribution to the covariance is normally simple enough to treat analytically: if $C_\ell$ is the total power spectrum (including shot-noise effects) of the signal considered, the full-sky Gaussian covariance is
\ba
\Cov_G(C_\ell,C_{\ell'})= \frac{2 \ C_\ell^2}{2\ell+1} \ \delta_{\ell,\ell'}
\ea
Super-sample covariance (SSC) \citep{Hu2003,Takada2013} is currently thought to be the dominant non-Gaussian contribution to the covariance. The effect comes from the non-linear modulation of local observables by long wavelength density fluctuations. In other words, the survey can be non-representative of the universe by probing a region denser (or less dense) than average, as all observables react to such background density change.  SSC has already had an impact on cosmological constraints from current surveys, with \cite{Hildebrandt2017} finding that failure to include it would lead to 1-$\sigma$ shift in their constraint on $S_8$.

Numerical investigation by the author in the case of photometric galaxy clustering have shown that indeed SSC is the dominant effect beyond the Gaussian covariance for specifications of ongoing survey such as DES \citep[slide 14]{Lacasa2017-LAL}. However when implementing specifications of future surveys, I found that other non-Gaussian terms (1-halo and 2-halo1+3, which will be introduced in the article's main text) have an impact on the covariance which is as important as SSC, sometimes even more important depending on redshift and scale \citep[slide 8]{Lacasa2017-LAL}. As these terms become significant, the question arises of the importance of all the other non-Gaussian terms. It thus becomes timely to carry out an exhaustive derivation of all possible non-Gaussian covariance terms, within the current modelling framework that is the halo model.

Here, I undertake the task of carrying out this exhaustive derivation in the case of the angular power spectrum of galaxies $C_\ell^\mr{gal}$. The choice of the harmonic basis is the one underlying current covariance implementations, even when data are in the other popular basis: real space \citep[e.g.][]{Joachimi2008,Krause2017b}, indeed results can be converted straightforwardly through \citep[e.g.][]{Crocce2011}
\ba
w(\theta) = \sum_\ell \frac{2\ell+1}{4\pi} \ C_\ell \ P_\ell(\cos\theta) ,
\ea
\ba
\mathcal{C}_{\theta,\theta'} = \sum_{\ell,\ell'} \frac{(2\ell+1)(2\ell'+1)}{(4\pi)^2} \ \mathcal{C}_{\ell,\ell'} \ P_\ell(\cos\theta) \ P_{\ell'}(\cos\theta') .
\ea
The choice of signal is relevant, though most of the theoretical framework developed should adapt straightforwardly to another observable. I focus here on galaxy clustering using the halo model together with standard perturbation theory at tree-level. This is the most complex signal in the sense that galaxy discreteness (shot-noise) yields more covariance terms. I have  left the application to other observables (clusters, weak-lensing, secondary anisotropies of the CMB) to future works.

The methods used in the article are presented in Sect.~\ref{Sect:methods}. This includes in particular a diagrammatic approach to galaxy polyspectra with the halo model, and projection from 3D quantities to 2D observables or covariances. This should be of interest to theoreticians of large-scale structure tracers. Sections \ref{Sect:1halo} to \ref{Sect:shotnoise} contain the main calculation of the article: all non-Gaussian covariance terms are derived one by one; firstly in the most general case then simplified with relevant approximations. An explanation of the origin of these terms and their ordering will be given in Sect.~\ref{Sect:cov-terms}. A regrouping of terms, comparison with derivation of previous literature, and analytical discussion of the potential importance of these non-Gaussian terms will be performed in Sect.~\ref{Sect:discu}. This should be of interest to give more physical interpretation of the derivation, and intuition on when and why it should be considered of importance. Finally, a self-consistent summary of the results for the busy reader is given in Sect.~\ref{Sect:summary}, using simplifications of relevance to current data analysis. This is the section of reference for numerical implementations of the formulae, for inclusion in analysis of present and future surveys.


\section{Methods}\label{Sect:methods}


\subsection{Conventions}

\subsubsection*{2.1.1. Cosmological notations}
I use the following notations for cosmological quantities : $r(z)$ is the comoving distance to redshift z, $G(z)$ is the growth factor and $\dd V = r^2(z) \frac{\dd r}{\dd z} \dd z$ is the comoving volume element per unit solid angle. Unit vectors have an upper hat, for example, $\hn$ is a direction on the celestial sphere. In the Limber approximation, the peak of the spherical bessel function $j_\ell(kr)$ is $k_\ell=(\ell+1/2)/r(z)$ and depends on an implicit redshift.\\
I also make use of quantities in the halo model \citep{Cooray2002}, such as the halo mass function $\frac{\dd n_h}{\dd M}$, halo spherical profile $u(k|M,z)$ and halo bias $b_\beta(M,z)$ where $\beta=1,2,3$ for the local bias terms used here, and $\beta=s2$ for the quadratic tidal tensor bias \citep{Chan2012,Baldauf2012}.


\subsubsection*{2.1.2. Shortenings}
I have used the following shortenings in order to keep long equations as readable as can be possible.
Spherical harmonics indices are shortened through $i\equiv(\ell_i,m_i)$, including in the case of indices in sums.
The sum of Fourier wavevectors is shortened through $\kk_{i+j}\equiv\kk_i+\kk_j$, implying in particular $k_{i+j}=\left|\kk_i+\kk_j\right|$.
When unambiguous, arguments of multivariate functions are shortened through $f(z_{1234})\equiv f(z_1,z_2,z_3,z_4)$. For example for polyspectra,
\ba
\mathcal{P}^{(n)}(\kk_{1\cdots n},z_{1\cdots n}) \equiv \mathcal{P}^{(n)}(\kk_1,\cdots, \kk_n | z_1,\cdots, z_n)
\ea
Outside of function arguments, repetition of indices is used to note multiplication : $X_{ij}\equiv X_i \, X_j$, for example, in integration elements $\dd M_{\alpha\beta}\equiv\dd M_\alpha \, \dd M_\beta$.\\
The number of galaxy n-tuples (pairs, triplets...) is shortened to
\ba
N_\mr{gal}^{(n)} \equiv N_\mr{gal} \ \left(N_\mr{gal}-1\right) \ \cdots \ \left(N_\mr{gal}-(n-1)\right)
\ea


\subsubsection*{2.1.3. Definitions}
Inspired by the notations of \cite{Takada2013}, I defined the following integrals for galaxies :
\ba
\nonumber I_\mu^\beta(k_1,\cdots,k_\mu|z) = \int \dd M \ & \frac{\dd n_h}{\dd M} \ \lbra N_\mr{gal}^{(\mu)}\rbra \ b_\beta(M,z) \\ 
& \times u(k_1|M,z) \cdots u(k_\mu|M,z)  
\ea
where $\mu$ is an integer (the galaxy tuple power) and $\beta$ is the bias type. I note that $I_\mu^\beta \to cst$ when $k_1,\cdots,k_\mu \rightarrow 0$, as $u(k)\underset{k\to 0}{\longrightarrow}1$. $I_\mu^\beta$ becomes scale-dependent only on small scales, of order of the halo sizes.\\
I also introduced the following integrals, useful later for angular quantities projected in a redshift bin
\ba
\mathcal{I}_\ell^{\mu,\beta}(k;k_1,\cdots,k_\mu|i_z) = \int_{z\in i_z} \dd V \ j_\ell(kr) \ G(z) \ I_\mu^\beta(k_1,\cdots,k_\mu|z)
\ea
and its generalisation to multiple Bessel functions
\ba
\nonumber \mathcal{I}_{n;\ell_1,\cdots,\ell_n}^{\mu,\beta}(k_1,\cdots,k_n;k'_1,\cdots,k'_\mu|i_z) = \int_{z\in i_z} \dd V \ j_{\ell_1}(k_1 r)\cdots j_{\ell_n}(k_n r) \\
\times G(z)^n \ I_\mu^\beta(k'_1,\cdots,k'_\mu|z)
\ea
In the following, when unambiguous I will leave redshift integration bounds implicit for simplicity of notation.

\subsection{Diagrammatic}\label{Sect:diagrammatic}

I modelled the galaxy density field using the halo model \citep{Cooray2002} coupled with tree-level perturbation theory, allowing first-principle description of all galaxy statistics. In this context, the galaxy number density is written as \citep{Lacasa2014}:
\be 
n_\mr{gal}(\xx) = \sum_i \sum_{j=1}^{N_\mr{gal}(i)} \delta^{(3)}(\xx-\xx_j)
\ee
where the first sum runs over halos and the second over galaxies inside that halo.\\
In Fourier space, the (absolute) galaxy polyspectrum of order $n$ is defined by\footnote{With this absolute convention, all 3D polyspectra have dimension Mpc$^{-3}$.  I note also that all galaxy angular polyspectra have dimension sr$^{-1}$, and angular covariances have dimension sr$^{-2}$.}
\ba
\nonumber\lbra n_\mr{gal}(\kk_1,z_1)\cdots n_\mr{gal}(\kk_n,z_n)\rbra_c = (2\pi)^3 \ &\delta^{(3)}(\kk_1+\cdots+\kk_n) \\
& \times \mathcal{P}^{(n)}_\mathrm{gal}(\kk_{1\cdots n},z_{1\cdots n})
\ea

\cite{Lacasa2014} introduced a diagrammatic method to compute the different terms of the galaxy polyspectrum with the halo model. This method was illustrated in more detail, including a trispectrum example, in Sect. 3.4.4 of \cite{Lacasa2014b}.

For the polyspectrum of order $n$, the first step is to draw in diagrams all the possibilities of putting $n$ galaxies in halo(s). Potentially two or more galaxies can lie at the same point (`contracted') for the shot-noise terms. Then for each diagram, the galaxies should be labelled from 1 to $n$, as well as the halos for example, with $\alpha_1$ to $\alpha_p$.\\
Each diagram produces a polyspectrum term which is an integral over the halo masses $\int \dd M_{\alpha_1 \cdots \alpha_p}$ of several factors:
\begin{itemize}
\item for each halo $\alpha_j$ there is a corresponding :
\begin{itemize}
\item halo mass function $\orange{\frac{\dd n_h}{\dd M}(M_{\alpha_j})}$
\item average of the number of galaxy tuples in that halo.\\
for example, $\darkgreen{\lbra N_\mr{gal}\rbra}$ for a single galaxy in that halo, $\darkgreen{\lbra N_\mr{gal} (N_\mr{gal} - 1)\rbra}$ for a pair etc.
\item as many halo profile ${\color{red} u(k|M_{\alpha_j})}$ as different points, where $k = \left| \sum_{i\in \mr{point}} \kk_i \right|$.\\
for example, $k=k_i$ for a non-contracted galaxy, while $k=k_{i_1+i_2}=\left|\kk_{i_1}+\kk_{i_2}\right|$ for a galaxy contracted twice.
\end{itemize}
\item the halo polyspectrum of order p, conditioned to the masses of
the corresponding haloes :
$${\color{blue} \mathcal{P}^{(p)}_\mathrm{halo}\left(\sum_{i\in \alpha_1} \kk_i\, , \cdots ,
\sum_{i\in \alpha_p} \kk_i \,\Bigg| M_{\alpha_1} , \cdots , M_{\alpha_p} \right)
}$$
where the sum $\sum_{i\in \alpha_j} \kk_i$ runs over the indexes i of the galaxies inside the halo $\alpha_j$ .
\end{itemize}
Finally one should account for all the possible permutations of the galaxy labels 1 to $n$ in the diagram.

Additionally, if one is interested in the polyspectrum of the relative density fluctuations $\delta_\mr{gal} \equiv \frac{\delta n_\mr{gal}}{\nbargal}$, instead of the absolute fluctuations, one should add a $1/\nbargal^n$ prefactor to the preceding expression. This can prove useful for 3D observables; however for 2D projected observables, as the angular power spectrum studied in this article, it is the absolute fluctuations which naturally enter the equations.

\subsection{Projection to 2D observables}\label{Sect:2Dproj}

The projected galaxy density in the direction $\hn$ in a redshift bin $i_z$, $n_\mr{gal}(\hn,i_z)$ is the line-of-sight integral:
\be
n_\mr{gal}(\hn,i_z) 
= \int \dd V \ n_\mr{gal}(\xx=r\hn,z)
\ee
with $\dd V=r^2\dd r$ being the comoving volume per unit solid angle.\\
This projection neglects redshift-space distortion and other general relativistic effects \citep[for example,][]{Bonvin2011}, whose impacts are left for future studies.

In full sky, after spherical harmonic decomposition the harmonic coefficients are \citep[for example,][]{Lacasa2016}
\ba 
a_{\ell m}^\mr{gal}(i_z)  &= \int \dd^2\hn \; n_\mr{gal}(\hn,i_z)\; Y^*_{\ell m}(\hn)\\
&= \int \dd V \; \dd^2\hn \; \frac{\dd^3\kk}{(2\pi)^3} \; n_\mr{gal}(\kk,z) \; \mre^{\ii \kk \cdot r\hn} \; Y^*_{\ell m}(\hn)\\
&= \ii^\ell \int \dd V \; \frac{\dd^3\kk}{2\pi^2} \; j_\ell(k r) \; n_\mr{gal}(\kk,z) \; Y^*_{\ell m}(\hk)
\ea
The galaxy power spectrum estimator is then
\ba
\hat{C}_\ell^\mr{gal}(i_z,j_z) &= \frac{1}{2\ell+1} \sum_m \ a_{\ell m}^\mr{gal}(i_z) \ \left(a_{\ell m}^\mr{gal}(j_z)\right)^* ,\\
\nonumber &= \int \dd V_{12} \frac{\dd^3\kk_{12}}{(2\pi^2)^2} \, j_\ell(k_1 r_1) \,  j_\ell(k_2 r_2) \,  n_\mr{gal}(\kk_1,z_1) \, n^*_\mr{gal}(\kk_2,z_2) \\
& \qquad \times \frac{1}{2\ell+1}\sum_m Y^*_{\ell m}(\hk_1) \, Y_{\ell m}(\hk_2) ,\\
\label{Eq:Clgal-estimator-Lagrange} \nonumber &= 4\pi \, (-1)^\ell \int \dd V_{12} \frac{\dd^3\kk_{12}}{(2\pi)^6} \; j_\ell(k_1 r_1) \,  j_\ell(k_2 r_2) \ P_\ell(\hk_1 \cdot \hk_2)\\
& \qquad \times n_\mr{gal}(\kk_1,z_1) \; n_\mr{gal}(\kk_2,z_2)
\ea
with the $(-1)^\ell$ coming from the change $\kk_2\rightarrow -\kk_2$ and the parity of the Legendre polynomial $P_\ell$ .\\
Its expectation value is
.\ba
C_\ell^\mr{gal}(i_z,j_z) &= \frac{2}{\pi} \int \dd V_{12} \ k^2 \, \dd k \ j_\ell(k r_1) \,  j_\ell(k r_2) \ P_\mr{gal}(k|z_{12})
\ea
The non-Gaussian part of the galaxy spectrum covariance is thus:
\ba
\nonumber \mathcal{C}_{\ell,\ell'} = (4\pi)^2 \, (-1)^{\ell+\ell'} &\int \dd V_{1234} \frac{\dd^3\kk_{1234}}{(2\pi)^{12}} \; j_\ell(k_1 r_1) \,  j_\ell(k_2 r_2) \, j_{\ell'}(k_3 r_3) \\
\nonumber & j_{\ell'}(k_4 r_4) \ P_\ell(\hk_1 \cdot \hk_2) \ P_{\ell'}(\hk_3 \cdot \hk_4) \\
& \times (2\pi)^3 \, \delta^{(3)}\left(\kk_1+\cdots+\kk_4\right) \, T_\mr{gal}(\kk_{1234},z_{1234}) \label{Eq:CovClgal-Lagrange}
\ea
where I used the abbreviation
$$\mathcal{C}_{\ell,\ell'}\equiv \Cov\left(\hat{C}_\ell^\mr{gal}(i_z,j_z),\hat{C}_{\ell'}^\mr{gal}(k_z,l_z)\right)$$
leaving redshift bins implicit hereafter

\begin{figure}[htbp]
\begin{center}
\begin{tikzpicture}
\draw [->, very thick,teal] (0,0) -- node[below] {$\vec{k}_1$} (-4,2);
\draw [->, very thick,teal] (-4,2.1) -- node[above] {$\vec{k}_2$} (0,1);
\draw [->, very thick,brown] (0,1) -- node[above] {$\vec{k}_3$} (4,3);
\draw [->, very thick,brown] (4,2.9) -- node[below] {$\vec{k}_4$} (0,0);
\draw [->, very thick, dashed,red] (0,0) -- node[left] {$\vec{k}_{1+2}$} (0,1);
\end{tikzpicture}
\caption{3D trispectrum in the squeezed diagonal limit.}
\label{Fig:squeezed-trispectrum}
\end{center}
\end{figure}
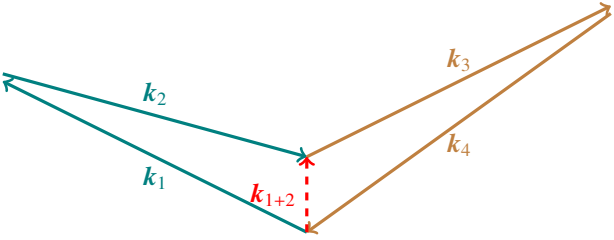

As a power spectrum estimator, most of the contribution to $\hat{C}_\ell$ Eq. \ref{Eq:Clgal-estimator-Lagrange} will come from $\kk_1\!\approx\!-\kk_2$, that is, $k_{1+2} \ll k_1 \!\approx\! k_2$. Thus, similarly to the case of the 3D power spectrum estimator \citep{Takada2013}, the covariance Eq. \ref{Eq:CovClgal-Lagrange} probes the trispectrum in the squeezed diagonal configuration represented in Fig.~\ref{Fig:squeezed-trispectrum}~: $k_{1+2}=k_{3+4} \ll k_1 \!\approx\! k_2,\, k_3\!\approx\! k_4$. Contrary to \citep{Takada2013} however, the present derivation does not rely on any approximation or Taylor expansion in terms of $k_{1+2}$.

The 3D trispectrum generally depends on six degrees of freedom (d.o.f.) that fix the shape of the quadrilateron $\kk_1+\kk_2+\kk_3+\kk_4=0$. There is however no unique natural choice for these six d.o.f. \footnote{contrary to the power spectrum which has one d.o.f. : $k$, and the bispectrum which has three d.o.f. that it is natural to chose as the sides $k_1,k_2,k_3$ of the triangle, although some studies prefer to take one or two length(s) and one or two angle(s), or use lengths ratios.} : the choice will depend on the trispectrum term considered. For all the terms considered here (see Appendix \ref{App:3Dhalopolysp}), four d.o.f. will be the quadrilateron sides $k_1,k_2,k_3,k_4$, then the trispectrum may also depend on the length of one of the diagonals $k_{1+2},k_{1+3},k_{1+4}$ and/or on angles either between base wavevectors $\hk_i\cdot\hk_j$ or between a diagonal and a base wavevector $\hk_{i+j}\cdot\hk_{l}$. Deriving the projection Eq. \ref{Eq:CovClgal-Lagrange} analytically in all the necessary trispectrum cases proves a complex task, and is the subject of Appendices \ref{App:2Dproj-trisp-angindep} \& \ref{App:2Dproj-trisp-angdep}. I list below the three less complex cases where the trispectrum does not depend on any angle, solely on lengths of base wavevectors or diagonals.

Firstly, the easiest case is of a diagonal-independent trispectrum, that is, $T_\mr{gal}(\kk_{1234},z_{1234})=T_\mr{gal}(k_{1234},z_{1234})$. This case was treated by \cite{Lacasa2014,Lacasa2014b} for a general diagonal-independent polyspectrum. In the present case, one finds (see Appendix \ref{App:2Dproj-trisp-diagindep} for a derivation):
\ba
\nonumber \mathcal{C}_{\ell,\ell'} &= \frac{\left(\frac{2}{\pi}\right)^4}{4\pi} \int x^2 \, \dd x \; \dd V_{1234} \; k^2_{1234} \, \dd k_{1234} \ j_\ell(k_1 r_1) \,  j_\ell(k_1 x) \\
\nonumber & \qquad j_\ell(k_2 r_2) \, j_\ell(k_2 x) \ j_{\ell'}(k_3 r_3)  \, j_{\ell'}(k_3 x) \  j_{\ell'}(k_4 r_4) \, j_{\ell'}(k_4 x) \\ 
& \qquad \times T_\mr{gal}(k_{1234},z_{1234}) \label{Eq:Cll'-diagindep-nolimber}
\ea
Using Limber's approximation (see Appendix \ref{App:2Dproj-trisp-diagindep}), this simplifies to
\ba\label{Eq:Cll'-diagindep-limber}
\mathcal{C}_{\ell,\ell'} = \frac{\delta_{i_z,j_z,k_z,l_z}}{4\pi} \int \dd V \ T_\mr{gal}(k_{\ell_{1234}},z).
\ea
This case will be relevant for a large part of the covariance terms later on.

The second case of interest is of a trispectrum depending on the length of the squeezed diagonal, $K=k_{1+2}$, additionally to the length of the four sides $k_{1234}$. This case is treated in Appendix \ref{App:2Dproj-trisp-sqzdiag}, giving: 
\ba\label{Eq:Cll'-sqzdiag-nolimber}
\nonumber \mathcal{C}_{\ell,\ell'} &= \frac{\left(\frac{2}{\pi}\right)^5}{4\pi} \int \dd V_{1234} \ k^2_{1234} \,\dd k_{1234} \ K^2\,\dd K \ \dd V_{ab} \\
\nonumber & \qquad j_\ell(k_1 r_1) \, j_\ell(k_1 x_a) \ j_\ell(k_2 r_2) \, j_\ell(k_2 x_a) \ j_0(K x_a) \\ 
\nonumber & \qquad j_0(K x_b) \ j_{\ell'}(k_3 r_3) \, j_{\ell'}(k_3 x_b) \  j_{\ell'}(k_4 r_4) \, j_{\ell'}(k_4 x_b) \\
& \qquad \times T_\mr{gal}(k_{1234},K,z_{1234}).
\ea
Using Limber's approximation on $k_{1234}$ (but not on $K$, since this would be a poor approximation for $j_0$ which has a large support and peaks at $K=0$), this simplifies to
\ba\label{Eq:Cll'-sqzdiag-limber}
\nonumber \mathcal{C}_{\ell,\ell'} &= \frac{\frac{2}{\pi} \ \delta_{i_z,j_z} \, \delta_{k_z,l_z} }{4\pi} \int K^2\dd K \; \dd V_{ab} \; j_0(K x_a) \, j_0(K x_b) \\
& \qquad \times T_\mr{gal}(k_{\ell},k_{\ell},k_{\ell'},k_{\ell'},K,z_{aabb}).
\ea
This case will be relevant for super-sample covariance terms later on (Sect.~\ref{Sect:discu-SSC}).

The third case of interest is of a trispectrum depending on the length of one of the other diagonal, $K=k_{1+3}$ (with $K=k_{1+4}$ giving a symmetric result), additionally to the length of the four sides $k_{1234}$. This case is treated in Appendix \ref{App:2Dproj-trisp-altdiag}, giving:  
\ba\label{Eq:Cll'-altdiag-nolimber}
\nonumber \mathcal{C}_{\ell,\ell'} &= \sum_{\ell_a} \frac{2\ell_a+1}{4\pi} \threeJz{\ell}{\ell'}{\ell_a}^2 \left(\frac{2}{\pi}\right)^5 \int \dd V_{1234} \ k^2_{1234} \, \dd k_{1234} \\
\nonumber & \qquad K^2\dd K \; \dd V_{ab} \ j_\ell(k_1 r_1) \, j_\ell(k_1 x_a) \ j_\ell(k_2 r_2) \, j_\ell(k_2 x_b) \\ 
\nonumber & \qquad j_{\ell_a}(K x_a) \, j_{\ell_a}(K x_b) \ j_{\ell'}(k_3 r_3) \, j_{\ell'}(k_3 x_a) \  j_{\ell'}(k_4 r_4) \, j_{\ell'}(k_4 x_b) \\
& \qquad \times T_\mr{gal}(k_{1234},K,z_{1234}).
\ea
Using Limber's approximation on $k_{1234}$ but not on the diagonal, the covariance simplifies to
\ba
\nonumber \mathcal{C}_{\ell,\ell'} = \delta_{i_z,k_z} \ \delta_{j_z,l_z} \sum_{\ell_a} \frac{2\ell_a+1}{4\pi} \threeJz{\ell}{\ell'}{\ell_a}^2 \frac{2}{\pi} \int K^2\dd K \; \dd V_{ab} \\
\label{Eq:Cll'-altdiag-partial-limber} j_{\ell_a}(K x_a) \, j_{\ell_a}(K x_b) \times T_\mr{gal}(k_{\ell_{1234}},K,z_{abab}).
\ea
This case will be relevant for braiding terms later on (Sect.~\ref{Sect:discu-braiding}).

There are furthermore eight cases of trispectra depending on angles between wavevectors : four cases where the trispectrum depends on one angle, tackled in Appendix~\ref{App:2Dproj-trisp-angdep-1angle}, and four cases where the trispectrum depends on two angles, tackled in Appendix~\ref{App:2Dproj-trisp-angdep-2angles}. Due to the complexity of these expressions, they are left to their respective appendices for the clarity of the main text. These formulae involve geometric coefficients which are shown in Appendix~\ref{App:3nJ-symbols} to be related to Wigner 3n-J symbols : 6J, 9J and even the case of a reduced 12J symbol of the second kind. Reduction checks are performed in Appendix~\ref{App:reductions} to assure the robustness of the results.

These 2D projection formulae, although not the main aim of this article, can be viewed as standalone results that should be of interest for other analyses, for example, interested in the covariance of other signals or using a different modelling framework such as a different flavour of perturbation theory.

\subsection{Example of the power spectrum}\label{Sect:power-spectrum}

In this article, I am interested in the covariance of the galaxy power spectrum. Before coming to the covariance, the diagrammatic formalism and the 2D projection explained above can already be illustrated at the power spectrum level. This will already uncover technical details of later interest.

\begin{figure}[!th]
\begin{center}
\includegraphics[width=\linewidth]{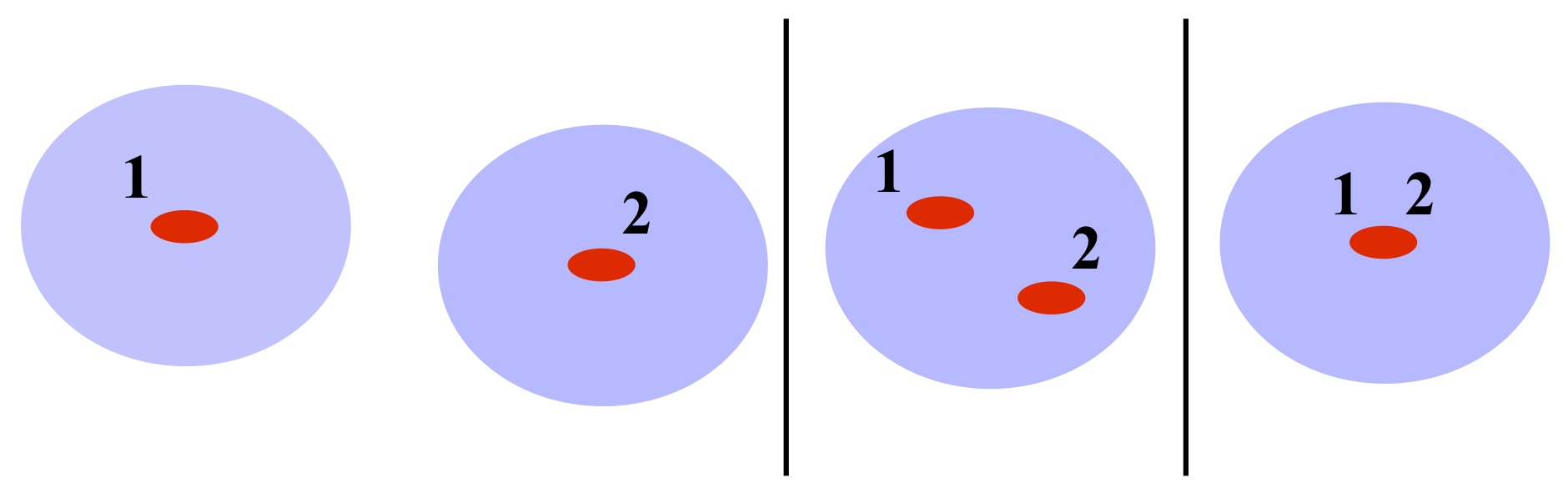}
\caption{Diagrams for the galaxy power spectrum. From left to right: two-halo (2h), one-halo (1h) and shot-noise (shot).}
\label{Fig:diagrams-spectrum}
\end{center}
\end{figure}

Figure~\ref{Fig:diagrams-spectrum} shows the power spectrum diagrams, recovering the well known fact that the spectrum decomposes in a two-halo term, a one-halo term and shot-noise. One immediate advantage, already underlined in \cite{Lacasa2014,Lacasa2014b}, is that shot-noise is described consistently and does not need a separate formalism like counts in cells \citep[e.g.][]{Peebles1980}.

Applying the diagrammatic rules yields for example the following expression for the two-halo term of the power spectrum between two possibly different redshifts :
\ba
\nonumber P^\mr{2h}_\mr{gal}(k|z_{12}) =\int &\dd M_{\alpha\beta} \ \orange{\frac{\dd n_h}{\dd M}(M_\alpha,z_1) \, \frac{\dd n_h}{\dd M}(M_\beta,z_2)} \\
\nonumber & \darkgreen{\lbra N_\mr{gal}\rbra(M_\alpha,z_1) \, \lbra N_\mr{gal}\rbra(M_\beta,z_2)} \\ 
& {\color{red} u(k|M_\alpha,z_1) \, u(k|M_\beta,z_2)} \ {\color{blue} P_\mr{hh}(k|M_{\alpha\beta},z_{12})}
\ea
In the following, I will shorten the argument of mass and redshift to its indices:
$$\left.\frac{\dd n_h}{\dd M}\right|_{\alpha,1}\equiv \frac{\dd n_h}{\dd M}(M_\alpha,z_1)$$
$$\lbra N_\mr{gal}\rbra_{\alpha,1}\equiv\lbra N_\mr{gal}\rbra(M_\alpha,z_1)$$
$$u(k|\alpha,1)\equiv(k|M_\alpha,z_1)$$

At tree level the halo power spectrum takes the form 
\be
P_\mr{halo}(k|M_{\alpha\beta},z_{12}) = b_1(M_\alpha,z_1) \; b_1(M_\beta,z_2) \ P_\mr{lin}(k|z_{12})
\ee
so that one recovers the familiar equation
\ba\label{Eq:P2h-treelevel}
P^\mr{2h}_\mr{gal}(k|z_{12}) &= \nbargal(z_1) \, \nbargal(z_2) \ b_1^\mr{gal}(k,z_1) \, b_1^\mr{gal}(k,z_2) \ P_\mr{lin}(k|z_{12})\\
&= I_1^1(k|z_1) \ I_1^1(k|z_2) \ P_\mr{lin}(k|z_{12})
\ea
with the (scale-dependent) galaxy first order bias
\be
b_1^\mr{gal}(k,z) = \left. \int \dd M \; \frac{\dd n_h}{\dd M} \, \lbra N_\mr{gal}\rbra \, u(k|M,z) \, b_1(M,z) \ \middle/ \ \nbargal(z) \right.
\ee
I note however that things become more complex at 1-loop order, where one would get additional contributions from higher order perturbation theory and halo biases, with a form more complex than Eq.~\ref{Eq:P2h-treelevel}. Since I will be working only at tree level, in the following I note $P(k,z)$ instead of $P_\mr{lin}(k,z)$ for simplicity.\\
The 2-halo part of the angular power spectrum is then given by
\ba
\Cl^\mr{2h} = \frac{2}{\pi} \int k^2 \dd k \ P(k,z=0) \ \mathcal{I}_\ell^{1,1}(k;k|i_z) \ \mathcal{I}_\ell^{1,1}(k;k|j_z)
\ea
In the following I note $P(k)\equiv P(k,z=0)$ for shortening.\\
Limber's approximation simplifies $\Cl^\mr{2h}$ to
\ba
\Cl^\mr{2h} = \int \dd V \ \left(I_1^1(k_\ell|z)\right)^2 \ P(k_\ell|z)
\ea

For the shot-noise power spectrum term, the diagrammatic rules give
\ba
\nonumber P^\mr{shot}_\mr{gal}(k|z_{12}) &=\int \dd M \ \orange{\frac{\dd n_h}{\dd M}}
\ \darkgreen{\lbra N_\mr{gal}\rbra(M,z)} \ {\color{red} u(0|M,z)} \ {\color{blue} \times 1}\\
&= \nbargal(z) \label{Eq:Pshot}
\ea
At this point I seem to face a slight incoherence : which redshift am I talking about, $z_1$ or $z_2$ ? This would be a real issue if one were computing the power spectra between slices of the universe at the same location but different times, in which case our correlation function could hit the same galaxy at two different redshifts. However real observables are located on the past light cone, so the two redshifts have to coincide. Whatever value is given to $P^\mr{shot}_\mr{gal}(k|z_{12})$ when $z_1\neq z_2$ should get canceled when the 2D projection of Sect. \ref{Sect:2Dproj} is carried out. So I can take
\ba\label{Eq:Pshot-possib}
P^\mr{shot}_\mr{gal}(k|z_{12}) &= \nbargal(z_1) \quad \mr{or} \quad \nbargal(z_2) \quad \mr{or} \quad \delta_{z_1,z_2} \ \nbargal(z_1)
\ea
and it should give the same angular power spectrum.\\
Indeed one can check for instance with the first possibility:
\ba
\nonumber C_\ell^\mr{shot} &=  \frac{2}{\pi} \int \dd V_{12} \ k^2 \, \dd k \ j_\ell(k r_1) \,  j_\ell(k r_2) \ P^\mr{shot}_\mr{gal}(k|z_{12}) \\
\nonumber &= \frac{2}{\pi} \int \dd V_{12} \ \nbargal(z_1) \underbrace{\int k^2 \, \dd k \ j_\ell(k r_1) \,  j_\ell(k r_2)}_{=\frac{\pi}{2 r_1^2} \delta(r_1-r_2)} \\
&= \delta_{i_z,j_z} \int \dd V \ \nbargal(z) = N_\mr{gal}(i_z) \ \delta_{i_z,j_z} \label{Eq:Clshot}
\ea
and one would get the same results with the two other possibilities given in Eq.~\ref{Eq:Pshot-possib}. I note the appearance of a Kronecker between redshift bins, assuming they are not overlapping (see Appendix \ref{App:shot-overlapping} for the case of overlapping bins).\\
In the following, I adopt notation from the third possibility (i.e. $P^\mr{shot}_\mr{gal}=\delta_{z_1,z_2} \ \nbargal(z_1)$, with $\delta_{z_1,z_2}$ being a Kronecker symbol) as it makes explicit that the redshift have to coincide.
I also note that the Limber approximation is exact for this power spectrum term, as $P^\mr{shot}_\mr{gal}$ is constant with $k$.\\
Finally, the issue of Poissonian or non-Poissonian shot-noise is discussed in Appendix \ref{App:shotvsPoisson}.

For the one-halo power spectrum term, the diagrammatic rules give
\ba
\nonumber P^\mr{1h}_\mr{gal}(k|z_{12}) &=\int \dd M \ \orange{\frac{\dd n_h}{\dd M}}
\ \darkgreen{\lbra N_\mr{gal}^{(2)}\rbra(M,z)} \ {\color{red} u(k|M,z)^2} \ {\color{blue} \times 1}\\
&= I_2^0(k,k|z)
\ea
again I am faced with an apparent redshift incoherence. But since the two galaxies hit by the correlation function are in the same halo, and since observations are located on the past light cone, the two redshifts must be close, limited by $\delta r < 2 R(M,z)$ where $R(M,z)$ is the virial radius of the halo. In this limited redshift interval there will be no appreciable evolution of the mass function, halo profiles etc. So all redshifts can be taken to be equal, finding
\ba
P^\mr{1h}_\mr{gal}(k|z_{12}) &= \delta_{z_1,z_2} \ I_2^0(k,k|z_1)
\ea
One can note that the Limber's approximation is particularly well adapted to this power spectrum term. Indeed, at low $\ell$ / low $k$ where the Limber's approximation often gets wrong, $P^\mr{1h}_\mr{gal}$ goes to a constant so that Limber becomes exact, and $P^\mr{1h}_\mr{gal}$ starts to have a scale-dependence only on halo size scales - small scales where Limber's approximation works well. Thus one gets the angular power spectrum
\ba
C_\ell^\mr{1h} &= \delta_{i_z,j_z} \int_{z\in i_z} \dd V \ I_2^0(k_\ell,k_\ell|z)
\ea
Again I note the presence of a Kronecker over redshift bins, and that other forms of $P^\mr{1h}_\mr{gal}$ coinciding for $z_1=z_2$ would have given the same answer for the observable.

In general for higher order polyspectra, Limber's approximation will be exact for wavevectors on which the polyspectrum does not depend, well justified for wavectors for which the dependence is only through halo profiles $u(k)$, justified only at high $\ell$ when there is a power spectrum dependence $P(k)$, and unjustified when there is a dependence on the angles between wavevectors (e.g. through perturbation theory kernels as will be seen later).\\
In the following, I thus apply Limber's approximation on wavevectors for which there is no dependence or only $u(k)$ dependence, and will provide both the no-Limber and Limber equations in the other cases.

\subsection{Power spectrum covariance terms}\label{Sect:cov-terms}

\begin{figure}[!th]
\begin{center}
\includegraphics[width=\linewidth]{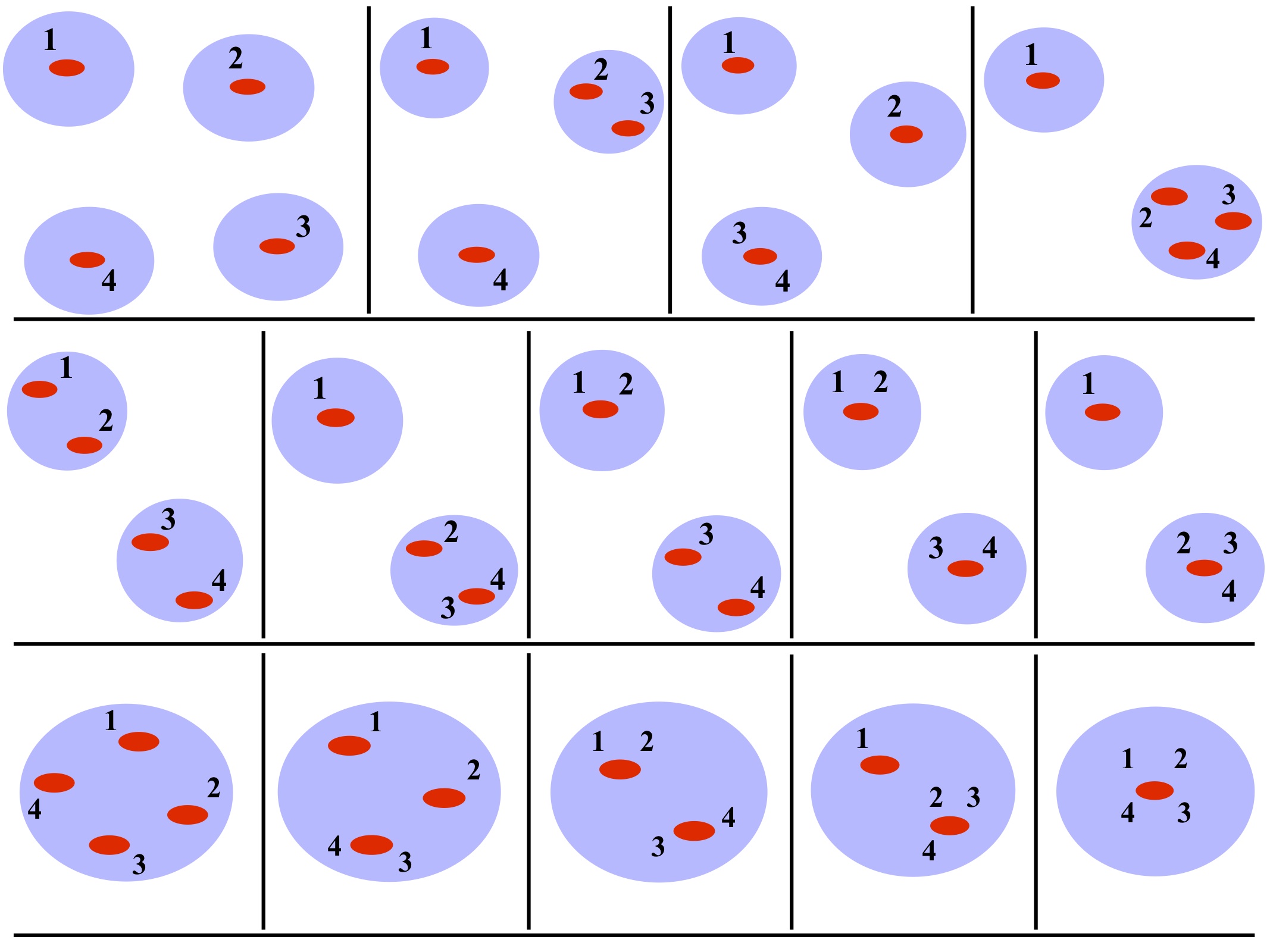}
\caption{Diagrams for the galaxy trispectrum. From left to right,
\textit{top row}: four-halo (4h), three-halo (3h), three-halo shot-noise (3h-shot3g), two-halo 1+3 (2h1+3).
\textit{Middle row}: two-halo 2+2 (2h2+2), two-halo three-galaxy shot-noise a (2ha-shot3g), two-halo three-galaxy shot-noise b (2hb-shot3g), two-halo two-galaxy shot-noise a (2ha-shot2g), two-halo two-galaxy shot-noise b (2hb-shot2g).
\textit{Bottom row}: one-halo (1h), one-halo three-galaxy shot-noise (1h-shot3g), one-halo two-galaxy shot-noise a (1ha-shot2g), one-halo two-galaxy shot-noise b (1hb-shot2g), one-galaxy shot-noise (shot1g).}
\label{Fig:diagrams-trispectrum}
\end{center}
\end{figure}

Section \ref{Sect:2Dproj} gave the projection equations from 3D to 2D. I now need the 3D trispectrum equations. Using the diagrammatic approach of Sect. \ref{Sect:diagrammatic}, the involved diagrams are shown in Fig.~\ref{Fig:diagrams-trispectrum}

This justifies the organisation of the next few sections of this article, as I derive the covariance terms in order of increasing complexity. I will start with the clustering terms: one-halo in Sect.~\ref{Sect:1halo}, two-halo (both 2+2 and 1+3) in Sect.~\ref{Sect:2halo}, three-halo in Sect.~\ref{Sect:3halo} and finally four-halo in Sect.~\ref{Sect:4halo}. I then move to shot-noise (all terms) in Sect.~\ref{Sect:shotnoise}.

\section{One-halo term}\label{Sect:1halo}
The one-halo term is the tenth diagram of Fig.~\ref{Fig:diagrams-trispectrum} (bottom row, first from the left), while the other diagrams with a single halo (remainder of the bottom row in Fig.~\ref{Fig:diagrams-trispectrum}) are shot-noise terms which will be treated in Sect.~\ref{Sect:shotnoise}. Applying the diagrammatic rules from Sect.~\ref{Sect:diagrammatic} and the notes in Sect.~\ref{Sect:power-spectrum} about coincident redshifts, the corresponding trispectrum part is:
\ba
\nonumber T^\mr{1h}_\mr{gal}(\kk_{1234},z_{1234}) &= \delta_{z_1,z_2,z_3,z_4} \int  \dd M \ \orange{\frac{\dd n_h}{\dd M}}
\ \darkgreen{\lbra N_\mr{gal}^{(4)}\rbra} \ {\color{red} u(k_1|M,z) }\\ 
& \qquad {\color{red} u(k_2|M,z) \, u(k_3|M,z) \, u(k_4|M,z)} \ {\color{blue} \times 1}\\
& = \delta_{z_1,z_2,z_3,z_4} \ I_4^0(k_{1234}|z)
\ea
where $\delta_{z_1,z_2,z_3,z_4}=\delta_{z_1,z_2}\,\delta_{z_2,z_3}\,\delta_{z_3,z_4}$, i.e. it is equal to 1 when all redshifts are equal, and 0 otherwise.\\
This trispectrum term is diagonal-independent, according to the nomenclature of Sect. \ref{Sect:2Dproj}.

When projecting onto the angular covariance, as argued in Sect.~\ref{Sect:power-spectrum}, Limber's approximation is justified on all wavevectors, since no $P(k)$ factors are present. One thus obtains:
\ba\label{Eq:Cll'-1h}
\mathcal{C}_{\ell,\ell'}^\mr{1h} &= \frac{\delta_{i_z,j_z,k_z,l_z}}{4\pi} \int \dd M \, \dd V \ \frac{\dd n_h}{\dd M}
\ \lbra N_\mr{gal}^{(4)}\rbra u(k_{\ell}|M,z)^2 \,  \, u(k_{\ell'}|M,z)^2 \\
&= \frac{\delta_{i_z,j_z,k_z,l_z}}{4\pi} \int \dd V \ I_4^0(k_{\ell},k_{\ell},k_{\ell'},k_{\ell'}|z).
\ea

\section{Two-halo terms}\label{Sect:2halo}
\subsection{Two-halo 1+3 term}\label{Sect:2halo1+3}
This term is the fourth diagram of Fig.~\ref{Fig:diagrams-trispectrum} (upper right corner). Applying the diagrammatic rules from Sect.~\ref{Sect:diagrammatic}, the corresponding trispectrum part is:
\ba
\nonumber T^\mr{2h1+3}_\mr{gal}(\kk_{1234},z_{1234}) &= \delta_{z_2,z_3,z_4} \int  \dd M_{\alpha\beta} \ \orange{\left.\frac{\dd n_h}{\dd M}\right|_{\alpha,1} \left.\frac{\dd n_h}{\dd M}\right|_{\beta,2}} \darkgreen{\lbra N_\mr{gal}\rbra_{\alpha,1}} \\ 
\nonumber & \qquad \darkgreen{\lbra N_\mr{gal}^{(3)}\rbra_{\beta,2}} {\color{red} u(k_1|\alpha,1) \, u(k_2|\beta,2) \, u(k_3|\beta,2)} \\
& \qquad \times {\color{red} u(k_4|\beta,2)} {\color{blue} \ P_\mr{hh}(k_1|M_{\alpha\beta},z_{12})} + 3 \ \mr{perm.} \\
\nonumber & = \delta_{z_2,z_3,z_4} \ I_1^1(k_1|z_1) \ I_3^1(k_{234}|z_2) \ P(k_1|z_{12})\\
& \qquad + 3 \ \mr{perm.}
\ea
This trispectrum term is diagonal-independent following the nomenclature of Sect. \ref{Sect:2Dproj}.

For the permutation presented above, Limber's approximation is justified on $k_2,k_3,k_4,$ but may not be justified on $k_1$ for low $\ell$. One finds the covariance term
\ba\label{Eq:Cll'-2h1+3-nolimber}
\nonumber \mathcal{C}_{\ell,\ell'}^\mr{2h1+3} = \frac{\frac{2}{\pi} \ \delta_{j_z,k_z,l_z}}{4\pi} \int k_1^2 \ \dd k_1 \ P(k_1,z=0) \ \mathcal{I}_\ell^{1,1}(k_1;k_1|i_z) \\
\times \mathcal{I}_\ell^{3,1}(k_1;k_\ell,k_{\ell'},k_{\ell'}|j_z) + 3 \ \mr{perm.}
\ea
Using the Limber's approximation also on $k_1$, one finds:
\ba\label{Eq:Cll'-2h1+3-limber}
\nonumber \mathcal{C}_{\ell,\ell'}^\mr{2h1+3} = \frac{\delta_{i_z,k_z,l_z}+\delta_{j_z,k_z,l_z}}{4\pi} & \int \dd M_{\alpha\beta} \, \dd V \ \frac {\dd n_h}{\dd M}(M_\alpha) \ \frac{\dd n_h}{\dd M}(M_\beta)
\ \lbra N_\mr{gal}\rbra_{\alpha} \\ 
\nonumber & \quad \lbra N_\mr{gal}^{(3)}\rbra_{\beta} \ u(k_{\ell}|M_\alpha) \, u(k_{\ell}|M_\beta) \, u(k_{\ell'}|M_\beta)^2 \\
\nonumber & \quad \times P_\mr{halo}(k_{\ell}|M_{\alpha\beta},z) \\
& + (\ell \leftrightarrow \ell')\\
\nonumber = \frac{\delta_{i_z,k_z,l_z}+\delta_{j_z,k_z,l_z}}{4\pi} & \int \dd V \ I_1^1(k_\ell|z) \ I_3^1(k_\ell,k_{\ell'},k_{\ell'}|z) \ P(k_\ell|z) \\
& + (\ell \leftrightarrow \ell')
\ea

\subsection{Two-halo 2+2 term}\label{Sect:2halo2+2}
This term is the fifth diagram of Fig.~\ref{Fig:diagrams-trispectrum}: middle row, first from the left. The other diagrams in the middle row with two halos are shot-noise terms which will be treated in Sect.~\ref{Sect:shotnoise}. Applying the diagrammatic rules from Sect.~\ref{Sect:diagrammatic}, the corresponding trispectrum part is:
\ba
\nonumber T^\mr{2h2+2}_\mr{gal}(\kk_{1234},z_{1234}) = \delta_{z_1,z_2} \ \delta_{z_3,z_4} \int \dd M_{\alpha\beta} \ \orange{\left.\frac{\dd n_h}{\dd M}\right|_{\alpha,1} \left.\frac{\dd n_h}{\dd M}\right|_{\beta,3}} \darkgreen{\lbra N_\mr{gal}^{(2)}\rbra_{\alpha,1}} \\ 
\nonumber \darkgreen{\lbra N_\mr{gal}^{(2)}\rbra_{\beta,3}} \ {\color{red} u(k_1|\alpha,1) \, u(k_2|\alpha,1) \, u(k_3|\beta,3) \, u(k_4|\beta,3)} \\
{\color{blue} \times P_\mr{hh}(k_{1+2}|M_{\alpha\beta},z_{13})} + \mr{2 \ perm.} \\
\label{Eq:T2h2+2-sqzdiag} = \delta_{z_1,z_2} \ \delta_{z_3,z_4} \ I_2^1(k_1,k_2|z_1) \ I_2^1(k_3,k_4|z_3) \ P(k_{1+2}|z_{13})\\
\label{Eq:T2h2+2-altdiag13} + \delta_{z_1,z_3} \ \delta_{z_2,z_4} \ I_2^1(k_1,k_3|z_1) \ I_2^1(k_2,k_4|z_2) \ P(k_{1+3}|z_{12}) \\
\label{Eq:T2h2+2-altdiag14} + \delta_{z_1,z_4} \ \delta_{z_2,z_3} \ I_2^1(k_1,k_4|z_1) \ I_2^1(k_2,k_3|z_2) \ P(k_{1+4}|z_{12})
\ea
The three permutations can be viewed respectively as flat rhyme (aabb), alternate rhyme (abab) and enclosed rhyme (abba).

For the first permutation (flat rhyme, Eq. \ref{Eq:T2h2+2-sqzdiag}), Limber's approximation is justified on $k_1,k_2,k_3,k_4$ but not on the squeezed diagonal $k_{1+2}$, especially since the later aliases into the monopole. Eq. \ref{Eq:Cll'-sqzdiag-limber} must then be used for the projection. One finds
\ba\label{Eq:Cll'-2h2+2-sqz}
\nonumber \mathcal{C}_{\ell,\ell'}^\mr{2h2+2-sqz} = \frac{\frac{2}{\pi} \ \delta_{i_z,j_z} \ \delta_{k_z,l_z}}{4\pi} \int K^2 \dd K \ \dd V_{ab} \ j_0(K x_a) \, j_0(K x_b) \\
\times I_2^1(k_\ell,k_\ell|z_a) \ I_2^1(k_{\ell'},k_{\ell'}|z_b) \ P(K|z_{ab})
\ea
where explicitely $z_a\in i_z$ and $z_b\in k_z$. There are two possible ways to compute this equation numerically, depending whether one first goes with the wavevector integral or the redshift integrals. Respectively
\ba
\mathcal{C}_{\ell,\ell'}^\mr{2h2+2-sqz} =& \frac{\delta_{i_z,j_z} \ \delta_{k_z,l_z}}{4\pi} \int \dd V_{ab} \ I_2^1(k_\ell,k_\ell|z_a) \ I_2^1(k_{\ell'},k_{\ell'}|z_b) \ C_0^m(z_a,z_b) \label{Eq:Cov-2h2+2-sqz} \\
\nonumber =& \frac{\delta_{i_z,j_z} \ \delta_{k_z,l_z}}{4\pi} \ \frac{2}{\pi} \int K^2 \, \dd K \ P(K) \ \mathcal{I}_{0}^{2,1}(K;k_\ell,k_{\ell}|i_z) \\
& \qquad \qquad \times \mathcal{I}_{0}^{2,1}(K;k_{\ell'},k_{\ell'}|k_z)
\ea
where
\ba
C_\ell^m(z_1,z_2) = \frac{2}{\pi} \int k^2 \ \dd k \ j_\ell(k r_1) \, j_\ell(k r_2) P(k|z_{12})
\ea
is the matter angular power spectrum (that would be measure if one could directly see total matter instead of galaxies), that is involved in several other equations below.

For the two other permutations (alternate and enclosed rhymes, Eqs. \ref{Eq:T2h2+2-altdiag13} \& \ref{Eq:T2h2+2-altdiag14}), Limber's approximation is justified on $k_1,k_2,k_3,k_4$ but may not be justified on the alternate diagonal $k_{1+3}$ (resp. $k_{1+4}$). Eq. \ref{Eq:Cll'-altdiag-partial-limber} must then be used for the projection.\\
Then one finds
\ba
\nonumber \mathcal{C}_{\ell,\ell'}^\mr{2h2+2-alt} = \delta_{i_z,k_z} \ \delta_{j_z,l_z} \sum_{\ell_a} \frac{2\ell_a+1}{4\pi} \threeJz{\ell}{\ell'}{\ell_a}^2 \frac{2}{\pi} \int K^2\dd K \; \dd V_{ab} \\
\nonumber j_{\ell_a}(K x_a) \ j_{\ell_a}(K x_b) \ I_2^1(k_{\ell},k_{\ell'}|z_a) \ I_2^1(k_{\ell},k_{\ell'}|z_b) \ P(K|z_{ab})\\
+ \mr{term}(k_{1+4}).
\ea
Again, there are two possible ways to compute this equation numerically, depending whether one first goes for the wavevector integral or the redshift integrals. Respectively
\ba
\nonumber \mathcal{C}_{\ell,\ell'}^\mr{2h2+2-alt} =& \ \left(\delta_{i_z,k_z} \ \delta_{j_z,l_z} + \delta_{i_z,l_z} \ \delta_{j_z,k_z}\right) \sum_{\ell_a} \frac{2\ell_a+1}{4\pi} \threeJz{\ell}{\ell'}{\ell_a}^2 \\
& \quad \times \int \dd V_{ab} \ I_2^1(k_{\ell},k_{\ell'}|z_a) \ I_2^1(k_{\ell},k_{\ell'}|z_b) \ C_{\ell_a}^m(z_a,z_b) \label{Eq:Cov-2h2+2-alt} \\
\nonumber =& \left(\delta_{i_z,k_z} \ \delta_{j_z,l_z} + \delta_{i_z,l_z} \ \delta_{j_z,k_z}\right) \sum_{\ell_a} \frac{2\ell_a+1}{4\pi} \threeJz{\ell}{\ell'}{\ell_a}^2  \\
& \frac{2}{\pi} \int K^2 \, \dd K \ P(K) \ \mathcal{I}_{\ell_a}^{2,1}(K;k_\ell,k_{\ell'}|i_z) \  \mathcal{I}_{\ell_a}^{2,1}(K;k_\ell,k_{\ell'}|j_z)
\ea
where $z_a\in i_z$ and $z_b\in j_z$.\\

\section{Three-halo term}\label{Sect:3halo}
This section is longer than the previous ones due to the increased complexity of the term, and is thus split into smaller subsections for clarity.

\subsection{Trispectrum}\label{Sect:3halo-trispectrum}
This term is the second diagram of Fig.~\ref{Fig:diagrams-trispectrum}. The other diagram containing three different halos (third diagram of Fig.~\ref{Fig:diagrams-trispectrum}) is a shot-noise term which will be treated in Sect.~\ref{Sect:shotnoise}. Applying the diagrammatic rules from Sect.~\ref{Sect:diagrammatic}, the corresponding trispectrum part is:
\ba
\nonumber T^\mr{3h}_\mr{gal}(\kk_{1234},z_{1234}) = \delta_{z_2,z_3} \int & \dd M_{\alpha\beta\gamma} \ \orange{\left.\frac{\dd n_h}{\dd M}\right|_{\alpha,1} \left.\frac{\dd n_h}{\dd M}\right|_{\beta,2} \left.\frac{\dd n_h}{\dd M}\right|_{\gamma,4}} \\
\nonumber & \darkgreen{\lbra N_\mr{gal}\rbra_{\alpha,1} \lbra N_\mr{gal}^{(2)}\rbra_{\beta,2} \lbra N_\mr{gal}\rbra_{\gamma,4}} \\ 
\nonumber & {\color{red} u(k_1|\alpha,1) \ u(k_2|\beta,2) \, u(k_3|\beta,2) \ u(k_4|\gamma,4)}\\
\nonumber & {\color{blue} B_\mr{hhh}(k_1,k_{2+3},k_4|M_{\alpha\beta\gamma},z_{124})} \\
& + \mr{5 \ perm.}
\ea
The halo bispectrum splits into three terms (b2, s2 and 2PT, see Appendix \ref{App:3Dhalopolysp}), and thus the 3h galaxy trispectrum too. Using notations from Appendix \ref{App:3Dhalobispec}, one finds:
\ba
\nonumber T^\mr{3h}_\mr{gal}(\kk_{1234},z_{1234}) &= \delta_{z_2,z_3} \sum_{X\in\{\mr{b2,s2,2PT}\}} \\
\nonumber & \Big[2! \ I_1^1(k_1|z_1) \ I_2^1(k_2,k_3|z_2) \ I_1^X(k_4|z_4) \\
\nonumber & \times K_X(\kk_1,\kk_{2+3}) \ P(k_1|z_{14}) \ P(k_{2+3}|z_{24}) \\
\nonumber & +2! \ I_1^1(k_1|z_1) \ I_2^X(k_2,k_3|z_2) \ I_1^1(k_4|z_4) \\
\nonumber & \times  K_X(\kk_1,\kk_{4}) \ P(k_1|z_{12}) \ P(k_{4}|z_{42}) \\
\nonumber & +2! \ I_1^X(k_1|z_1) \ I_2^1(k_2,k_3|z_2) \ I_1^1(k_4|z_4) \\
\nonumber & \times  K_X(\kk_{2+3},\kk_{4}) \ P(k_{2+3}|z_{21}) \ P(k_{4}|z_{41}) \Big] \\
& \quad + \mr{5 \ perm.}
\ea
This can be rewritten to regroup terms and explicit the 18 permutations, grouping first the six terms involving angles between base wavevectors ($\hk_\alpha\cdot\hk_\beta$), and then the twelve terms involving angles with a diagonal $\kk_{\alpha+\beta}$ :
\ba
\nonumber T^\mr{3h-X}_\mr{gal}(\kk_{1234},z_{1234}) =& \sum_{\{\alpha,\beta\}\in\{1,2,3,4\}}\!\!\!\!\!\! 2 \,\delta_{z_\gamma,z_\delta} \ I_1^1(k_\alpha|z_\alpha) \, I_2^X(k_\gamma,k_\delta|z_\gamma) \, I_1^1(k_\beta|z_\beta) \\
\nonumber & \times K_X(\kk_\alpha,\kk_{\beta}) \, P(k_\alpha|z_{\alpha\gamma}) \, P(k_{\beta}|z_{\beta\gamma}) \\
\nonumber & +\!\!\!\!\!\!\sum_{\{\alpha,\beta\}\in\{1,2,3,4\},\gamma}\!\!\!\!\!\!\!\! 2 \,\delta_{z_\alpha,z_\beta}  \ I_1^X(k_\delta|z_\delta) \ I_2^1(k_\alpha,k_\beta|z_\alpha) \ I_1^1(k_\gamma|z_\gamma) \\
& \times K_X(\kk_{\alpha+\beta},\kk_{\gamma}) \ P(k_{\alpha+\beta}|z_{\alpha\delta}) \ P(k_{\gamma}|z_{\gamma\delta}) \\
\nonumber =& T^\mr{3h-Xbase}_\mr{gal} + T^\mr{3h-Xdiag}_\mr{gal}.
\ea
After Legendre decomposition of the angle dependence, the 2PT term yields three subterms ($n=0,1,2$), while the b2 and s2 terms yield one subterm each ($n=0$ and $n=2$ respectively). Accounting for all permutations I thus have a total of 90 subterms.

\subsection{Covariance}
Let us first compute the contribution coming from the six terms with an angle between base wavevectors $\kk_\alpha\cdot\kk_{\beta}$. Using results from Appendix \ref{App:2Dproj-trisp-angdep-1angle-base}, for $X\in\{\mr{b2,s2,2PT}\}$, one finds
\ba
\nonumber \mathcal{C}_{\ell,\ell'}^\mr{3h-Xbase} =& \sum_{n=0}^2 (-1)^n \sum_{\ell_a} \frac{2\ell_a+1}{4\pi} \threeJz{\ell}{\ell_a}{n}^2 \\
\nonumber & \delta_{k_z,l_z} \left(\frac{2}{\pi}\right)^2 \int k^2_{12} \,\dd k_{12} \ 2 \ \mathcal{I}_\ell^{1,1}(k_1;k_1|i_z) \ \mathcal{I}_\ell^{1,1}(k_2;k_2|j_z) \\
\nonumber & \times \mathcal{I}_{2;\ell_a,\ell_a}^{2,X}(k_1,k_2;k_{\ell'},k_{\ell'}|k_z) \ K_X^n(k_1,k_2) \ P(k_1)\,P(k_2) \\
\label{Eq:Cov-3h-X-base-12} & + (\ell,i_z,j_z\leftrightarrow\ell',k_z,l_z) \\
\nonumber & +(-1)^{\ell+\ell'} \sum_{n=0}^2 (2n+1) \Bigg[\sum_{\lu,\lt} \ii^{\lu+\ell+\lt+\ell'} \ \frac{(2\ell+1)_{13}}{4\pi} \\
\nonumber & \threeJz{\ell}{\lu}{n}^2 \threeJz{\ell'}{\lt}{n}^2 \ \delta_{j_z,l_z} \left(\frac{2}{\pi}\right)^2 \int k^2_{13}\,\dd k_{13} \\
\nonumber & 2 \ \mathcal{I}_\ell^{1,1}(k_1;k_1|i_z) \ \mathcal{I}_{\ell'}^{1,1}(k_3;k_3|k_z) \ \mathcal{I}_{2;\lu,\lt}^{2,X}(k_1,k_3;k_{\ell},k_{\ell'}|j_z) \\
\nonumber & K_X^n(k_1,k_3) \ P(k_1)\,P(k_3) + ({}_3,k_z,l_z\leftrightarrow {}_4,l_z,k_z)\Bigg] \\
\label{Eq:Cov-3h-X-base-13}  & \qquad + ({}_1,i_z,j_z\leftrightarrow {}_2,j_z,i_z)
\ea
where the first term (Eq.~\ref{Eq:Cov-3h-X-base-12}) comes from the pairs (1,2) and (3,4), while the second term (Eq.~\ref{Eq:Cov-3h-X-base-13}) comes from the pairs (1,3), (1,4), (2,3) and (2,4).\\
Limber's approximation can be applied on the first term only if $\ell_a=\ell$, which is the sole contribution only for $n=0$. Similarly, Limber's approximation can be applied to the second term only for $n=0$. I return to the $n=0$ case later.

I tackle here the contribution coming from the twelve terms with an angle with a diagonal $\kk_{\alpha+\beta}\cdot\kk_{\gamma}$. Using results from Appendix \ref{App:2Dproj-trisp-angdep-1angle-diag}, for $X\in\{\mr{b2,s2,2PT}\}$, one finds
\ba
\nonumber \mathcal{C}_{\ell,\ell'}^\mr{3h-Xdiag} =& \sum_{n=0}^2 \Bigg[(-1)^{\ell'} \sum_{\lt} \ii^{\ell'+\lt+n} \frac{2\lt+1}{4\pi} \threeJz{\ell'}{\lt}{n}^2 \\
\nonumber & \delta_{i_z,j_z} \left(\frac{2}{\pi}\right)^2 \int k_3^2\,\dd k_3 \ K^2\,\dd K \ 2 K_X^n(K,k_3) \ \mathcal{I}_{0}^{2,1}(K;k_\ell,k_\ell|i_z) \\
\nonumber & \mathcal{I}_{\ell'}^{1,1}(k_3;k_3|k_z) \ \mathcal{I}_{2;\lt,n}^{1,X}(k_3,K;k_{\ell'}|l_z) \ P(K)\,P(k_3) \\
\label{Eq:Cov-3h-X-diag-sqz} & + ({}_3,k_z,l_z\leftrightarrow {}_4,l_z,k_z) \Bigg] + \left(\substack{{}_1,{}_2,i_z,j_z\leftrightarrow {}_3,{}_4,k_z,l_z \\ {}_3,{}_4,k_z,l_z\leftrightarrow {}_1,{}_2,i_z,j_z}\right) \\
\nonumber &+\sum_{n=0}^2 \Bigg[(-1)^{\ell+\ell'+n} \sum_{\ell_{ab2}} \ii^{\ell_a-\ell+\ld-\ell_b} \frac{(2\ell+1)_{ab2}}{4\pi} \\
\nonumber & \sixJ{\ell_a}{\ell_b}{n}{\ell_2}{\ell}{\ell'} \ \delta_{i_z,k_z} \left(\frac{2}{\pi}\right)^2 \int k_2^2\,\dd k_2 \ K^2\,\dd K \ 2 \ K_X^n(K,k_2) \\
\nonumber & \mathcal{I}_{\ell_a}^{2,1}(K;k_{\ell},k_{\ell'}|i_z) \ \mathcal{I}_{\ell}^{1,1}(k_2;k_2|j_z) \ \mathcal{I}_{2;\ld,\ell_b}^{1,X}(k_2,K;k_{\ell'}|l_z) \\
& P(K)\,P(k_2) + ({}_2,j_z,l_z,\ell,\ell'\leftrightarrow {}_4,l_z,j_z,\ell',\ell)\Bigg] + 3 \ \mr{perm.}
\label{Eq:Cov-3h-X-diag-alt}
\ea
where the first term (\ref{Eq:Cov-3h-X-diag-sqz}) comes from the permutations involving the squeezed diagonal (12-3,12-4,34-1,34-2), while the second term (\ref{Eq:Cov-3h-X-diag-alt}) comes from the permutations involving an alternate diagonal (13-2,13-4,14-2,14-3,23-1,23-4,24-1,24-3)\footnote{the first two permutations (13-2,13-4) are explicited, while the other six are in the `+3 perm.' term.}.

\subsection{Simplifications}
In the $n=0$ case, the covariance gets simpler, as it corresponds to an angle-independent trispectrum. The covariance becomes:
\ba
\nonumber \mathcal{C}_{\ell,\ell'}^\mr{3h-Xbase0} =& \frac{\delta_{k_z,l_z}}{4\pi} \left(\frac{2}{\pi}\right)^2 \int k^2_{12} \,\dd k_{12} \ 2 \ \mathcal{I}_\ell^{1,1}(k_1;k_1|i_z) \ \mathcal{I}_\ell^{1,1}(k_2;k_2|j_z) \\
\nonumber & \times \mathcal{I}_{2;\ell_a,\ell_a}^{2,X}(k_1,k_2;k_{\ell'},k_{\ell'}|k_z) \ K_X^0 \ P(k_1)\,P(k_2) \\
& + (\ell,i_z,j_z\leftrightarrow\ell',k_z,l_z) \\
\nonumber & + \Bigg[\frac{\delta_{j_z,l_z}}{4\pi} \left(\frac{2}{\pi}\right)^2 \int k^2_{13}\,\dd k_{13} \ 2 \ \mathcal{I}_\ell^{1,1}(k_1;k_1|i_z) \\
\nonumber & \mathcal{I}_{\ell'}^{1,1}(k_3;k_3|k_z) \ \mathcal{I}_{2;\lu,\lt}^{2,X}(k_1,k_3;k_{\ell},k_{\ell'}|j_z) \ K_X^0 \ P(k_1)\,P(k_3) \\
& + ({}_3,k_z,l_z\leftrightarrow {}_4,l_z,k_z)\Bigg] + ({}_1,i_z,j_z\leftrightarrow {}_2,j_z,i_z)
\ea
and
\ba
\nonumber \mathcal{C}_{\ell,\ell'}^\mr{3h-Xdiag0} =& \frac{\delta_{i_z,j_z}}{4\pi} \Bigg[\left(\frac{2}{\pi}\right)^2 \int k_3^2\,\dd k_3 \ K^2\,\dd K \ 2 \ K_X^0 \ \mathcal{I}_{0}^{2,1}(K;k_\ell,k_\ell|i_z) \\
\nonumber & \mathcal{I}_{\ell'}^{1,1}(k_3;k_3|k_z) \ \mathcal{I}_{2;\ell',0}^{1,X}(k_3,K;k_{\ell'}|l_z) \ P(K)\,P(k_3) \\
& + ({}_3,k_z,l_z\leftrightarrow {}_4,l_z,k_z) \Bigg] + \left(\substack{{}_1,{}_2,i_z,j_z\leftrightarrow {}_3,{}_4,k_z,l_z \\ {}_3,{}_4,k_z,l_z\leftrightarrow {}_1,{}_2,i_z,j_z}\right) \\
\nonumber &+ \sum_{\ell_a} \frac{2\ell_a+1}{4\pi} \threeJz{\ell}{\ell'}{\ell_a}^2 \Bigg\{ \delta_{i_z,k_z} \Bigg[\left(\frac{2}{\pi}\right)^2 \\
\nonumber & \int k_2^2\,\dd k_2 \ K^2\,\dd K \ 2 \ K_X^0 \ \mathcal{I}_{\ell_a}^{2,1}(K;k_{\ell},k_{\ell'}|i_z) \  \\
\nonumber & \times \mathcal{I}_{\ell}^{1,1}(k_2;k_2|j_z) \ \mathcal{I}_{2;\ell,\ell_a}^{1,X}(k_2,K;k_{\ell'}|l_z) \ P(K)\,P(k_2) \\
& + ({}_2,j_z,l_z,\ell,\ell'\leftrightarrow {}_4,l_z,j_z,\ell',\ell)\Bigg] + 3 \ \mr{perm.} \Bigg\}.
\ea
Limber's approximation can also be used (except on the squeezed diagonal), if one first performs the wavenumber integrals before the redshift integrals (instead of the opposite, that was used in the previous no-Limber equations). The resulting covariance is
\ba
\nonumber \mathcal{C}_{\ell,\ell'}^\mr{3h-Xbase0} = \frac{\delta_{i_z,j_z,k_z,l_z}}{4\pi} \int & \dd V \ 2 \ K_X^0 \ \left(I_1^{1}(k_\ell|z) \ P(k_\ell|z)\right)^2 I_2^{X}(k_{\ell'},k_{\ell'}|z) \\
\nonumber & + \ (\ell\leftrightarrow\ell') \\
\nonumber +\frac{4 \ \delta_{i_z,j_z,k_z,l_z}}{4\pi} \int & \dd V \ 2 \ K_X^0 \ I_1^{1}(k_\ell|z) \ I_1^{1}(k_{\ell'}|z) \ I_2^{X}(k_{\ell},k_{\ell'}|z) \\
& \times P(k_\ell|z) \ P(k_{\ell'}|z)
\ea
and
\ba
\nonumber \mathcal{C}_{\ell,\ell'}^\mr{3h-Xdiag0} =& \frac{\delta_{i_z,j_z} \ \delta_{k_z,l_z}}{4\pi} \int \dd V_{ab} \ 4 \ K_X^0 \ I_1^X(k_{\ell'}|z_b) \ I_1^1(k_{\ell'}|z_b) \\
& \times I_2^1(k_{\ell},k_{\ell}|z_a) \ P(k_{\ell'}|z_b) \ C_0^m(z_a,z_b) + (\ell,{}_a,{}_b\leftrightarrow\ell',{}_b,{}_a) \label{Eq:Cov-3h-Xdiag0-sqz-Limber} \\
\nonumber & + \sum_{\ell_a} \frac{2\ell_a+1}{4\pi} \threeJz{\ell}{\ell'}{\ell_a}^2 \ \left(\delta_{i_z,k_z} \ \delta_{j_z,l_z} + \delta_{i_z,l_z} \ \delta_{j_z,k_z}\right) \\
\nonumber & \int \dd V_{ab} \ I_2^1(k_{\ell},k_{\ell'}|z_a) \ \Big[ 2 \ K_X^0 \ I_1^X(k_{\ell'}|z_b) \ I_1^1(k_{\ell}|z_b) \ P(k_{\ell}|z_b) \\
& + (\ell\leftrightarrow \ell') \Big]\times C_{\ell_a}^m(z_a,z_b) + (z_a\leftrightarrow z_b) \label{Eq:Cov-3h-Xdiag0-alt-Limber}
\ea
where in Eq.~\ref{Eq:Cov-3h-Xdiag0-sqz-Limber} $z_a\in i_z$ and $z_b\in k_z$, while in Eq.~\ref{Eq:Cov-3h-Xdiag0-alt-Limber} $z_a\in i_z$ and $z_b\in j_z$.

One can perform the summation over $X \in \{\mr{b2,s2,2PT}\}$ by introducing the notation
\ba
I_i^{\Sigma_2} \equiv \sum_{X \in \{\mr{b2,s2,2PT}\}} K_X^0 \ I_i^X \label{Eq:def-I^Sigma} = \frac{17}{21} \ I_i^1 + \frac{1}{2} \ I_i^2
\ea
This yields
\ba
\nonumber \mathcal{C}_{\ell,\ell'}^\mr{3h-base0} = \frac{\delta_{i_z,j_z,k_z,l_z}}{4\pi} \int & \dd V \ 2 \;  \left(I_1^{1}(k_\ell|z) \ P(k_\ell|z)\right)^2 I_2^{\Sigma_2}(k_{\ell'},k_{\ell'}|z) \\
\nonumber & + \quad (\ell\leftrightarrow\ell') \\
\nonumber +\frac{4 \ \delta_{i_z,j_z,k_z,l_z}}{4\pi} \int & \dd V \ 2 \ I_1^{1}(k_\ell|z) \ I_1^{1}(k_{\ell'}|z) \ I_2^{\Sigma_2}(k_{\ell},k_{\ell'}|z) \\
& \times P(k_\ell|z) \ P(k_{\ell'}|z)
\ea
and
\ba
\nonumber \mathcal{C}_{\ell,\ell'}^\mr{3h-diag0} =& \frac{\delta_{i_z,j_z} \ \delta_{k_z,l_z}}{4\pi} \int \dd V_{ab} \ 4 \ I_1^{\Sigma_2}(k_{\ell'}|z_b) \ I_1^1(k_{\ell'}|z_b) \, P(k_{\ell'}|z_b) \\
& \times I_2^1(k_{\ell},k_{\ell}|z_a) \ C_0^m(z_a,z_b) + (\ell,{}_a,{}_b\leftrightarrow\ell',{}_b,{}_a) \label{Eq:Cov-3h-diag0-sqz-Limber} \\
\nonumber & + \sum_{\ell_a} \frac{2\ell_a+1}{4\pi} \threeJz{\ell}{\ell'}{\ell_a}^2 \ \left(\delta_{i_z,k_z} \ \delta_{j_z,l_z} + \delta_{i_z,l_z} \ \delta_{j_z,k_z}\right) \\
\nonumber & \int \dd V_{ab} \ I_2^1(k_{\ell},k_{\ell'}|z_a) \ \Big[ 2 \ I_1^{\Sigma_2}(k_{\ell'}|z_b) \ I_1^1(k_{\ell}|z_b) \, P(k_{\ell}|z_b) \\
& + (\ell\leftrightarrow \ell') \Big]\times C_{\ell_a}^m(z_a,z_b) + (z_a\leftrightarrow z_b) \label{Eq:Cov-3h-diag0-alt-Limber}
\ea

\section{Four-halo terms}\label{Sect:4halo}

\subsection{Trispectrum}
This term is the first diagram of Fig.~\ref{Fig:diagrams-trispectrum}. Applying the diagrammatic rules from Sect.~\ref{Sect:diagrammatic}, the corresponding trispectrum part is:
\ba
\nonumber T^\mr{4h}_\mr{gal}(\kk_{1234},z_{1234}) = \int & \dd M_{\alpha\beta\gamma\delta} \ \orange{\left.\frac{\dd n_h}{\dd M}\right|_{\alpha,1} \left.\frac{\dd n_h}{\dd M}\right|_{\beta,2} \left.\frac{\dd n_h}{\dd M}\right|_{\gamma,3} \left.\frac{\dd n_h}{\dd M}\right|_{\delta,4}} \\
\nonumber & \darkgreen{\lbra N_\mr{gal}\rbra_{\alpha,1} \lbra N_\mr{gal}\rbra_{\beta,2} \lbra N_\mr{gal}\rbra_{\gamma,3} \lbra N_\mr{gal}\rbra_{\delta,4}} \\ 
\nonumber &   {\color{red} u(k_1|\alpha,1) u(k_2|\beta,2) \, u(k_3|\gamma,3) \, u(k_4|\delta,4)} \\
\nonumber & {\color{blue} \times \ T_\mr{hhhh}(\kk_{1234}|M_{\alpha\beta\gamma\delta},z_{1234}).}
\ea

Following Appendix \ref{App:3Dhalopolysp}, the halo trispectrum splits into three terms, and thus the 4h galaxy trispectrum too :
\ba
\nonumber T^\mr{4h}_\mr{gal}(\kk_{1234},z_{1234}) = T^\mr{4h-b3} + T^\mr{4h-3PT} + T^\mr{4h-2\times 2}
\ea
where
\ba\label{Eq:T4h-b3}
\nonumber T^\mr{4h-b3}(\kk_{1234},z_{1234}) =& I_1^1(k_1,z_1) \ I_1^1(k_2,z_2) \ I_1^1(k_3,z_3) \ I_1^3(k_4,z_4)\\
& \times P(k_1|z_{14}) \ P(k_2|z_{24}) \ P(k_3|z_{34}) + 3 \ \mr{perm.}
\ea
\ba\label{Eq:T4h-3PT}
\nonumber T^\mr{4h-3PT}(\kk_{1234},z_{1234}) = I_1^1(k_1,z_1) \ I_1^1(k_2,z_2) \ I_1^1(k_3,z_3) \ I_1^1(k_4,z_4) \\
\times \Big[3! \ F_3(\kk_1,\kk_2,\kk_3) \ P(k_1|z_{14}) \ P(k_2|z_{24}) \ P(k_3|z_{34}) + 3 \ \mr{perm.}\Big]
\ea
and
\ba
\nonumber T^\mr{4h-2\times 2}(\kk_{1234},z_{1234}) = \sum_{X,Y\in \{\mr{b2,s2,2PT}\} } I_1^1(k_1,z_1) \ I_1^1(k_2,z_2) \ I_1^X(k_3,z_3) \\
\nonumber I_1^Y(k_4,z_4) \ 4\ K_X(\kk_{1+3},-\kk_1) \ K_Y(\kk_{1+3},\kk_2) \\
P(k_{1+3}|z_{34}) \, P(k_1|z_{13}) \, P(k_2|z_{24}) + 11 \ \mr{perm.}
\ea
where the permutations are explicited for example in Appendix \ref{App:3Dhalotrispec}.\\
After Legendre decomposition of the angles and accounting for all permutations, the b3 term splits into four subterms, the 3PT term into $9\times3\times4=108$ subterms, and the $2\times 2$ term into $25\times 12= 300$ subterms. I thus have a total of 412 subterms to compute.

\subsection{Covariance}
First, the term from third order halo bias is
\ba\label{Eq:Cov-4h-b3}
\nonumber \mathcal{C}_{\ell,\ell'}^\mr{4h-b3} =& \frac{1}{4\pi} \left(\frac{2}{\pi}\right)^3\int k^2_{123} \, \dd k^2_{123} \  \mathcal{I}_\ell^{1,1}(k_1;k_1|i_z) \ \mathcal{I}_\ell^{1,1}(k_2;k_2|j_z) \\
\nonumber & \mathcal{I}_{\ell'}^{1,1}(k_3;k_3|k_z) \ \mathcal{I}_{3;\ell,\ell,\ell'}^{1,3}(k_1,k_2,k_3;k_{\ell'}|l_z) \ P(k_1) \, P(k_2) \, P(k_3)\\
& +3 \ \mr{perm.}
\ea
This term is the simplest as the trispectrum does not have an angle dependence. As such, if valid, the Limber approximation may easily be applied, and gives
\ba
\nonumber \mathcal{C}_{\ell,\ell'}^\mr{4h-b3} =& \frac{2 \ \delta_{i_z,j_z,k_z,l_z}}{4\pi} \int \dd V \ \left(I_1^1(k_{\ell},z)\right)^2 \ I_1^1(k_{\ell'},z) \ I_1^3(k_{\ell'},z) \\
& \times P(k_{\ell}|z) \ P(k_{\ell}|z) \ P(k_{\ell'}|z) \quad + \quad (\ell\leftrightarrow\ell').
\ea

Next, the term from third order perturbation theory splits in two parts: one coming from trispectrum permutations involving the squeezed diagonal $\kk_{1+2}$, and one coming from permutations involving an alternate diagonal $\kk_{1+3}$ or $\kk_{1+4}$.\\
The first part is
\ba
\nonumber \mathcal{C}_{\ell,\ell'}^\mr{4h-3PT-sqz} = \sum_{n,n'} (-1)^{\ell+\ell'} \sum_{\lu,\lt} \ii^{2\lu+\ell'+\lt+n'} \ \frac{(2\ell+1)_{13}}{4\pi} \threeJz{\ell}{\lu}{n}^2 \\
\nonumber \threeJz{\ell'}{\lt}{n'}^2 \left(\frac{2}{\pi}\right)^4 \int k^2_{123} \,\dd k_{123} \ 3! \ F_{3;n,n'}^{1,2;3,1+2}(k_1,k_2,k_3) \\
\nonumber \mathcal{I}_\ell^{1,1}(k_1;k_1|i_z) \ \mathcal{I}_\ell^{1,1}(k_2;k_2|j_z) \ \mathcal{I}_{\ell'}^{1,1}(k_3;k_3|k_z) \ P(k_1) \, P(k_2) \, P(k_3) \\
\nonumber \int \dd V_4 \ G(z_4)^3 \ j_{\lt}(k_3 r_4) \ I_1^1(k_{\ell'},z_4) \int K^2\,\dd K \ \dd V_a \ j_{\lu}(k_1 x_a) \\
\times \ j_{\lu}(k_2 x_a) \ j_{0}(K x_a) \ j_{n'}(K r_4) \quad + \quad 3 \ \mr{perm.}
\ea
I note that the last integral (over $K$ and $x_a$) is purely analytic, however I did not find a closed form expression for it, except in the case $n'=0$ which will be tackled later.\\
The second part of the 3PT term is
\ba
\mathcal{C}_{\ell,\ell'}^\mr{4h-3PT-alt} = \sum_{n,n'} (-1)^{\ell+\ell'} \sum_{\ell_{123ab}} \ii^{\lu+\lt-\ell_a+\ell_b+\ld+\ell'} \ \frac{(2\ell+1)_{123ab}}{4\pi} \\
\nonumber K_{n',\ell_b,\ld}^{\lu,\lt,n;\ell,\ell_a,\ell'} \left(\frac{2}{\pi}\right)^4 \int k^2_{123} \,\dd k_{123} \ 3! \ F_{3;n,n'}^{1,3;2,1+3}(k_1,k_2,k_3) \\
\nonumber \mathcal{I}_\ell^{1,1}(k_1;k_1|i_z) \ \mathcal{I}_\ell^{1,1}(k_2;k_2|j_z) \ \mathcal{I}_{\ell'}^{1,1}(k_3;k_3|k_z) \ P(k_1) \, P(k_2) \, P(k_3) \\
\nonumber \int \dd V_4 \ G(z_4)^3 \ j_{\ld}(k_2 r_4) \ I_1^1(k_{\ell'},z_4) \int K^2\,\dd K \ \dd V_a \ j_{\lu}(k_1 x_a) \\
\times \ j_{\ell_a}(K x_a) \ j_{\ell_b}(K r_4) \ j_{\lt}(k_3 x_a) \quad + \quad 7 \ \mr{perm.}
\ea
where the $K_{n',\ell_b,\ld}^{\lu,\lt,n;\ell,\ell_a,\ell'}$ symbol is related to a contraction of a 12J symbol of the second kind in Appendix \ref{App:K-and-12J}. Again the last integral (over $K$ and $x_a$) is purely analytic with no known closed form expression, except in the case $n'=0$ which will be tackled later.\\

Finally the $2\times2$ trispectrum term also splits in two parts : one coming from trispectrum permutations involving the squeezed diagonal $\kk_{1+2}$, and one coming from permutations involving an alternate diagonal $\kk_{1+3}$ or $\kk_{1+4}$.\\
The first part is
\ba
\nonumber \mathcal{C}_{\ell,\ell'}^\mr{4h-2\times 2-sqz} = \sum_{\substack{X,Y\in \{\mr{b2,s2,2PT}\} \\ n,n'}} (-1)^{\ell+\ell'} \sum_{\ell_{13}} \ii^{\lu+\ell+n+n'+\lt+\ell'} \ \frac{(2\ell+1)_{13}}{4\pi} \\ 
\nonumber \threeJz{\ell}{\lu}{n}^2 \ \threeJz{\ell'}{\lt}{n'}^2 \ \left(\frac{2}{\pi}\right)^3 \int k^2_{13}\,\dd k_{13} \ K^2\,\dd K \\
\nonumber 4 \, K_{X,n}(K,k_1) \, K_{Y,n'}(K,k_3) \ \mathcal{I}_\ell^{1,1}(k_1;k_1|i_z) \ \mathcal{I}_{2;\lu,n}^{1,X}(k_1,K;k_\ell|j_z) \\
\nonumber \times \ \mathcal{I}_{\ell'}^{1,1}(k_3;k_3|k_z) \ \mathcal{I}_{2;n',\lt}^{1,Y}(K,k_3;k_{\ell'}|l_z) \ P(K) \, P(k_1) \, P(k_3) \\
+ \ 3 \ \mr{perm.}
\ea
The second part of the $2\times2$ term is
\ba
\nonumber \mathcal{C}_{\ell,\ell'}^\mr{4h-2\times 2-alt} = \sum_{\substack{X,Y\in \{\mr{b2,s2,2PT}\} \\ n,n'}} (-1)^{\ell+n} \sum_{\ell_{12abc}} \ii^{\lu-\ell_a+\ell_b+\ld} \frac{(2\ell+1)_{12abc}}{4\pi} \\
\nonumber J_{\ell',\lu,\ell_a}^{\ld,\ell,n';\ell_b,n,\ell_c} \ \left(\frac{2}{\pi}\right)^3 \int k^2_{12}\,\dd k_{12} \ K^2\,\dd K \ 4 \, K_{X,n}(K,k_1) \\
\nonumber K_{Y,n'}(K,k_2) \ P(K) \, P(k_1) \, P(k_2) \ \mathcal{I}_\ell^{1,1}(k_1;k_1|i_z) \\
\nonumber \mathcal{I}_\ell^{1,1}(k_2;k_2|j_z) \ \mathcal{I}_{2;\lu,\ell_a}^{1,X}(k_1,K;k_{\ell'}|k_z) \ \mathcal{I}_{2;\ld,\ell_b}^{1,Y}(k_2,K;k_{\ell'}|l_z) \\
+ \ 7 \ \mr{perm.}
\ea

\subsection{Simplifications}
For some cases of the Legendre decomposition, the covariance equations get simpler; I tackle these cases here.\\\ 

First, beginning with 3PT terms, when $n'=0$, the 3PT-squeezed term simplifies to
\ba
\nonumber \mathcal{C}_{\ell,\ell'}^\mr{4h-3PT-sqz0} = \sum_{n} (-1)^{\ell} \sum_{\lu} \ii^{2\lu} \ \frac{(2\lu+1)}{4\pi} \threeJz{\ell}{\lu}{n}^2 \qquad \\
\nonumber \left(\frac{2}{\pi}\right)^3 \int k^2_{123} \,\dd k_{123} \ 3! \ F_{3;n,0}^{1,2;3,1+2}(k_1,k_2,k_3) \ \mathcal{I}_\ell^{1,1}(k_1;k_1|i_z) \\
\nonumber \mathcal{I}_\ell^{1,1}(k_2;k_2|j_z) \ \mathcal{I}_{\ell'}^{1,1}(k_3;k_3|k_z) \ \mathcal{I}_{3;\lu,\lu,\ell'}^{1,1}(k_1,k_2,k_3;k_{\ell'}|l_z) \\
\times \ P(k_1) \, P(k_2) \, P(k_3) \quad + \quad 3 \ \mr{perm.}
\ea
If I further set $n=0$, I have
\ba
\nonumber \mathcal{C}_{\ell,\ell'}^\mr{4h-3PT-sqz00} = \left(\frac{2}{\pi}\right)^3 \int k^2_{123} \,\dd k_{123} \ 3! \ F_{3;0,0}^{1,2;3,1+2} \ \mathcal{I}_\ell^{1,1}(k_1;k_1|i_z) \qquad \\
\nonumber \mathcal{I}_\ell^{1,1}(k_2;k_2|j_z) \ \mathcal{I}_{\ell'}^{1,1}(k_3;k_3|k_z) \ \mathcal{I}_{3;\ell,\ell,\ell'}^{1,1}(k_1,k_2,k_3;k_{\ell'}|l_z) \\
\times \ P(k_1) \, P(k_2) \, P(k_3) \quad + \quad 3 \ \mr{perm.}
\ea
and Limber's approximation can be used on all wavevectors to yield:
\ba
\nonumber \mathcal{C}_{\ell,\ell'}^\mr{4h-3PT-sqz00} = \frac{2 \ \delta_{i_z,j_z,k_z,l_z}}{4\pi} \int \dd V \ 3! \ F_{3;0,0} \left(I_1^1(k_{\ell},z)\right)^2 \left(I_1^1(k_{\ell'},z)\right)^2 \\
\times \ P(k_{\ell}|z) \ P(k_{\ell}|z) \ P(k_{\ell'}|z) \quad + \quad (\ell\leftrightarrow\ell')
\ea
where I used the fact that $F_{3;0,0}$ is independent of its arguments and superscripts (see Appendix \ref{App:3Dhalotrispec}).\\
If $n'=0$, the 3PT-alternate term also simplifies:
\ba
\nonumber \mathcal{C}_{\ell,\ell'}^\mr{4h-3PT-alt0} = \sum_{n} (-1)^{\ell+\ell'} \sum_{\ell_{13a}} \ii^{\lu+\lt+\ell+\ell'} \ \frac{(2\ell+1)_{13a}}{4\pi} \sixJ{\ell}{\ell'}{\ell_a}{\lt}{\lu}{n} \\
\nonumber \left(\frac{2}{\pi}\right)^3 \int k^2_{123}\,\dd k_{123} \ 3! \ F_{3;n,0}^{1,3;2,1+3}(k_1,k_2,k_3) \ \mathcal{I}_\ell^{1,1}(k_1;k_1|i_z) \\
\nonumber \mathcal{I}_\ell^{1,1}(k_2;k_2|j_z) \ \mathcal{I}_{\ell'}^{1,1}(k_3;k_3|k_z) \ \mathcal{I}_{3;\lu,\ell,\lt}^{1,1}(k_1,k_2,k_3;k_{\ell'}|l_z)  \\
\times \ P(k_1) \, P(k_2) \, P(k_3) \quad + \quad 7 \ \mr{perm.}
\ea
when I further set $n=0$
\ba
\nonumber \mathcal{C}_{\ell,\ell'}^\mr{4h-3PT-alt00} = \frac{1}{4\pi} \left(\frac{2}{\pi}\right)^3 \int k^2_{123}\,\dd k_{123} \ 3! \ F_{3;0,0}^{1,3;2,1+3} \ \mathcal{I}_\ell^{1,1}(k_1;k_1|i_z) \\
\nonumber  \ \mathcal{I}_\ell^{1,1}(k_2;k_2|j_z) \ \mathcal{I}_{\ell'}^{1,1}(k_3;k_3|k_z) \ \mathcal{I}_{3;\ell,\ell,\ell'}^{1,1}(k_1,k_2,k_3;k_{\ell'}|l_z) \\
\times \ P(k_1) \, P(k_2) \, P(k_3) \quad + \quad 7 \ \mr{perm.}
\ea
and Limber's approximation can be used on all wavevectors to yield:
\ba
\nonumber \mathcal{C}_{\ell,\ell'}^\mr{4h-3PT-alt00} = \frac{4 \ \delta_{i_z,j_z,k_z,l_z}}{4\pi} \int \dd V \ 3! \ F_{3;0,0} \left(I_1^1(k_{\ell},z)\right)^2 \left(I_1^1(k_{\ell'},z)\right)^2 \\
\times \ P(k_{\ell}|z) \ P(k_{\ell}|z) \ P(k_{\ell'}|z) \quad + \quad (\ell\leftrightarrow\ell')
\ea
This result is exactly twice that of the squeezed term\footnote{this can already be seen to hold before the Limber approximation} : $\mathcal{C}_{\ell,\ell'}^\mr{4h-3PT-alt00}=2\times\mathcal{C}_{\ell,\ell'}^\mr{4h-3PT-sqz00}$. As such, the two terms can be grouped together :
\ba
\nonumber \mathcal{C}_{\ell,\ell'}^\mr{4h-3PT-00} = \frac{6 \ \delta_{i_z,j_z,k_z,l_z}}{4\pi} \int \dd V \ 3! \ F_{3;0,0} \left(I_1^1(k_{\ell},z)\right)^2 \left(I_1^1(k_{\ell'},z)\right)^2 \\
\times \ P(k_{\ell}|z) \ P(k_{\ell}|z) \ P(k_{\ell'}|z) \quad + \quad (\ell\leftrightarrow\ell')
\ea
Another straighter way to get to this equation is that taking $(n,n')=(0,0)$ is equivalent to inputting $F_3(\kk_1,\kk_2,\kk_3)= 3 \, F_{3;0,0}$ in Eq.~\ref{Eq:T4h-3PT} and realising the analogy with the 4h-b3 case. Thus Eq.~\ref{Eq:Cov-4h-b3} can be used with the replacements $I_1^3 \rightarrow 3! \ 3 \, F_{3;0,0} \times I_1^1 $. This remarks allows us to unify the b3 and 3PT terms into a single equation:
\ba
\nonumber \mathcal{C}_{\ell,\ell'}^\mr{4h-3} = \frac{2 \ \delta_{i_z,j_z,k_z,l_z}}{4\pi} \int \dd V \ 3! \ \left(I_1^1(k_{\ell},z)\right)^2 \ I_1^1(k_{\ell'},z) \ I_1^{\Sigma_3}(k_{\ell'},z) \\
\times \ P(k_{\ell}|z) \ P(k_{\ell}|z) \ P(k_{\ell'}|z) \quad + \quad (\ell\leftrightarrow\ell')
\ea
where
\ba
I_1^{\Sigma_3}(k,z) \equiv 3 \, F_{3;0,0} \ I_1^{1}(k,z) + \frac{1}{3!} \ I_1^{3}(k,z)
\ea
with $3 \, F_{3;0,0}=\frac{1023}{1701}$.\\\

Second, let's look now at $2\times 2$ terms which give symmetric roles to $n$ and $n'$ and thus significantly simplify only when both are zero. When $n=n'=0$, the $2\times 2$-squeezed term simplifies to
\ba
\nonumber \mathcal{C}_{\ell,\ell'}^\mr{4h-2\times 2-sqz00} =& \frac{1}{4\pi} \!\sum_{X,Y\in \{\mr{b2,s2,2PT}\} }\! \left(\frac{2}{\pi}\right)^3 \int k^2_{13}\,\dd k_{13} \ K^2\,\dd K \ 4 \ K_{X,0} \ K_{Y,0} \\
\nonumber & \mathcal{I}_\ell^{1,1}(k_1;k_1|i_z) \ \mathcal{I}_{2;\ell,0}^{1,X}(k_1,K;k_\ell|j_z) \ \mathcal{I}_{\ell'}^{1,1}(k_3;k_3|k_z) \\
& \mathcal{I}_{2;0,\ell'}^{1,Y}(K,k_3;k_{\ell'}|l_z) \ P(K) \ P(k_1) \ P(k_3) \ + \ 3 \ \mr{perm.}
\ea
Using Limber's approximation on $k_1,k_3$, and definition \ref{Eq:def-I^Sigma} of $I_i^{\Sigma_2}$, one finds:
\ba
\nonumber \mathcal{C}_{\ell,\ell'}^\mr{4h-2\times 2-sqz00} = \frac{\delta_{i_z,j_z} \ \delta_{k_z,l_z}}{4\pi} \int \dd V_{ab} \ 4 \ I_1^{\Sigma_2}(k_{\ell},z_a) \ I_1^1(k_{\ell},z_a) \, P(k_{\ell},z_a) \\
\times \ 4 \ I_1^{\Sigma_2}(k_{\ell'},z_b) \ I_1^1(k_{\ell'},z_b) \, P(k_{\ell'},z_b) \ C_{0}^m(z_a,z_b) \label{Eq:Cov-4h-2X2-sqz00-Limber}
\ea
where $z_a\in i_z$, $z_b\in k_z$.\\
The $2\times 2$-alternate term also simplifies when $n=n'=0$
\ba
\nonumber \mathcal{C}_{\ell,\ell'}^\mr{4h-2\times 2-alt00} = \!\!\!\!\!\!\!\!\sum_{\substack{\ell_a \\ X,Y\in \{\mr{b2,s2,2PT}\} }}\!\!\!\!\!\!\!\! \frac{(2\ell_a+1)}{4\pi} \threeJz{\ell}{\ell'}{\ell_a}^2 \left(\frac{2}{\pi}\right)^3 \int k^2_{12}\ \dd k_{12} \\
\nonumber K^2\ \dd K \ 4 \ K_{X,0} \ K_{Y,0} \ \mathcal{I}_\ell^{1,1}(k_1;k_1|i_z) \ \mathcal{I}_\ell^{1,1}(k_2;k_2|j_z) \\
\nonumber \mathcal{I}_{2;\ell,\ell_a}^{1,X}(k_1,K;k_{\ell'}|k_z) \, \mathcal{I}_{2;\ell,\ell_b}^{1,Y}(k_2,K;k_{\ell'}|l_z) \ P(K) \, P(k_1) \, P(k_2) \\
+ \quad 7 \ \mr{perm.}
\ea
Using Limber's approximation on $k_1,k_2$, and definition \ref{Eq:def-I^Sigma} of $I_i^{\Sigma_2}$, one finds:
\ba
\nonumber \mathcal{C}_{\ell,\ell'}^\mr{4h-2\times 2-alt00} =& \left(\delta_{i_z,k_z}\ \delta_{j_z,l_z}+\delta_{i_z,l_z}\ \delta_{j_z,k_z}\right) \ \sum_{\ell_a} \frac{(2\ell_a+1)}{4\pi} \threeJz{\ell}{\ell'}{\ell_a}^2 \\
\nonumber & \int \dd V_{ab} \ \left[2 \ I_1^{\Sigma_2}(k_{\ell'},z_a) \ I_1^1(k_{\ell},z_a) \  P(k_{\ell},z_a) + (\ell\leftrightarrow\ell')\right] \\
\nonumber & \times \left[2 \ I_1^{\Sigma_2}(k_{\ell},z_b) \ I_1^1(k_{\ell'},z_b) \  P(k_{\ell'},z_b) + (\ell\leftrightarrow\ell')\right]  \\
& \times \ C_{\ell_a}^m(z_a,z_b) \label{Eq:Cov-4h-2X2-alt00-Limber}
\ea
where $z_a\in i_z$, $z_b\in j_z$.

\section{Shot-noise}\label{Sect:shotnoise}

\subsection{one-galaxy shot-noise}\label{Sect:shot1g}
This term is the last diagram of Fig. \ref{Fig:diagrams-trispectrum}. Applying the diagrammatic rules from Sect. \ref{Sect:diagrammatic}, the corresponding trispectrum part is:
\ba
\nonumber T^\mr{shot1g}_\mr{gal}(\kk_{1234},z_{1234}) &= \int \dd M \ \orange{\frac{\dd n_h}{\dd M}} \ \darkgreen{\lbra N_\mr{gal}\rbra} \ {\color{red} u(k_{1+\cdots+4}|M,z)} \  {\color{blue} \times 1}\\
\nonumber &= \delta_{z_1,z_2,z_3,z_4} \ I_1^0(k_{1+\cdots+4}|z) \\
&= \delta_{z_1,z_2,z_3,z_4} \  \nbargal(z)
\ea
since $\kk_1+\cdots+\kk_4=0$ and the halo profile is normalised to $u(0|M,z)=1$.

The corresponding covariance term is 
\ba
\mathcal{C}_{\ell,\ell'}^\mr{shot1g} = \frac{\delta_{i_z,j_z,k_z,l_z}}{4\pi} \ n_\mr{gal}(i_z)
\ea

\subsection{Two-galaxy shot-noise}\label{Sect:shot2g}
These terms are the 8th, 9th, 12th, and 13th diagrams of Fig. \ref{Fig:diagrams-trispectrum}. One could apply the diagrammatic rules from Sect. \ref{Sect:diagrammatic} to write down the corresponding trispectrum part; however it was realised by \cite{Lacasa2014} that these diagrams are identical to ones of lower order polyspectra. For instance for the 8th diagram (2ha-shot2g)
\be
T^\mr{2ha-shot2g}_\mr{gal}(\kk_{1234},z_{1234}) = P^\mr{2h}_\mr{gal}(k_{1+2},z_{13}) + 2 \ \mr{perm.}
\ee
These diagrams can then be resummed to reveal the clustering part of the lower order polyspectrum (here the power spectrum). Noting $P^\mr{clust}_\mr{gal}=P^\mr{2h}_\mr{gal}+P^\mr{1h}_\mr{gal}$ and writing down explicitely all involved permutations of (1234), one finds:
\ba
\nonumber T^\mr{shot2g}_\mr{gal}(\kk_{1234},z_{1234}) =& \ T^\mr{2hb-shot2g} + T^\mr{1hb-shot2g} \\
\nonumber & + T^\mr{2ha-shot2g} + T^\mr{1ha-shot2g} \\
\nonumber =& \ \delta_{z_2,z_3,z_4} \, P^\mr{clust}_\mr{gal}(k_1,z_{12}) + \delta_{z_1,z_3,z_4} \, P^\mr{clust}_\mr{gal}(k_2,z_{12}) \\
\nonumber & + \ \delta_{z_1,z_2,z_4} \, P^\mr{clust}_\mr{gal}(k_3,z_{13}) + \delta_{z_1,z_2,z_3} \, P^\mr{clust}_\mr{gal}(k_4,z_{14})\\
\nonumber & + \delta_{z_1,z_3} \ \delta_{z_2,z_4} \ P^\mr{clust}_\mr{gal}(k_{1+3},z_{12}) \\
\nonumber & + \delta_{z_1,z_4} \ \delta_{z_2,z_3} \ P^\mr{clust}_\mr{gal}(k_{1+4},z_{12}) \\
&+ \ \delta_{z_1,z_2} \ \delta_{z_3,z_4} \ P^\mr{clust}_\mr{gal}(k_{1+2},z_{13}) \label{Eq:Tshot2g-resummed}
\ea
where the first two lines in Eq. \ref{Eq:Tshot2g-resummed} come from "1+3" diagrams (2hb-shot2g and 1hb-shot2g, respectively 9th and 13th diagrams in Fig. \ref{Fig:diagrams-trispectrum}), and the last three lines come from "2+2" diagrams (2ha-shot2g and 1ha-shot2g, respectively 8th and 12th diagrams in Fig. \ref{Fig:diagrams-trispectrum}).

The corresponding covariance is 
\ba
\nonumber \mathcal{C}_{\ell,\ell'}^\mr{shot2g} =& \frac{\delta_{j_z,k_z,l_z}+\delta_{i_z,k_z,l_z}}{4\pi} \ C_\ell^\mr{gal,clust}(i_z,j_z) \\
\nonumber & + \frac{\delta_{j_z,k_z,l_z}+\delta_{i_z,k_z,l_z}}{4\pi} \ C_{\ell'}^\mr{gal,clust}(k_z,l_z) \\
\nonumber & + \left(\delta_{i_z,k_z} \ \delta_{j_z,l_z}+ \delta_{i_z,l_z} \ \delta_{j_z,k_z}\right) \sum_{\ell_a} \frac{2\ell_a+1}{4\pi} \threeJz{\ell}{\ell'}{\ell_a}^2 \\
& \times \ C_{\ell_a}^\mr{gal,clust}(i_z,j_z) \label{Eq:Cov-shot2g-alt} \\
& + \frac{\delta_{i_z,j_z} \ \delta_{k_z,l_z}}{4\pi} \ C_{0}^\mr{gal,clust}(i_z,k_z) \label{Eq:Cov-shot2g-sqz}
\ea
where $C_{\ell}^\mr{gal,clust}$ is the clustering part (i.e. without shot-noise) of the galaxy angular power spectrum

\subsection{Three-galaxy shot-noise}\label{Sect:shot3g}
These terms are the 3rd, 6th and 11th diagrams of Fig. \ref{Fig:diagrams-trispectrum} (3h-shot3g, 2h-shot3g and 1h-shot3g). As in the previous subsection, I remark that these diagrams are identical to ones of the galaxy bispectrum and can be resummed. Noting $B^\mr{clust}_\mr{gal}=B^\mr{3h}_\mr{gal}+B^\mr{2h}_\mr{gal}+B^\mr{1h}_\mr{gal}$ and writing explicitly all involved permutations of (1234), one finds:
\ba
\nonumber T^\mr{shot3g}_\mr{gal}(\kk_{1234},z_{1234}) =
\delta_{z_1,z_2} \ B^\mr{clust}_\mr{gal}(k_{1+2},k_3,k_4,z_{134})\\
\nonumber + \delta_{z_1,z_3} \ B^\mr{clust}_\mr{gal}(k_{1+3},k_2,k_4,z_{124})\\
\nonumber + \delta_{z_1,z_4} \ B^\mr{clust}_\mr{gal}(k_{1+4},k_2,k_3,z_{123})\\
\nonumber + \delta_{z_2,z_3} \ B^\mr{clust}_\mr{gal}(k_1,k_{2+3},k_4,z_{124})\\
\nonumber + \delta_{z_2,z_4} \ B^\mr{clust}_\mr{gal}(k_1,k_{2+4},k_3,z_{123})\\
+ \delta_{z_3,z_4} \ B^\mr{clust}_\mr{gal}(k_1,k_2,k_{3+4},z_{123})
\ea

The corresponding covariance is (first with terms in the same order)
\ba
\nonumber \mathcal{C}_{\ell,\ell'}^\mr{shot3g} &= \frac{\delta_{i_z,j_z}}{4\pi} \ b_{0,\ell',\ell'}^\mr{gal,clust}(i_z,k_z,l_z) \\
\nonumber &+ \delta_{i_z,k_z} \sum_{\ell_a} \frac{2\ell_a+1}{4\pi} \threeJz{\ell}{\ell'}{\ell_a}^2 \ b_{\ell_a,\ell,\ell'}^\mr{gal,clust}(i_z,j_z,l_z) \\
\nonumber &+ \delta_{i_z,l_z} \sum_{\ell_a} \frac{2\ell_a+1}{4\pi} \threeJz{\ell}{\ell'}{\ell_a}^2 \ b_{\ell_a,\ell,\ell'}^\mr{gal,clust}(i_z,j_z,k_z) \\
\nonumber &+ \delta_{j_z,k_z} \sum_{\ell_a} \frac{2\ell_a+1}{4\pi} \threeJz{\ell}{\ell'}{\ell_a}^2 \ b_{\ell,\ell_a,\ell'}^\mr{gal,clust}(i_z,j_z,l_z) \\
\nonumber &+ \delta_{j_z,l_z} \sum_{\ell_a} \frac{2\ell_a+1}{4\pi} \threeJz{\ell}{\ell'}{\ell_a}^2 \ b_{\ell,\ell_a,\ell'}^\mr{gal,clust}(i_z,j_z,k_z) \\
\nonumber &+ \frac{\delta_{k_z,l_z}}{4\pi} \ b_{0,\ell,\ell}^\mr{gal,clust}(i_z,j_z,k_z) \\
&= \frac{\delta_{i_z,j_z}}{4\pi} \ b_{0,\ell',\ell'}^\mr{gal,clust}(i_z,k_z,l_z) + \frac{\delta_{k_z,l_z}}{4\pi} \ b_{0,\ell,\ell}^\mr{gal,clust}(i_z,j_z,k_z) \label{Eq:Cov-shot3g-sqz} \\
\nonumber &+ \sum_{\ell_a} \frac{2\ell_a+1}{4\pi} \threeJz{\ell}{\ell'}{\ell_a}^2 \Big[
\delta_{i_z,k_z} \ b_{\ell_a,\ell,\ell'}^\mr{gal,clust}(i_z,j_z,l_z) \\
\nonumber & + \delta_{i_z,l_z} \ b_{\ell_a,\ell,\ell'}^\mr{gal,clust}(i_z,j_z,k_z) + \delta_{j_z,k_z} \ b_{\ell_a,\ell,\ell'}^\mr{gal,clust}(j_z,i_z,l_z) \\
& + \delta_{j_z,l_z} \ b_{\ell_a,\ell,\ell'}^\mr{gal,clust}(j_z,i_z,k_z)
\Big] \label{Eq:Cov-shot3g-alt}
\ea
where $b_{\ell_1,\ell_2,\ell_3}^\mr{gal,clust}$ is the clustering part (i.e. without shot noise) of the galaxy angular bispectrum.

\subsection{Shot-noise subtraction}\label{Sect:shot-subs}

The shot-noise contribution to the power spectrum is $C_\ell^\mr{shot}(i_z,j_z) = \delta_{i_z,j_z} \ n_\mr{gal}^\mr{obs}(i_z)$ where $n_\mr{gal}^\mr{obs}$ is the actual number of galaxies in the survey (in a redshift bin, and per steradian). This number is perfectly known, and thus it can be subtracted from the measurement in order to reveal power up to smaller scales. This subtraction is indeed actually applied in power spectrum measurement by past and current surveys working with the relative fluctuation $\delta_\mr{gal}$: the corrected spectrum is $\tilde{C}_\ell = C_\ell^\mr{obs}-1/\nbargal$.

Naively this subtraction shouldnt affect the covariance, since covariances are invariant under addition of a constant. However it does affect it, because one is subtracting the actual number of galaxies, not the model-predicted number. The actual number of galaxies is itself a random variable, so subtracting it will add covariance terms. This random number is, in fact, positively correlated with the galaxy power spectrum measurement, thus the spectrum covariance will be reduced by the shot-noise subtraction.

Explicitly, the shot-noise subtraction removes several of the power spectrum covariance terms. There are two equivalent ways to see which terms are going to be canceled.
The first way uses the fact that the shot-noise contribution to the power spectrum corresponds to the diagram with coinciding galaxies (right most diagram in Fig.~\ref{Fig:diagrams-spectrum}), thus subtracting shot-noise corresponds to only taking the two other spectrum diagrams, that is, forbidding galaxies 1 and 2 to coincide. At the covariance level, galaxies 1 and 2 correspond to $C_\ell$ and galaxies 3 and 4 correspond to $C_{\ell'}$ ; shot-noise subtraction (for both $C_\ell$ and $C_{\ell'}$) thus corresponds to forbidding diagrams with a coincidence 1=2 and/or a coincidence 3=4. This removes many of the terms presented in the previous sections \ref{Sect:shot1g}, \ref{Sect:shot2g} and \ref{Sect:shot3g}, however there are still terms remaining, corresponding to the coincidences 1=3 or instance.
The second equivalent way goes through real-space: in $w(\theta)$, shot-noise corresponds to a dirac at $\theta=0$. So shot-noise subtraction will kill all terms that yield a dirac at $\theta=0$ and/or $\theta'=0$ in the real-space covariance $C_{\theta,\theta'}$. Harmonic transforming this back to $\mathcal{C}_{\ell,\ell'}$, this corresponds to all terms which have no dependence on at least one of the multipole. For example, in the shot3g term (Sect.~\ref{Sect:shot3g}) $\frac{\delta_{i_z,j_z}}{4\pi} b^\mr{gal,clust}_{0,\ell',\ell'}$ has no dependence on $\ell$, and thus will be canceled by the shot-noise subtraction.

Through either equivalent way, one finds that the following shot-noise covariance terms are canceled by shot-noise subtraction:
\begin{itemize}
\item $\mathcal{C}_{\ell,\ell'}^\mr{shot1g}$
\item the "1+3" part of $\mathcal{C}_{\ell,\ell'}^\mr{shot2g}$ : $\frac{\delta_{j_z,k_z,l_z}+\delta_{i_z,k_z,l_z}}{4\pi} \ C_\ell^\mr{gal,clust}(i_z,j_z)$ and the symmetric term in $\ell'$
\item the squeezed part of $\mathcal{C}_{\ell,\ell'}^\mr{shot2g}$ : $\frac{\delta_{i_z,j_z} \ \delta_{k_z,l_z}}{4\pi} \ C_{0}^\mr{gal,clust}(i_z,k_z)$
\item the squeezed part of $\mathcal{C}_{\ell,\ell'}^\mr{shot3g}$ : $\frac{\delta_{k_z,l_z}}{4\pi} \ b_{0,\ell,\ell}^\mr{gal,clust}(i_z,j_z,k_z)$ and the symmetric term in $\ell'$
\end{itemize}
The following covariance terms are however still present: 
\begin{itemize} 
\item the alternate part of $\mathcal{C}_{\ell,\ell'}^\mr{shot2g}$ :
$$\left(\delta_{j_z,l_z}+ \delta_{i_z,l_z} \ \delta_{j_z,k_z}\right) \sum_{\ell_a} \frac{2\ell_a+1}{4\pi} \threeJz{\ell}{\ell'}{\ell_a}^2 \ C_{\ell_a}^\mr{gal,clust}(i_z,j_z)$$
\item the alternate part of $\mathcal{C}_{\ell,\ell'}^\mr{shot3g}$ : 
$$\delta_{i_z,k_z} \sum_{\ell_a} \frac{2\ell_a+1}{4\pi} \threeJz{\ell}{\ell'}{\ell_a}^2 \ b_{\ell_a,\ell,\ell'}^\mr{gal,clust}(i_z,j_z,l_z) \ + \ 3 \ \mr{perm.}$$
\end{itemize}

In summary, most shot-noise effects are canceled out. Most, but not all: a small group of terms, later participating in the braiding covariance (see Sect.~\ref{Sect:discu-braiding}), resist because one cannot erase the discreteness of the galaxy density field. Inference from a discretely sampled field cannot be the same as from the underlying continuous field.

\section{Discussion of the results}\label{Sect:discu}

\subsection{Super-sample covariance}\label{Sect:discu-SSC}

Super-sample covariance (SSC) has been studied in the past literature, mostly for 3D surveys \citep[e.g.][]{Takada2013} but then also in spherical harmonics: \cite{Lacasa2016} derived its impact on the cross-covariance between cluster counts and the galaxy angular power spectrum. However its impact on the auto-covariance of an angular power spectrum has never been derived rigorously, but was surmised using the 3D results or the cross-covariance result. Here I show that SSC does emerge naturally from the halo model derivation, and does recover the postulation from \cite{Lacasa2016} based on the cross-covariance result. I use only equations after Limber's approximation, leaving the issue of no-Limber SSC to be tackled in future works. Also, I first tackle in Sect. \ref{Sect:discu-SSC-angindep} the simpler case of SSC terms coming from angle-independent trispectra ($n=0$), which will be the terms going into the summary Sect.~\ref{Sect:summary}. Then in Sect. \ref{Sect:discu-SSC-angdep} I will discuss the case of angle-dependent terms ($n=1,2$) and how to generalise SSC to other modelisation and partial sky coverage.

\subsubsection{Angle-independent terms}\label{Sect:discu-SSC-angindep}

We see SSC emerging when grouping all covariance terms where the trispectrum has a dependence on the squeezed diagonal through a $P(k_{1+2})$, making a $C_{0}^m(z_a,z_b)$ appear in the covariance. Specifically, there is such dependence in the 2h2+2-sqz term, Eq.~\ref{Eq:Cov-2h2+2-sqz}, in the 3h-diag-sqz0 term, Eq.~\ref{Eq:Cov-3h-Xdiag0-sqz-Limber}, and in the 4h-2X2-sqz00 term, Eq.~\ref{Eq:Cov-4h-2X2-sqz00-Limber}. There are also shot-noise contributions which need a bit more work to yield a form unifiable with the clustering terms.\\
The first shot-noise contribution comes from the two-halo part of Eq.~\ref{Eq:Cov-shot2g-sqz} 
\ba
\nonumber \mathcal{C}_{\ell,\ell'}^\mr{shot2g-sqz-2h} = & \frac{\delta_{i_z,j_z} \ \delta_{k_z,l_z}}{4\pi} \ C_{0}^\mr{gal,2h}(i_z,k_z) \\
\simeq & \frac{\delta_{i_z,j_z} \ \delta_{k_z,l_z}}{4\pi} \ \int \dd V_{ab} \ I_1^1(0|z_a) \ I_1^1(0|z_b) \ C_{0}^m(z_a,z_b)
\ea
where I assumed that $k_{1+2}$ is sufficiently small to neglect the scale dependence of $I_1^1$, meaning that galaxy bias can be considered constant for super-survey modes, a reasonable approximation.
The second shot-noise contribution comes from the two- and three-halo parts of Eq.~\ref{Eq:Cov-shot3g-sqz}.
\ba
\nonumber \mathcal{C}_{\ell,\ell'}^\mr{shot3g-sqz-23h} =& \frac{\delta_{i_z,j_z}}{4\pi} \left( b_{0,\ell',\ell'}^\mr{gal,2h}(i_z,k_z,l_z) + b_{0,\ell',\ell'}^\mr{gal,3h}(i_z,k_z,l_z) \right) + \mr{perm.} \\
\simeq & \frac{\delta_{i_z,j_z}}{4\pi} \int \dd V_{ab} \ I_1^1(0|z_a) \ I_2^1(k_{\ell'},k_{\ell'}|z_b) \ C_{0}^m(z_a,z_b) \\
\nonumber & + \frac{\delta_{i_z,j_z}}{4\pi} \sum_{X\in \{\mr{b2,s2,2PT}\} } \int \dd V_{ab} \ I_1^1(0|z_a) \ 4 \ K_X^0 \ I_1^X(k_{\ell'}|z_b) \\
\nonumber & \qquad \times \ I_1^1(k_{\ell'}|z_b) \ C_{0}^m(z_a,z_b) \\
& + \mr{perm.}
\ea

Combining all these terms (2h2+2-sqz, 3h-diag-sqz, 4h-2X2-sqz, shot2g-sqz-2h, shot3g-sqz-23h), one finds:
\ba\label{Eq:SSC-unified-wshot}
\mathcal{C}_{\ell,\ell'}^\mr{SSC} = \frac{\delta_{i_z,j_z} \ \delta_{k_z,l_z}}{4\pi} \int \dd V_{ab} \ \Psi_\ell^\mr{sqz}(z_a) \ \Psi_{\ell'}^\mr{sqz}(z_b) \ C_0^m(z_a,z_b)  
\ea
where
\ba
\nonumber \Psi_\ell^\mr{sqz}(z) =& \sum_{X\in \{\mr{b2,s2,2PT}\} } 4 \ K_X^0 \ I_1^X(k_{\ell}|z) \ I_1^1(k_{\ell}|z) \ P(k_{\ell}|z) \\
\nonumber & \qquad + I_2^1(k_{\ell},k_{\ell}|z) + I_1^1(0|z) \\
=& 4 \ I_1^{\Sigma_2}(k_{\ell}|z) \ I_1^1(k_{\ell}|z) \ P(k_{\ell}|z) + I_2^1(k_{\ell},k_{\ell}|z) + I_1^1(0|z) \label{Eq:def-Psi^sqz}
\ea

More familiar equations can be found by expliciting the sum over $X\in \{\mr{b2,s2,2PT}\}$, using that
\ba
K_\mr{2PT}^0 = \frac{17}{21} \qquad K_\mr{s2}^0 = 0 \qquad K_\mr{b2}^0 = \frac{1}{2} \ ,
\ea
introducing the effective galaxy bias
\ba
I_1^X(k|z) = b_X^\mr{gal}(k,z) \ \nbargal(z)
\ea
and the variance of the background matter density
\ba
\sigma^2(z_a,z_b) = \frac{C_0^m(z_a,z_b)}{4\pi}
\ea
With all these notations, one finds
\ba
\nonumber \mathcal{C}_{\ell,\ell'}^\mr{SSC} = \delta_{i_z,j_z} \ \delta_{k_z,l_z} \int \dd V_{ab} \ \nbargal(z_a)^2 \ \nbargal(z_b)^2 \ \frac{\partial P_\mr{gal}(k_{\ell},z_a)}{\partial \delta_b} \\
\times \ \frac{\partial P_\mr{gal}(k_{\ell'},z_b)}{\partial \delta_b} \ \sigma^2(z_a,z_b)
\ea
with
\ba
\nonumber \frac{\partial P_\mr{gal}(k,z)}{\partial \delta_b} = \left[\frac{68}{21}\left(b_1^\mr{gal}(k,z)\right)^2 + 2 \ b_1^\mr{gal}(k,z) \ b_2^\mr{gal}(k,z) \right] P(k|z) \\
+ \ \frac{I_2^1(k,k|z)}{\nbargal(z)^2} \ + \ \frac{b_1^\mr{gal}(k=0,z)}{\nbargal(z)} \label{Eq:dPgalddeltab}
\ea
I thus nicely recover the same SSC equation as the one derived in \cite{Lacasa2016} in the cross-covariance case.

In the literature, for example \cite{Takada2013}, the first term in Eq.~\ref{Eq:dPgalddeltab} is called beat-coupling (BC) and the third term is called halo sample variance (HSV) ; the second and fourth terms were discovered by \cite{Lacasa2016}, they come respectively from the non-linear response of halos to matter density and from shot noise.\\
However the SSC shot-noise terms will be canceled by the shot-noise subtraction described in Sect.~\ref{Sect:shot-subs}, the shot-noise corrected SSC then becomes:
\ba
\nonumber \mathcal{C}_{\ell,\ell'}^\mr{SSC,corr} = \delta_{i_z,j_z} \ \delta_{k_z,l_z} \int \dd V_{ab} \ \nbargal(z_a)^2 \ \nbargal(z_b)^2 \ \frac{\partial P_\mr{gal}^\mr{corr}(k_{\ell},z_a)}{\partial \delta_b} \\
\times \ \frac{\partial P_\mr{gal}^\mr{corr}(k_{\ell'},z_b)}{\partial \delta_b} \ \sigma^2(z_a,z_b) \label{Eq:SSC-unified-shotcorr} 
\ea
with
\ba
\nonumber \frac{\partial P_\mr{gal}^\mr{corr}(k,z)}{\partial \delta_b} =& \left[\frac{68}{21}\left(b_1^\mr{gal}(k,z)\right)^2 + 2 \ b_1^\mr{gal}(k,z) \ b_2^\mr{gal}(k,z) \right] P(k|z) \\
& \ + \ \frac{I_2^1(k,k|z)}{\nbargal(z)^2} \label{Eq:dPgalcorrddeltab}
\ea

\subsubsection{Angle-dependent terms}\label{Sect:discu-SSC-angdep}

Two subleading SSC effects found in the recent literature are not present in the previous subsection. Here I show that is because they come from the $n=1$ and $n=2$ terms.

The first effect is the so-called dilation effect found by \cite{Li2014} in the 3D $P(k)$ case. From \cite{Li2014}, one sees that this term comes from the 
\ba
\frac{1}{2}\left(\frac{k_1}{k_2}+\frac{k_2}{k_1}\right) \cos\theta_{12}
\ea
part of the 2PT kernel $F_2(\kk_1,\kk_2)$.\\
In my derivation this part of $F_2$ yields the 2PT $n=1$ Legendre term, cf Appendix \ref{App:3Dhalopolysp}. Hence the dilation effect is present here if one considers $n=1$.

The second effect is from super-survey tidal fields. This has been first uncovered by \cite{Li2017,Akitsu2017} for the redshift space power spectrum of galaxies. Then \cite{Barreira2017b}, which appeared the same day as this article v1 on arXiv, showed that it also affects the isotropic power spectrum of weak-lensing. To discus this issue, I consider matter only, not galaxies, and adopt the same notations as \cite{Barreira2017b}. The central notion is that of the large scale structure response to long wavelength (=soft) modes. The first order response $\mathcal{R}_1$, for one such soft mode, is defined through the squeezed limit of the bispectrum
\ba
\nonumber \lim_{\mathbf{p}\rightarrow 0} \lbra \delta(\kk) \ \delta(\kk') \ \delta(\mathbf{p}) \rbra = (2\pi)^3 \ \delta_D(\kk+\kk'+\mathbf{p}) \ \mathcal{R}_1(\kk,\mu=\hk\cdot\hat{p}) \\
\times \ P(k) \ P(p)
\ea
and \cite{Barreira2017b} decompose this response into its isotropic and tidal field part
\ba
\mathcal{R}_1(\kk,\mu) = R_1(k) + \frac{2}{3} R_K(k) \ P_2(\mu)
\ea
where $P_2$ is the second order Legendre polynomial.

In this article, I use standard perturbation theory at tree-level. Then from Appendix \ref{App:3Dhalobispec} giving the 2PT bispectrum, one sees that the resulting power spectrum response is
\ba
\mathcal{R}_1(\kk,\mu) = 2 \ F_2(\kk,\mathbf{p}) + 2 \ F_2(-\kk+\mathbf{p},\mathbf{p}) \frac{P(|\kk-\mathbf{p}|)}{P(k)}.
\ea
From this one sees easily that the $n=2$ Legendre term in $F_2$ will source the tidal response. Taking appropriate limits when $\mathbf{p}\rightarrow 0$, one can further see that the $n=1$ Legendre term in $F_2$ will source both the isotropic and the tidal responses. The $n=0$ term considered in the previous subsection sources the so-called growth-only part of the isotropic response. Hence the decomposition of $F_2$ in $n=0,1,2$ Legendre terms is equivalent to including growth-only, dilation and tidal effects in the response approach to SSC.

Although the derivation presented here uses standard perturbation theory at tree-level, it is possible to generalise the SSC equations to another modeling of matter, simply by adapting the power spectrum response. For instance, this response can be fitted to numerical simulations and then fed in the equations \citep[see e.g.][]{Barreira2017,Barreira2017b}. This appears a solution for observables directly sensitive to the matter power spectrum (e.g. weak-lensing), but it may not be feasible for galaxies. Indeed for the galaxy spectrum, the one-halo response $I_2^1(k,k|z)$ is a critical element at intermediate to small scales, and it is fully non-perturbative and heavily dependent on the galaxy selection function. As such it necessitates a halo modelisation to be predicted correctly.

The advantage of the present derivation is that it does not rely on any particular soft mode limit $\mathbf{p}\rightarrow 0$ or any Taylor expansion as in \cite{Takada2013}. Instead, the present derivation is exact within the modelling assumption. Hence it remains valid even on large scales comparable to the survey size, while the response approach is limited to $k\ll p$, in other words, at small scales.

A side note is that I developed here all the covariance equations in the full-sky limit. However this is not a practical limitation, as \cite{Lacasa2016b} has recently developed the formalism to predict analytically SSC in the more realistic case of partial sky coverage with an arbitrary survey mask. One basically needs to change $\sigma^2$ in Eq.~\ref{Eq:SSC-unified-shotcorr} to account for the effect of the mask power spectrum.

Finally, there has been a lot of emphasis on SSC in literature, however the systematic derivation presented in this article finds a wealth of other non-Gaussian covariance terms which have never been considered before. The following subsections are devoted to these new terms and their potential importance.

\subsection{Braiding terms}\label{Sect:discu-braiding}

Braiding terms are those that arise when the trispectrum has a dependence on one of the alternate diagonal through a $P(k_{1+3})$ (or $P(k_{1+4})$), making a $C_{\ell_a}^m(z_a,z_b)$ appear in the covariance as well as a 3J symbol mixing the multipoles $\ell$ and $\ell'$. For any super-sample covariance term (studied in Sect.~\ref{Sect:discu-SSC}) there is a corresponding braiding term. Namely the clustering contributions are the 2h2+2-alt term, Eq.~\ref{Eq:Cov-2h2+2-alt}, the 3h-diag-alt term, Eq.~\ref{Eq:Cov-3h-diag0-alt-Limber}, and the 4h-2X2-alt term, Eq. \ref{Eq:Cov-4h-2X2-alt00-Limber}. Defining the braiding kernel
\ba
\mathcal{B}_{\ell,\ell'}(z_a,z_b) \equiv \sum_{\ell_a} \frac{2\ell_a+1}{4\pi} \threeJz{\ell}{\ell'}{\ell_a}^2 C_{\ell_a}^m(z_a,z_b)
\ea
these three clustering terms can be unified into:
\ba
\nonumber \mathcal{C}_{\ell,\ell'}^\mr{braid-clust} = \left(\delta_{i_z,k_z}\ \delta_{j_z,l_z}+\delta_{i_z,l_z}\ \delta_{j_z,k_z}\right) \int \dd V_{ab} \ \Psi^\mr{alt,clust}_{\ell,\ell'}(z_a) \\
\times \ \Psi^\mr{alt,clust}_{\ell,\ell'}(z_b) \ \mathcal{B}_{\ell,\ell'}(z_a,z_b) \label{Eq:braiding-unified-clust}
\ea
where
\ba
\Psi^\mr{alt,clust}_{\ell,\ell'}(z) = \Big[ 2 \ I_1^{\Sigma_2}(k_{\ell'}|z) \ I_1^1(k_{\ell}|z) \, P(k_{\ell}|z) + (\ell \leftrightarrow\ell') \Big] + I_2^1(k_{\ell},k_{\ell'}|z)
\ea
which is relatively similar to $\Psi^\mr{sqz}_{\ell}$ defined in Eq. \ref{Eq:def-Psi^sqz}, except for the shot-noise part and the multipole-coupling structure. In fact, one has the identity
\ba\label{Eq:identity-Psi-alt-sqz}
\Psi^\mr{alt,clust}_{\ell,\ell} = \Psi^\mr{sqz,clust}_{\ell}
\ea

Shot-noise contributions to the braiding covariance are given by Eq. \ref{Eq:Cov-shot2g-alt}
\ba
\nonumber \mathcal{C}_{\ell,\ell'}^\mr{shot2g-alt} =& \left(\delta_{i_z,k_z} \ \delta_{j_z,l_z}+ \delta_{i_z,l_z} \ \delta_{j_z,k_z}\right) \sum_{\ell_a} \frac{2\ell_a+1}{4\pi} \threeJz{\ell}{\ell'}{\ell_a}^2 \\
& \times \ C_{\ell_a}^\mr{gal,clust}(i_z,j_z)
\ea
and Eq. \ref{Eq:Cov-shot3g-alt}
\ba
\nonumber \mathcal{C}_{\ell,\ell'}^\mr{shot3g-alt} =& \sum_{\ell_a} \frac{2\ell_a+1}{4\pi} \threeJz{\ell}{\ell'}{\ell_a}^2 \Big[
\delta_{i_z,k_z} \ b_{\ell_a,\ell,\ell'}^\mr{gal,clust}(i_z,j_z,l_z) \\
\nonumber & + \delta_{i_z,l_z} \ b_{\ell_a,\ell,\ell'}^\mr{gal,clust}(i_z,j_z,k_z) + \delta_{j_z,k_z} \ b_{\ell_a,\ell,\ell'}^\mr{gal,clust}(j_z,i_z,l_z) \\
& + \delta_{j_z,l_z} \ b_{\ell_a,\ell,\ell'}^\mr{gal,clust}(j_z,i_z,k_z)
\Big]
\ea
In order to include these terms in a unified formula similar to Eq. \ref{Eq:braiding-unified-clust}, one has to add some approximations:
\ba
C_{\ell_a}^\mr{gal,clust}(i_z,j_z) \simeq \int \dd V_{ab} \ I_1^1(k_{\ell_a}|z_a) \ I_1^1(k_{\ell_a}|z_b) \ C_{\ell_a}^m(z_a,z_b)
\ea
which neglects $C_{\ell_a}^\mr{1h}$, and
\ba
b_{\ell_a,\ell,\ell'}^\mr{gal,clust}(i_z,j_z,k_z) \simeq \delta_{j_z,k_z} \int \dd V_{ab} \ I_1^1(k_{\ell_a}|z_a) \ \Psi^\mr{alt}_{\ell,\ell'}(z_b) \  C_{\ell_a}^m(z_a,z_b)
\ea
which neglects several bispectrum terms\footnote{all the one-halo bispectrum, two permutations of the two-halo, one permutation of the three-halo, and $n\geq 1$ in the Legendre decomposition of the kernels $K_X(\kk_\alpha,\kk_\beta)$.}, and uses Limber's approximation on $\ell$ and $\ell'$.\\
With these, one gets the unified formula
\ba
\nonumber \mathcal{C}_{\ell,\ell'}^\mr{braid} =& \left(\delta_{i_z,k_z} \ \delta_{j_z,l_z}+ \delta_{i_z,l_z} \ \delta_{j_z,k_z}\right) \sum_{\ell_a} \frac{2\ell_a+1}{4\pi} \threeJz{\ell}{\ell'}{\ell_a}^2 \\
& \times \int \dd V_{ab} \ \Psi^{\ell_a,\mr{alt}}_{\ell,\ell'}(z_a) \ \Psi^{\ell_a,\mr{alt}}_{\ell,\ell'}(z_b) \ C_{\ell_a}^m(z_a,z_b)
\ea
where
\ba
\Psi^{\ell_a,\mr{alt}}_{\ell,\ell'}(z) = \Psi^{\mr{alt,clust}}_{\ell,\ell'}(z) + I_1^1(k_{\ell_a}|z)
\ea

\subsection{Importance of terms}\label{Sect:discu-importance}

Among the non-Gaussian covariance terms, super-sample covariance is the main reference against which to compare the new terms discovered in this article. Indeed it has already been well studied in the literature, including for the galaxy angular power spectrum, for example, in combination with cluster number counts \citep{Lacasa2016} or with weak lensing \citep{Krause2017}. Its importance is already well recognised, and it is is indeed included in the analysis of current galaxy surveys \citep[e.g.][]{vanUitert2017,Krause2017b,DES2017} having an impact both on the cosmological error bars as well as central values \citep{Hildebrandt2017}.

As already mentionned in the introduction, numerical investigations I performed \citep{Lacasa2017-LAL} show that the 1h and 2h1+3 terms have an impact comparable to SSC on the signal to noise ratio of $C_\ell^\mr{gal}$, when using survey specifications representative of future missions like Euclid. As these terms become important, there is a priori no reason for others not to be, so I now turn to analytical arguments comparing all the other terms to SSC in adequate regimes where they can be compared.

The braiding terms (Sect.~\ref{Sect:discu-braiding}) are the easiest ones to be compared with SSC, as it has already been noted that they have some similarity to it\footnote{This similarity is not a coincidence, but a straightforward consequences that a trispectrum term with a $P(k_{1+2})$ is a particular permutation of a contribution which also yield a term with a $P(k_{1+3})$ and a term with a $P(k_{1+4})$.}. Indeed, using Eq.~\ref{Eq:identity-Psi-alt-sqz}, in the case $\ell=\ell'$ and $i_z=j_z=k_z=l_z$ one has the identity
\ba\label{Eq:braid-on-diag}
\nonumber \mathcal{C}_{\ell,\ell}^\mr{braid-clust} = 2 \sum_{\ell_a} \frac{2\ell_a+1}{4\pi} \threeJz{\ell}{\ell}{\ell_a}^2 \int \dd V_{ab} \ \Psi^\mr{sqz,clust}_{\ell}(z_a) \\
\times \ \Psi^\mr{sqz,clust}_{\ell}(z_b) \ C_{\ell_a}^m(z_a,z_b)
\ea
which can be compared with the corresponding SSC case
\ba
\mathcal{C}_{\ell,\ell}^\mr{SSC,corr} = \frac{1}{4\pi} \int \dd V_{ab} \ \Psi^\mr{sqz,clust}_{\ell}(z_a) \ \Psi^\mr{sqz,clust}_{\ell}(z_b) \ C_{0}^m(z_a,z_b)
\ea
in particular one sees that the term $\ell_a=0$ in the sum of Eq.~\ref{Eq:braid-on-diag} implies
\ba
\mathcal{C}_{\ell,\ell}^\mr{braid-clust} > \frac{2}{2\ell+1} \mathcal{C}_{\ell,\ell}^\mr{SSC,corr}
\ea
so at low multipoles, braiding must be non-negligible. To go further, assumptions need to be made. If one can assume that the matter power spectrum $C_{\ell_a}^m$ is slowly varying over the range of multipoles of interest, then
\ba
\nonumber \mathcal{C}_{\ell,\ell}^\mr{braid-clust} \simeq & 2 \sum_{\ell_a} \frac{2\ell_a+1}{4\pi} \threeJz{\ell}{\ell}{\ell_a}^2 \int \dd V_{ab} \ \Psi^\mr{sqz,clust}_{\ell}(z_a) \\
\nonumber & \qquad \qquad \qquad \qquad \times \ \Psi^\mr{sqz,clust}_{\ell}(z_b) \ C_{0}^m(z_a,z_b) \\
=& \  2 \ \mathcal{C}_{\ell,\ell}^\mr{SSC,corr}
\ea
More explicitely, if $C_{\ell}^m$ is an increasing function of $\ell$ over $[0,2\ell]$, then one finds
\ba
\frac{\mathcal{C}_{\ell,\ell}^\mr{braid-clust}}{\mathcal{C}_{\ell,\ell}^\mr{SSC,corr}} > 2
\ea
whereas if $C_{\ell}^m$ is decreasing
\ba
\frac{2}{2\ell+1} < \frac{\mathcal{C}_{\ell,\ell}^\mr{braid-clust}}{\mathcal{C}_{\ell,\ell}^\mr{SSC,corr}} < 2
\ea
The first situation occurs when the survey probes scales larger than the matter-radiation equality where $P(k)$ has a maximum, i.e. $\ell \lesssim \ell_\mr{eq} = k_\mr{eq} r(z)$, which will be the case for future surveys covering large portions of the sky. The second situation occurs at smaller scales. From current constraints $C_{\ell}^m \propto 1/\ell$ in the cosmologically interesting domain ; then from power counting argument one gets that the covariance ratio is $\mathcal{O}\left(\frac{\ln \ell}{\ell}\right)$. So in summary
\ba
\frac{\mathcal{C}_{\ell,\ell}^\mr{braid-clust}}{\mathcal{C}_{\ell,\ell}^\mr{SSC,corr}} &> 2 \quad \mr{if} \quad \ell \lesssim \ell_\mr{eq} \\
\frac{\mathcal{C}_{\ell,\ell}^\mr{braid-clust}}{\mathcal{C}_{\ell,\ell}^\mr{SSC,corr}} &= \mathcal{O}\left(1\right) \quad \mr{if} \quad C_{\ell}^m \sim C_{0}^m \\
\frac{\mathcal{C}_{\ell,\ell}^\mr{braid-clust}}{\mathcal{C}_{\ell,\ell}^\mr{SSC,corr}} &= \mathcal{O}\left(\frac{\ln \ell}{\ell}\right) \quad \mr{at \ high \ \ell}
\ea

Another regime where braiding is important is cross-redshifts: from Eq.~\ref{Eq:SSC-unified-wshot} one sees that SSC vanishes for cross-spectra $i_z\neq j_z$ as a consequence of Limber's approximation ; however from Eq.~\ref{Eq:braiding-unified-clust} one sees that braiding covariance does not vanish in this regime. Hence braiding will be of importance for effects producing non-vanishing cross-spectra, for example when dropping Limber's approximation or accounting for general relativistic effects \citep[e.g.][]{Cardona2016}.

The next terms which may be important are the third order ones involved in the 4-halo term, either the third order bias term (4h-b3) or the term from third perturbation theory (4h-3PT). For these terms, it is simpler to argue at the level of the 3D trispectrum: both these terms give a trispectrum
$$ T \propto P(k_{\ell}|z) \ P(k_{\ell}|z) \ P(k_{\ell'}|z) \quad + \quad (\ell\leftrightarrow\ell') $$
whereas the SSC from 4h-2X2-sqz (beat coupling BC-BC in the literature) comes from a trispectrum
$$ T \propto P(k_{1+2}|z) \ P(k_{\ell}|z) \ P(k_{\ell'}|z) $$
where the prefactor of both trispectra are of the same order.\\
Thus these terms will be more important than BC-BC in the range of multipoles where $P(k_{\ell}|z) \gtrsim P(k_{1+2}|z)$. In this full sky derivation, $k_{1+2}$ becomes aliased in the monopole $\ell=0$, but in general $k_{1+2}$ is a super-survey mode, so for a general survey covering a fraction $f_\mr{SKY}$ of the sky, one gets the rule of thumb that these terms are going to be important for multipoles where
.\ba\label{Eq:thumbrule-gtSSC}
P(k_{\ell}) \gtrsim P(1 / f_\mr{SKY} \, r(z))
\ea
This certainly happens for $\ell \lesssim \ell_\mr{eq}$ if the survey is large enough to see the matter-radiation equality scale. Interestingly, Eq.~\ref{Eq:thumbrule-gtSSC} is basically equivalent to the condition for braiding to be important with respect to SSC, although in the former case this was argued only on the diagonal $\ell=\ell'$ whereas here it suffices that one of the multipoles fullfills the condition, either $\ell$ or $\ell'$.

The penultimate terms are coming from the three-halo term, explicitely 3h-base which has two contributions whose trispectra follow
$$ T \propto P(k_\ell|z)^2 \ I_2^{\Sigma_2}(k_{\ell'},k_{\ell'}|z) \quad + \quad (\ell\leftrightarrow\ell')$$
and
$$ T \propto P(k_\ell|z) \ P(k_{\ell'}|z) \ I_2^{\Sigma_2}(k_{\ell},k_{\ell'}|z)$$
whereas the SSC from 3h-diag-sqz (BC-HSV in the literature) comes from a trispectrum
$$ T \propto P(k_{1+2}|z) \ P(k_{\ell'}|z) \ I_2^{1}(k_{\ell},k_{\ell}|z) \quad + \quad (\ell\leftrightarrow\ell')$$
where the prefactor of both trispectra are of the same order.\\
Thus these terms will be more important than BC-HSV in the range of multipoles where $P(k_{\ell}|z) \gtrsim P(k_{1+2}|z)$. Hence the condition Eq.~\ref{Eq:thumbrule-gtSSC} is again the one that rules the importance of these terms.

Finally, shot-noise terms impact the measurements in a manner inversely proportional to the number of observed galaxies. More precisely the impact on the signal to noise of $C_\ell^\mr{gal}$ is of order $\mathcal{O}\left(N_\ell/N_\mr{gal}\right)$, where $N_\ell$ is the number of multipoles considered. For future surveys, this effect can thus be expect to be well below the percent level, unless targeting high multipoles with thin redshift bins at the lowest and highest redshifts, where galaxy numbers decrease.
In summary, most terms have a chance to be of importance if the survey considered probes scales comparable to or larger than the matter-radiation equality $k_\mr{eq}$, and shot-noise can be of importance for a spectroscopic survey targeting information on small scales.

\section{Summary}\label{Sect:summary}

This section summarises covariance terms that should be considered in the simplest case where one uses Limber's approximation and shot-noise subtraction (Sect.~\ref{Sect:shot-subs}), as is usually done in current galaxy surveys, and further considering only $n=0$ for angle-dependent kernels. The more general equations can be found in the main body of the text. This section can thus be considered by the busy reader as the reference summary containing the first order equations to be implemented numerically.

\subsection{Notations and remarks}
In order for this section to be self-contained, I recapitulate here the particular notations which are used in the covariance terms.\\
To begin with, as discussed in Sect.~\ref{Sect:methods}, I consider the power spectrum of the absolute fluctuations $\delta n_\mr{gal}(\xx)$ and not the relative fluctuations $\delta_\mr{gal}=\delta n_\mr{gal}/\nbargal$. One can convert my absolute power spectrum into a relative one by dividing by the factor $N_\mr{gal}(i_z) \ N_\mr{gal}(j_z)$,  where $i_z$ and $j_z$ are the indices of the two redshift bins considered.\\
The power spectrum covariance is noted for simplicity
$$\mathcal{C}_{\ell,\ell'} \equiv \Cov\left(C_\ell^\mr{gal}(i_z,j_z),C_{\ell'}^\mr{gal}(k_z,l_z)\right)$$
and needs to be divided by a factor
$$N_\mr{gal}(i_z) \ N_\mr{gal}(j_z) \ N_\mr{gal}(k_z) \ N_\mr{gal}(l_z)$$
if one wants relative fluctuations instead of absolute ones.

Most importantly, I use the following definitions:\\
$k_\ell=(\ell+1/2)/r(z)$ is the comoving wavenumber given by Limber's approximation at multipole $\ell$ and redshift $z$,
\ba
\nonumber I_\mu^\beta(k_1,\cdots,k_\mu|z) \equiv \int \dd M \ & \frac{\dd n_h}{\dd M} \ \lbra N_\mr{gal}^{(\mu)}\rbra \ b_\beta(M,z) \\ 
& \times u(k_1|M,z) \cdots u(k_\mu|M,z)  
\ea
is an integral that will appear frequently, 
\ba
I_\mu^{\Sigma_2} & \equiv  \sum_{X \in \{\mr{b2,s2,2PT}\}} K_X^0 \ I_\mu^X = \frac{17}{21} \ I_\mu^{1} + \frac{1}{2} \ I_\mu^{2}
\ea
is the sum of second order contributions, and
\ba
I_\mu^{\Sigma_3}(k,z) \equiv \frac{1023}{1701} \ I_\mu^{1}(k,z) + \frac{1}{3!} \ I_\mu^{3}(k,z)
\ea
is the sum of third order contributions.\\
The angular power spectrum of matter is
\ba
C_\ell^m(z_a,z_b)= \int k^2\,\dd k \ P(k|z_{ab}) \ j_\ell(k r_a) \ j_\ell(k r_b)
\ea
and in full-sky the SSC and braiding kernels are respectively
\ba
\sigma^2(z_a,z_b) = \frac{C_0^m(z_a,z_b)}{4\pi}
\ea
\ba
\mathcal{B}_{\ell,\ell'}(z_a,z_b) = \sum_{\ell_a} \frac{2\ell_a+1}{4\pi} \threeJz{\ell}{\ell'}{\ell_a}^2 \ C_{\ell_a}^m(z_a,z_b).
\ea

\subsection{Covariance terms}

This subsection simply lists the different covariance contributions, ordered in term of simplicity.\\
The first contribution is the one-halo term (Sect.~\ref{Sect:1halo})
\ba
\mathcal{C}_{\ell,\ell'}^\mr{1h} = \frac{\delta_{i_z,j_z,k_z,l_z}}{4\pi} \int \dd V \ I_4^0(k_{\ell},k_{\ell},k_{\ell'},k_{\ell'}|z)
\ea
then there is the two-halo 1+3 term (Sect.~\ref{Sect:2halo1+3})
\ba
\nonumber \mathcal{C}_{\ell,\ell'}^\mr{2h1+3} = \frac{\delta_{i_z,k_z,l_z}+\delta_{j_z,k_z,l_z}}{4\pi} & \int \dd V \ I_1^1(k_\ell|z) \ I_3^1(k_\ell,k_{\ell'},k_{\ell'}|z) \ P(k_\ell|z) \\
& + (\ell \leftrightarrow \ell')
\ea
the three-halo base term (Sect.~\ref{Sect:3halo})
\ba
\nonumber \mathcal{C}_{\ell,\ell'}^\mr{3h-base0} = \frac{\delta_{i_z,j_z,k_z,l_z}}{4\pi} \int & \dd V \ 2 \;  \left(I_1^{1}(k_\ell|z) \ P(k_\ell|z)\right)^2 I_2^{\Sigma_2}(k_{\ell'},k_{\ell'}|z) \\
\nonumber & + \quad (\ell\leftrightarrow\ell') \\
\nonumber +\frac{4 \ \delta_{i_z,j_z,k_z,l_z}}{4\pi} \int & \dd V \ 2 \ I_1^{1}(k_\ell|z) \ I_1^{1}(k_{\ell'}|z) \ I_2^{\Sigma_2}(k_{\ell},k_{\ell'}|z) \\
& \times P(k_\ell|z) \ P(k_{\ell'}|z)
\ea
and the four-halo term from third order contributions (Sect.~\ref{Sect:4halo})
\ba
\nonumber \mathcal{C}_{\ell,\ell'}^\mr{4h-3} = \frac{2 \ \delta_{i_z,j_z,k_z,l_z}}{4\pi} \int \dd V \ 3! \ \left(I_1^1(k_{\ell},z)\right)^2 \ I_1^1(k_{\ell'},z) \ I_1^{\Sigma_3}(k_{\ell'},z) \\
\times \ P(k_{\ell}|z) \ P(k_{\ell}|z) \ P(k_{\ell'}|z) \quad + \quad (\ell\leftrightarrow\ell')
\ea
Then there are groups of terms unifying contributions with different numbers of halos.\\
First is the super-sample covariance (Sect.~\ref{Sect:discu-SSC})
\ba
\nonumber \mathcal{C}_{\ell,\ell'}^\mr{SSC} = \delta_{i_z,j_z} \ \delta_{k_z,l_z} \int \dd V_{ab} \ \Psi_\ell^\mr{sqz,clust}(z_a) \ \Psi_{\ell'}^\mr{sqz,clust}(z_b) \ \sigma^2(z_a,z_b) 
\ea
where $z_a \in i_z$, $z_b \in k_z$, and
\ba
\Psi_\ell^\mr{sqz,clust}(z) = 4 \ I_1^{\Sigma_2}(k_{\ell}|z) \ I_1^1(k_{\ell}|z) \ P(k_{\ell}|z) + I_2^1(k_{\ell},k_{\ell}|z) 
\ea
Second is the braiding covariance (Sect.~\ref{Sect:discu-braiding})
\ba
\nonumber \mathcal{C}_{\ell,\ell'}^\mr{braid-clust} = \left(\delta_{i_z,k_z}\ \delta_{j_z,l_z}+\delta_{i_z,l_z}\ \delta_{j_z,k_z}\right) \int \dd V_{ab} \ \Psi^\mr{alt,clust}_{\ell,\ell'}(z_a) \\
\times \ \Psi^\mr{alt,clust}_{\ell,\ell'}(z_b) \ \mathcal{B}_{\ell,\ell'}(z_a,z_b)
\ea
where $z_a \in i_z$, $z_b \in j_z$, and
\ba
\Psi^\mr{alt,clust}_{\ell,\ell'}(z) = \Big[ 2 \ I_1^{\Sigma_2}(k_{\ell'}|z) \ I_1^1(k_{\ell}|z) \, P(k_{\ell}|z) + (\ell \leftrightarrow\ell') \Big] + I_2^1(k_{\ell},k_{\ell'}|z)
\ea
Finally, shot-noise terms, where the only surviving shot-noise subtraction (Sect.~\ref{Sect:shot-subs}) are braiding ones (Sect.~\ref{Sect:discu-braiding})
\ba
\nonumber \mathcal{C}_{\ell,\ell'}^\mr{shot2g-alt} =& \left(\delta_{i_z,k_z} \ \delta_{j_z,l_z}+ \delta_{i_z,l_z} \ \delta_{j_z,k_z}\right) \sum_{\ell_a} \frac{2\ell_a+1}{4\pi} \threeJz{\ell}{\ell'}{\ell_a}^2 \\
& \times \ C_{\ell_a}^\mr{gal,clust}(i_z,j_z)
\ea
and
\ba
\nonumber \mathcal{C}_{\ell,\ell'}^\mr{shot3g-alt} =& \sum_{\ell_a} \frac{2\ell_a+1}{4\pi} \threeJz{\ell}{\ell'}{\ell_a}^2 \Big[
\delta_{i_z,k_z} \ b_{\ell_a,\ell,\ell'}^\mr{gal,clust}(i_z,j_z,l_z) \\
\nonumber & + \delta_{i_z,l_z} \ b_{\ell_a,\ell,\ell'}^\mr{gal,clust}(i_z,j_z,k_z) + \delta_{j_z,k_z} \ b_{\ell_a,\ell,\ell'}^\mr{gal,clust}(j_z,i_z,l_z) \\
& + \delta_{j_z,l_z} \ b_{\ell_a,\ell,\ell'}^\mr{gal,clust}(j_z,i_z,k_z)
\Big]
\ea

The importance of all these covariance terms is partially discussed in Sect.~\ref{Sect:discu-importance} with analytical arguments. The actual importance for a galaxy survey will strongly depend on the survey specifications, galaxy selection, choice of data vector (e.g. redshift and scale cuts) and so on, and as such cannot be forecast easily analytically. Instead numerical analysis must be carried out, which will be the subject of future works. I expect that at least several of these terms, if not most of them, will be of importance for cosmological constraints from future surveys such as Euclid and LSST.

\section{Conclusion}\label{Sect:conclusion}

I have carried out an exhaustive analytic derivation of all non-Gaussian covariance terms of the galaxy angular power spectrum $C_\ell^\mr{gal}$ when using the halo model at tree level. The calculation of the involved trispectrum is developed up to third order both in halo bias and standard perturbation theory, including non-local halo bias and all shot-noise terms.

The projection of the trispectrum into the angular covariance has been derived in all trispectra cases, including complex cases with several dependence on angle between wavenumbers, in two appendices (Appendices~\ref{App:2Dproj-trisp-angindep} \& \ref{App:2Dproj-trisp-angdep}) together with robustness checks of the formulae performed (Appendix~\ref{App:reductions}). These derivations, though not the original aim of the article, are standalone results that can be used in order to model the angular covariance of other signals or alter the modelling framework, for example, using a different flavour of perturbation theory.

A wealth of non-Gaussian covariance terms has been found, providing a rigorous derivation of the already known super-sample covariance (SSC) in the angular case, and more importantly discovering several new terms. A whole new class of terms, which I dub braiding covariance, stems from the same physical effects that lead to SSC but leads to different couplings between multipoles and redshift bins. Other terms (3h-base, third order contributions) furthermore exist, and I provide a unified treatment of shot-noise terms, including how they are affected by the popular habit of subtracting $1/\nbargal$ from the observed power spectrum.

A clean executive summary is provided in Sect.~\ref{Sect:summary} in the simplest case where one uses Limber's approximation, shot-noise subtraction (Sect.~\ref{Sect:shot-subs}), and retains only $n=0$ terms for angle-dependent kernels. This section can serve as a reference for the minimal number of non-Gaussian terms to quantify numerically for galaxy surveys.

The potential importance of the new non-Gaussian terms has been discussed with analytical arguments in Sect.~\ref{Sect:discu-importance}, in particular in comparison with super-sample covariance, as the latter has already shown to have an impact on constraint from current surveys. It was found that some terms (braiding, 3h-base, 4h-3) should become comparable to, if not bigger than, SSC on scales comparable to the matter-radiation equality. Other terms (1h, 2h1+3) can become important for deep surveys with a high portion of satellite galaxies, with only numerical calculation which can decide precisely on their actual impact. Finally shot-noise terms can become relevant when analysing small scales with spectroscopic (more sparse) surveys, such as when constraining neutrinos or non-cold dark matter.

Numerical codes computing SSC and the one-halo term already exist and have been used. The one by the author, for instance, can compute covariance on a standard laptop in a matter of seconds or tens of seconds depending on the number of multipoles. Including the new non-Gaussian terms presented here shall prove very feasible and will be studied in future works. It is expected that this inclusion will not alter the order of magnitude of the speed of the calculation. Hence the analytical approach to covariances will remain feasible, and in fact the most competitive, for current and future surveys.

\section*{Acknowledgements}
\vspace{0.2cm}

I thank Ruth Durrer, Vittorio Tansella and Alexandre Barreira for helpful discussions, Pierre Fleury for help with 9J symbols, and Elena Sellentin and Martin Kunz for proofreading and suggestions that improved this article.\\
I acknowledge support by the Swiss National Science Foundation.

\bibliographystyle{aa}
\bibliography{bibliography}

\appendix

\section{3D halo polyspectra}\label{App:3Dhalopolysp}

The density field of halos at a given mass is composed of a local bias hierarchy \citep{Fry1993} and a non-local bias term from tidal forces \citep{Chan2012,Baldauf2012}
\ba
\nonumber \delta_h(\xx|M,z) = \sum_{n=1}^3 \frac{b_n(M,z)}{n!} \left(\delta(\xx,z)^n-\lbra\delta(z)^n\rbra\right)\\
+ b_{s^2}(M,z) \left(s^2(\xx,z)-\lbra s^2(z)\rbra\right)
\ea
where I cut the local hierarchy at third order, and the quadratic tidal tensor field $s^2$ is related to the matter density field via \citep{Baldauf2012}
\ba
s^2(\kk) = \int \frac{\dd^3\kk'}{(2\pi)^3} \ S_2(\kk',\kk-\kk') \ \delta(\kk') \ \delta(\kk-\kk')
\ea
with 
\ba
S_2(\kk_1,\kk_2') = \frac{(\kk_1\cdot\kk_2)^2}{k_1^2 \, k_2^2} - \frac{1}{3}.
\ea

This bias model has been shown to reproduce adequately the halo clustering by \cite{Hoffmann2017b}. The bias parameters can be either taken as free parameters to be fit together with cosmology in observations, or theoretical inputs can be used. Indeed the peak-background split prescribes the local bias as \citep[e.g.][]{Mo1997}
\ba
b_n(\nu) = \frac{1}{f(\nu)} \frac{\dd^n f(\nu)}{d \nu^n}
\ea
where $\nu$ is the peak height and $f(\nu)$ the mass function. Alternatively a near-universal relation between $b_2$ and $b_1$ was found by \cite{Hoffmann2017a}.\\
For the tidal tensor bias, the local lagrangian model gives \citep{Baldauf2012}
\ba
b_{s2} = -\frac{2}{7}(b_1-1)
\ea

\subsection{Power spectrum}\label{App:3Dhalospec}
At tree level, the only contribution is that from first order bias and first order (linear) perturbation theory. Thus the halo power spectrum is given by
\ba
P_{hh}(k|M_{12},z_{12}) = b_1(M_1,z_1) \ b_1(M_2,z_2) \ P_\mr{lin}(k|z_{12}).
\ea

\subsection{Bispectrum}\label{App:3Dhalobispec}
For the bispectrum, the first order contribution is zero, so one needs to go at second order in both perturbation theory and local and non-local bias. The halo bispectrum thus splits into three terms
\ba
B_{hhh}(k_{123}|M_{123},z_{123}) =  B^\mr{b2} + B^\mr{s2} + B^\mr{2PT}
\ea
where the second order local halo bias term is
\ba
\nonumber B_{hhh}^\mr{b2}(k_{123}|M_{123},z_{123}) = b_1(M_1,z_1) \ b_1(M_2,z_2) \ b_2(M_3,z_3) \\
\times P(k_1|z_{13}) \ P(k_2|z_{23}) + 2 \ \mr{perm.}
\ea
the tidal tensor term is
\ba
B_{hhh}^\mr{s2}(k_{123}|M_{123},z_{123}) = b_1(M_1,z_1) \ b_1(M_2,z_2) \ b_\mr{s2}(M_3,z_3) \\
\times \left(2! \ S_2(\kk_1,\kk_2) \ P(k_1|z_{13}) \ P(k_2|z_{23}) + 2 \ \mr{perm.} \right)
\ea
and finally the second order perturbation theory term is
\ba
\nonumber B_{hhh}^\mr{2PT}(k_{123}|M_{123},z_{123}) = b_1(M_1,z_1) \ b_1(M_2,z_2) \ b_1(M_3,z_3) \\
\times \left(2! \ F_2(\kk_1,\kk_2) \ P(k_1|z_{13}) \ P(k_2|z_{23}) + 2 \ \mr{perm.} \right) 
\ea
with \citep{Fry1984}
\ba
F_2^s(\kk_1,\kk_2) = \frac{5}{7} + \frac{1}{2} \left(\frac{k_1}{k_2}+\frac{k_2}{k_1}\right) \frac{\kk_1\cdot\kk_2}{k_1 \, k_2} + \frac{2}{7} \frac{\left(\kk_1\cdot\kk_2\right)^2}{k_1^2 \, k_2^2}
\ea
These can be summarised in a general equation
\ba
\nonumber B_{hhh}^\mr{X}(k_{123}|M_{123},z_{123}) = 2! \ b_1(M_1,z_1) \ b_1(M_2,z_2) \ b_X(M_3,z_3) \\
\times K_X(\kk_1,\kk_2) \ P(k_1|z_{13}) \ P(k_2|z_{23}) + 2 \ \mr{perm.}
\ea
with $X\in \{\mr{b2,s2,2PT}\}$, the kernels are
\ba
K_\mr{b2} = \frac{1}{2!} \qquad K_\mr{s2} = S_2 \qquad K_\mr{2PT} = F_2^s
\ea
and the biases are the expected ones
\ba
b_\mr{b2} = b_2 \qquad b_\mr{s2} = b_\mr{s2} \qquad b_\mr{2PT} = b_1
\ea
These notations will become useful in the next subsection.

\subsection{Trispectrum}\label{App:3Dhalotrispec}
For the trispectrum, at tree-level one needs either to go at second other in two density contrasts simultaneously, or go to third order in one density contrast. The halo trispectrum thus splits into three terms
\ba
\nonumber T_{hhhh}(\kk_{1234}|M_{1234},z_{1234}) = T^\mr{2 \times 2} + T^\mr{b3} + T^\mr{3PT}
\ea
where, using the notations of Sect. \ref{App:3Dhalobispec}
\ba
\nonumber T_{hhhh}^\mr{2\times 2}(\kk_{1234}|M_{1234},z_{1234}) = \sum_{X,Y\in \{\mr{b2,s2,2PT}\}} b_1(M_1,z_1) \ b_1(M_2,z_2) \\
\nonumber b_X(M_3,z_3) \ b_Y(M_4,z_4) \ \Big[ (2!)^2 \ K_X(\kk_{1+3},-\kk_1)  \\ 
\nonumber \times K_Y(\kk_{1+3},\kk_2) \ P(k_{1+3}|z_{34}) \ P(k_1|z_{13}) \ P(k_2|z_{24}) \Big] \\
+ 11 \ \mr{perm.}
\ea
Because the involved permutations are not completely trivial, here is another writing of this equation:
\ba
\nonumber T_{hhhh}^\mr{2\times 2}(\kk_{1234}|M_{1234},z_{1234}) = \sum_{\substack{X,Y\in \{\mr{b2,s2,2PT}\} \\ \{\alpha,\beta\} \in \{1,2,3,4\}}} b_1(M_\alpha,z_\alpha) \ b_1(M_\beta,z_\beta) \\
\nonumber b_X(M_\gamma,z_\gamma) \ b_Y(M_\delta,z_\delta) \ \Big[4\ K_X(\kk_{\alpha+\gamma},-\kk_\alpha) \ K_Y(\kk_{\alpha+\gamma},\kk_\beta) \\ 
P(k_{\alpha+\gamma}|z_{\gamma\delta}) \ P(k_\alpha|z_{\alpha\gamma}) \ P(k_\beta|z_{\beta\delta}) + (\gamma \leftrightarrow \delta)\Big]
\ea
where the sums runs over all six possible pairs $\{\alpha,\beta\}$, and where $\{\gamma,\delta\}$ are the remaining two indices in $\{1,2,3,4\}$.

The trispectrum from third order halo bias is
\ba
\nonumber T_{hhhh}^\mr{b3}(\kk_{1234}|M_{1234},z_{1234}) = b_1(M_1,z_1) \ b_1(M_2,z_2) \ b_1(M_3,z_3) \\
b_3(M_4,z_4) \ P(k_1|z_{14}) \ P(k_2|z_{24}) \ P(k_3|z_{34}) + 3 \ \mr{perm.}
\ea
and finally the trispectrum from third order perturbation theory is
\ba
\nonumber T_{hhhh}^\mr{3PT}(\kk_{1234}|M_{1234},z_{1234}) = b_1(M_1,z_1) \ b_1(M_2,z_2) \ b_1(M_3,z_3) \\
\nonumber b_1(M_4,z_4) \ \Big[3! \ F_3(\kk_1,\kk_2,\kk_3) \ P(k_1|z_{14}) \ P(k_2|z_{24}) \ P(k_3|z_{34}) \\
+ 3 \ \mr{perm.}\Big].
\ea

For the third order perturbation theory kernel, \cite[Eq. 39-43-44]{Bernardeau2002} gives the recurrence equations which can be used to derive all $F_n$ kernel
\ba
\nonumber F_n(\kk_1,\cdots,\kk_n) = & \sum_{m=1}^{n-1} \frac{G_m(\kk_1,\cdots,\kk_m)}{(2n+3)(n-1)} \\
\nonumber & \times \big[(2n+1) \ \alpha(\kk'_1,\kk'_2) \ F_{n-m}(\kk_{m+1},\cdots,\kk_n) \\
& + 2 \ \beta(\kk'_1,\kk'_2) \ G_{n-m}(\kk_{m+1},\cdots,\kk_n)\big]
\ea
where $\kk'_1=\kk_1+\cdots+\kk_m$ and $\kk'_2=\kk_{m+1}+\cdots+\kk_n$.\\
Using this recurrence equation, after symmetrisation of the kernel I find
\ba\label{Eq:F_3^s-Lacasa}
\nonumber F_3^s(\kk_1,\kk_2,\kk_3) = & \frac{1}{54} \ G_2^s(\kk_1,\kk_2) \ \Bigg(7+2\frac{\kk_{3}\cdot\kk_{1+2}}{k_3^2}+9\ \frac{\kk_{3}\cdot\kk_{1+2}}{k_{1+2}^2} \\
\nonumber & + 4 \ \frac{\left(\kk_{3}\cdot\kk_{1+2}\right)^2}{k_3^2 \, k_{1+2}^2}\Bigg) + \frac{7}{54} \ F_2^s(\kk_1,\kk_2) \ \left(1+\frac{\kk_{3}\cdot\kk_{1+2}}{k_3^2}\right) \\
& + 2 \ \mr{perm.}
\ea
where
\ba
G_2^s(\kk_1,\kk_2) = \frac{3}{7} + \frac{1}{2} \left(\frac{k_1}{k_2}+\frac{k_2}{k_1}\right) \frac{\kk_1\cdot\kk_2}{k_1 \, k_2} + \frac{4}{7} \frac{\left(\kk_1\cdot\kk_2\right)^2}{k_1^2 \, k_2^2}.
\ea
This result agrees with \cite{Nielsen2017}.

\subsection{Legendre decomposition}
It will be necessary for Sect. \ref{App:2Dproj-trisp-angdep} to project the dependence of trispectra on wavevectors angles onto Legendre polynomials $P_n$. To this end, I give here the decomposition of the halo bispectrum and trispectrum in Legendre series.

Using results from Sect. \ref{App:3Dhalobispec}, the halo bispectrum writes
\ba
\nonumber B_{hhh}(k_{123}|M_{123},z_{123}) = \sum_{X\in \{\mr{b2,s2,2PT}\}} b_1(M_1,z_1) \ b_1(M_2,z_2) \ b_X(M_3,z_3) \\
\times \sum_{n=0}^\infty K_{X,n}(k_1,k_2) \ P_n(\hk_1\cdot\hk_2) \ P(k_1) \ P(k_2) + 2 \ \mr{perm.}
\ea
where I note $P_n^{i,j}\equiv P_n(\hk_i,\hk_j)$, and
\ba
K_{\mr{b2},n} &= \delta_{n,0} \\
K_{\mr{S2},n} &= \frac{2}{3}\delta_{n,2} \\
K_{\mr{2PT},n} &= \frac{17}{21}\delta_{n,0} + \frac{1}{2} \left(\frac{k_1}{k_2}+\frac{k_2}{k_1}\right)\delta_{n,1} + \frac{4}{21}\delta_{n,2}
\ea

Using results from Sect. \ref{App:3Dhalotrispec}, the halo trispectrum writes
\ba
\nonumber T_{hhhh}(\kk_{1234}|M_{1234},z_{1234}) = T^\mr{2 \times 2} + T^\mr{b3} + T^\mr{3PT}
\ea
with
\ba
\nonumber T_{hhhh}^\mr{2\times 2}(\kk_{1234}|M_{1234},z_{1234}) = \sum_{X,Y\in \{\mr{b2,s2,2PT}\}} \sum_{n,n'=0}^\infty K_{X,n}(k_{1+3},k_1) \\
\nonumber K_{Y,n'}(k_{1+3},k_2) \ P_n(-\hk_{1+3}\cdot\hk_1) \; P_{n'}(\hk_{1+3}\cdot\hk_2) \\
\times P(k_{1+3}) \; P(k_1) \; P(k_2) + 11 \ \mr{perm.}
\ea
The third order bias bispectrum has no dependence on wavevectors angles (equivalently it decomposes with just a $(n,n')=(0,0)$ term).\\
For the 3PT contribution, noting that
\ba
F_2(\kk_1,\kk_2) = & \frac{17}{21} \ P_0^{1,2} + \frac{1}{2} \left(\frac{k_1}{k_2}+\frac{k_2}{k_1}\right) \ P_1^{1,2} + \frac{4}{21} \ P_2^{1,2} \\
G_2(\kk_1,\kk_2) = & \frac{13}{21} \ P_0^{1,2} + \frac{1}{2} \left(\frac{k_1}{k_2}+\frac{k_2}{k_1}\right) \ P_1^{1,2} + \frac{8}{21} \ P_2^{1,2}
\ea
and
\ba
\nonumber F_3(\kk_1,\kk_2,\kk_3) =& \frac{1}{54} \ G_2(\kk_1,\kk_2) \ \Bigg[\frac{25}{3}P_0^{3,1+2}+\left(\frac{2}{k_3}+\frac{9}{k_{1+2}}\right)P_1^{3,1+2} \\
\nonumber & + \frac{8}{3} P_2^{3,1+2}\Bigg] + \frac{7}{54} \ F_2(\kk_1,\kk_2) \ \left[1+\frac{1}{k_3}P_1^{3,1+2}\right] \\
& + 2 \ \mr{perm.}
\ea
one finds
\ba
\nonumber F_3(\kk_1,\kk_2,\kk_3) =& \frac{341}{1701} \ P_0^{1,2} \ P_0^{3,1+2} \\
\nonumber & + \left(\frac{145}{1134}\frac{k_{1+2}}{k_3}+\frac{13}{126}\frac{k_{3}}{k_{1+2}}\right) \ P_0^{1,2} \ P_1^{3,1+2} \\
\nonumber & + \frac{52}{1701} \ P_0^{1,2} \ P_2^{3,1+2} \\
\nonumber & + \frac{23}{162} \left(\frac{k_1}{k_2}+\frac{k_2}{k_1}\right) \ P_1^{1,2} \ P_0^{3,1+2} \\
\nonumber & + \frac{1}{12} \left(\frac{k_1}{k_2}+\frac{k_2}{k_1}\right) \left(\frac{k_{1+2}}{k_3}+\frac{k_3}{k_{1+2}}\right) \ P_1^{1,2} \ P_1^{3,1+2} \\
\nonumber & + \frac{2}{81} \left(\frac{k_1}{k_2}+\frac{k_2}{k_1}\right) \ P_1^{1,2} \ P_2^{3,1+2} \\
\nonumber & + \frac{142}{1701} \ P_2^{1,2} \ P_0^{3,1+2} \\
\nonumber & + \left(\frac{22}{567}\frac{k_{1+2}}{k_3}+\frac{4}{63}\frac{k_{3}}{k_{1+2}}\right) \ P_2^{1,2} \ P_1^{3,1+2} \\
\nonumber & + \frac{32}{1701} \ P_2^{1,2} \ P_2^{3,1+2} \\
& + 2 \ \mr{perm.} \\
& \equiv  \sum_{n,n'=0}^2 F_{3;n,n'}^{1,2;3,1+2}(k_{123}) P_n^{1,2} \ P_{n'}^{3,1+2} + 2 \ \mr{perm.}
\ea
thus the trispectrum term from third order perturbation is
\ba
\nonumber T_{hhhh}^\mr{3PT}(\kk_{1234}|M_{1234},z_{1234}) = \left[b_1(M_i,z_i)\right]_{i=1234} \bigg(\sum_{n,n'} F_{3;n,n'}^{1,2;3,1+2}(k_{123}) \\
\times P_n^{1,2} \ P_{n'}^{3,1+2} \ P(k_1) \ P(k_2) \ P(k_3) + 11 \ \mr{perm.}\bigg)
\ea

\section{Projection of the 3D trispectrum to the covariance of the angular power spectrum I : angle-independent case}\label{App:2Dproj-trisp-angindep}

In both this section and Appendix \ref{App:2Dproj-trisp-angdep}, I derive the projection of the 3D trispectrum into the angular covariance $\mathcal{C}_{\ell,\ell'}$, examining the various cases of functionnal dependence of the trispectrum.
In all these cases, the underlying strategy is basically the same. Starting from Eq.~\ref{Eq:CovClgal-Lagrange}, the Legendre polynomials are expanded in spherical harmonics, and the wavevector dirac(s) is (are) Fourier transformed introducing one (diagonal-independent case) or two (diagonal-dependent case) auxiliary comoving volume integrand. Then integration of spherical harmonics over unit vectors $\hk_i$ is performed, yielding either Kroneckers of harmonic indices or Gaunt coefficients. Next, integration over unit vectors of the auxiliary comoving volume(s) is performed, yielding more Kroneckers or Gaunt coefficients, and any appearing simplification is carried out. Finally, sums over azimuthal parameters $m_i$ are performed and other simplifications may appear. I end up with a sum over potential auxiliary multipoles of a geometric coefficient multiplying an angular trispectrum.

That trispectrum is an integral over comoving volumes and wavenumbers, without dependence on any angle or unit wavevector. In the coming calculations, three type of functional form for these trispectra are going to appear.\\
The first type happens when there is a dependence on the squeezed diagonal $\mathbf{K}=\kk_{1+2}$ :
\ba
\nonumber T_{\ell,\ell,\ell',\ell',1+2}^{\lu,\ld,\ell_a,\ell_b,\lt,\ell_4} =& \left(\frac{2}{\pi}\right)^5 \int \dd V_{1234} \ k_{1234}^2\,\dd k_{1234} \ K^2\,\dd K \ x^2_{ab}\,\dd x_{ab} \\
\nonumber & j_\ell(k_1 r_1) \, j_{\lu}(k_1 x_a) \ j_\ell(k_2 r_2) \, j_{\ld}(k_2 x_a) \ j_{\ell_a}(K x_a) \\
\nonumber & j_{\ell_b}(K x_b) j_{\ell'}(k_3 r_3) \, j_{\ell_3}(k_3 x_b) \  j_{\ell'}(k_4 r_4) \, j_{\ell_4}(k_4 x_b) \\
& \times f(K,k_1,k_2,k_3,k_4)
\ea
The second type happens when there is a dependence on the alternate diagonal $\mathbf{K}=\kk_{1+3}$ :
\ba
\nonumber T_{\ell,\ell,\ell',\ell',1+3}^{\lu,\ld,\ell_a,\ell_b,\lt,\ell_4} =& \left(\frac{2}{\pi}\right)^5 \int \dd V_{1234} \ k_{1234}^2\,\dd k_{1234} \ K^2\,\dd K \ x^2_{ab}\,\dd x_{ab} \\
\nonumber & j_\ell(k_1 r_1) \, j_{\lu}(k_1 x_a) \ j_\ell(k_2 r_2) \, j_{\ld}(k_2 x_b) \ j_{\ell_a}(K x_a) \\
\nonumber & j_{\ell_b}(K x_b) j_{\ell'}(k_3 r_3) \, j_{\ell_3}(k_3 x_a) \  j_{\ell'}(k_4 r_4) \, j_{\ell_4}(k_4 x_b) \\
& \times f(K,k_1,k_2,k_3,k_4)
\ea
(another type would appear for the last diagonal $\mathbf{K}=\kk_{1+4}$, but this is just a symmetric version of the 1+3 type)\\
In the case $\ell_a=\ell_b$ and $f(K,k_1,k_2,k_3,k_4)$ independent of $K$, both trispectra reduce to the third type
\ba
\nonumber T_{\ell,\ell,\ell',\ell',\ci}^{\lu,\ld,\lt,\ell_4} =& \left(\frac{2}{\pi}\right)^4 \int \dd V_{1234} \ k_{1234}^2\,\dd k_{1234} \ x^2\,\dd x \ j_\ell(k_1 r_1) \, j_{\lu}(k_1 x) \\
\nonumber & j_\ell(k_2 r_2) \, j_{\ld}(k_2 x) \ j_{\ell'}(k_3 r_3) \, j_{\ell_3}(k_3 x) \  j_{\ell'}(k_4 r_4) \, j_{\ell_4}(k_4 x) \\
& \times f(k_1,k_2,k_3,k_4).
\ea
These notations prove useful in unifying the results of all the next subsections, both in this section and Appendix \ref{App:2Dproj-trisp-angdep}.\\
I now turn to examining each possible 3D trispectrum functional form and derive the corresponding covariance.

\subsection{Diagonal-independent trispectrum}\label{App:2Dproj-trisp-diagindep}
The trispectrum is said diagonal-independent if $T_\mr{gal}(\kk_{1234},z_{1234})=T_\mr{gal}(k_{1234},z_{1234})$, that is, it depends only on the length of the four Fourier wavevectors and not on their relative orientations. This is for example the case of the one-halo term, the two-halo 1+3 term and some shot-noise terms. In that case, the power spectrum covariance is
\ba
\nonumber \mathcal{C}_{\ell,\ell'} &= (4\pi)^2 \, (-1)^{\ell+\ell'}\int  \dd V_{1234} \frac{\dd^3\kk_{1234}}{(2\pi)^{12}} \; j_\ell(k_1 r_1) \,  j_\ell(k_2 r_2) \\
\nonumber & \qquad j_{\ell'}(k_3 r_3) \,  j_{\ell'}(k_4 r_4) \ P_\ell(\hk_1 \cdot \hk_2) \ P_{\ell'}(\hk_3 \cdot \hk_4) \\
& \qquad \times (2\pi)^3 \ \delta^{(3)}\left(\kk_1+\cdots+\kk_4\right) \ T_\mr{gal}(k_{1234},z_{1234}) \\
\nonumber &= \frac{(-1)^{\ell+\ell'}}{(2\ell+1)(2\ell'+1)} \int \dd V_{1234} \frac{\dd^3\kk_{1234}}{(2\pi^2)^{4}} j_\ell(k_1 r_1) \,  j_\ell(k_2 r_2) \\ 
\nonumber & \qquad j_{\ell'}(k_3 r_3) \,  j_{\ell'}(k_4 r_4) \sum_{m,m'} Y_{\ell m}(\hk_1) Y_{\ell m}^*(\hk_2) Y_{\ell' m'}(\hk_3) Y_{\ell' m'}^*(\hk_4)\\
& \qquad \times \int \dd^3\xx \ \mre^{\ii \xx \cdot (\kk_1+\kk_2+\kk_3+\kk_4)} \ T_\mr{gal}(k_{1234},z_{1234}) \\
\nonumber &= \frac{\left(\frac{2}{\pi}\right)^4 (-1)^{\ell+\ell'}}{(2\ell+1)(2\ell'+1)}  \int \dd V_{1234} \, \dd^3\kk_{1234} \, \dd^3\xx \, j_\ell(k_1 r_1) \,  j_\ell(k_2 r_2) \\ 
\nonumber & \qquad j_{\ell'}(k_3 r_3) \,  j_{\ell'}(k_4 r_4) \sum_{m,m'} Y_{\ell m}(\hk_1) Y_{\ell m}^*(\hk_2) Y_{\ell' m'}(\hk_3) Y_{\ell' m'}^*(\hk_4)\\
\nonumber & \qquad \times \sum_{1,2,3,4} \left[j_{\ell_i}(k_i x)\right]_{i=1234} Y_1^*(\hk_1) \, Y_1(\hx) \ Y_2(\hk_2) Y_2^*(\hx) \\ 
& \qquad \times \ii^{\lu+\ld+\lt+\lq} \ Y_3^*(\hk_3) \, Y_3(\hx) \ Y_4(\hk_4) \, Y_4^*(\hx) \ T_\mr{gal}(k_{1234},z_{1234}) \\
\nonumber &= \frac{\left(\frac{2}{\pi}\right)^4}{(2\ell+1)(2\ell'+1)}  \int \dd V_{1234} \, k^2_{1234} \dd k_{1234} \, \dd^3\xx\, j_\ell(k_1 r_1) \,  j_\ell(k_1 x) \\
\nonumber & \qquad j_\ell(k_2 r_2) j_\ell(k_2 x) \; j_{\ell'}(k_3 r_3)  j_{\ell'}(k_3 x) \;  j_{\ell'}(k_4 r_4) j_{\ell'}(k_4 x) \\
& \qquad \times \sum_{m,m'} Y_{\ell m}(\hx) Y_{\ell m}^*(\hx) Y_{\ell' m'}(\hx) Y_{\ell' m'}^*(\hx) \ T_\mr{gal}(k_{1234},z_{1234}) \\
\nonumber &= \frac{\left(\frac{2}{\pi}\right)^4}{(4\pi)^2}  \int \dd V_{1234} \, k^2_{1234} \dd k_{1234} \, \dd^3\xx\, j_\ell(k_1 r_1) \,  j_\ell(k_1 x) \\
\nonumber & \qquad j_\ell(k_2 r_2) j_\ell(k_2 x) \; j_{\ell'}(k_3 r_3)  j_{\ell'}(k_3 x) \;  j_{\ell'}(k_4 r_4) j_{\ell'}(k_4 x) \\
& \qquad \times P_\ell(\hx\cdot\hx) \, P_{\ell'}(\hx\cdot\hx) \ T_\mr{gal}(k_{1234},z_{1234}) \\
&= \frac{T_{\ell\ell\ell'\ell'}}{4\pi}
\ea
where the angular trispectrum is
\ba\label{Eq:angtrisp-diagindep}
\nonumber T_{\ell\ell\ell'\ell'} &= T_{\ell\ell\ell'\ell',\ci}^{\ell\ell\ell'\ell'} \\
\nonumber &=\left(\frac{2}{\pi}\right)^4 \int x^2 \, \dd x \; \dd V_{1234} \; k^2_{1234} \, \dd k_{1234} \ j_\ell(k_1 r_1) \,  j_\ell(k_1 x) \\
\nonumber & \qquad j_\ell(k_2 r_2) \, j_\ell(k_2 x) \ j_{\ell'}(k_3 r_3)  \, j_{\ell'}(k_3 x) \  j_{\ell'}(k_4 r_4) \, j_{\ell'}(k_4 x) \\ 
& \qquad \times T_\mr{gal}(k_{1234},z_{1234})
\ea
Using Limber's approximation \citep{Limber1953,LoVerde2008} on all wavevectors, the angular trispectrum simplifies to
\ba
\nonumber T_{\ell\ell\ell'\ell'} &= \delta_{i_z,j_z,k_z,l_z} \int \dd V \ T_\mr{gal}(k_{\ell_{1234}},z)
\ea
with $k_{\ell_i} = (\ell_i+1/2)/r(z)$ (where $\ell_1=\ell_2=\ell$ and $\ell_3=\ell_4=\ell'$) and $z\in i_z$

\subsection{Trispectrum depending on the squeezed diagonal}\label{App:2Dproj-trisp-sqzdiag}
In this case the trispectrum also depends on the length of the squeezed diagonal $k_{1+2}$
This case proceeds in a manner similar to that of section \ref{App:2Dproj-trisp-diagindep}, except that I force the diagonal to appear in the wavenumber dirac through the identity
\be
\delta^{(3)}\left(\kk_1+\cdots+\kk_4\right) = \int \dd^3\mathbf{K} \ \delta^{(3)}\left(\kk_1+\kk_2-\mathbf{K}\right) \ \delta^{(3)}\left(\mathbf{K}+\kk_3+\kk_4\right)
\ee
One then gets
\ba
\nonumber \mathcal{C}_{\ell,\ell'} &= \frac{(-1)^{\ell+\ell'}}{(2\ell+1)(2\ell'+1)} \int \dd V_{1234} \frac{\dd^3\kk_{1234}}{(2\pi^2)^4} \frac{\dd^3\mathbf{K}}{(2\pi)^3} j_\ell(k_1 r_1) \,  j_\ell(k_2 r_2) \\ 
\nonumber & \qquad j_{\ell'}(k_3 r_3) \,  j_{\ell'}(k_4 r_4) \sum_{m,m'} Y_{\ell m}(\hk_1) Y_{\ell m}^*(\hk_2) Y_{\ell' m'}(\hk_3) Y_{\ell' m'}^*(\hk_4)\\
& \qquad \times \int \dd^3\xx_{ab} \ \mre^{\ii \xx_a \cdot (\kk_1+\kk_2-\mathbf{K})} \ \mre^{\ii \xx_b \cdot (\mathbf{K}+\kk_3+\kk_4)} \ T_\mr{gal}(K,k_{1234},z_{1234}) \\
\nonumber &= \frac{\left(\frac{2}{\pi}\right)^5 (-1)^{\ell+\ell'}}{(2\ell+1)(2\ell'+1)} \int \dd V_{1234} \dd^3\kk_{1234} \dd^3\mathbf{K} \dd^3\xx_{ab} \ j_\ell(k_1 r_1) \,  j_\ell(k_2 r_2) \\ 
\nonumber & \qquad j_{\ell'}(k_3 r_3) \,  j_{\ell'}(k_4 r_4) \sum_{m,m'} Y_{\ell m}(\hk_1) Y_{\ell m}^*(\hk_2) Y_{\ell' m'}(\hk_3) Y_{\ell' m'}^*(\hk_4)\\
\nonumber & \qquad \sum_{1,2,a,3,4,b} \ii^{\lu+\ld-\ell_a+\ell_b+\lt+\lq} j_{\ell_1}(k_1 x_a) j_{\ell_2}(k_2 x_a) j_{\ell_a}(K x_a) \\
\nonumber & \qquad j_{\ell_b}(K x_b) j_{\ell_3}(k_3 x_b) j_{\ell_4}(k_4 x_b)\\
\nonumber & \qquad Y_1^*(\hk_1) \, Y_1(\hx_a) \ Y_2(\hk_2) \, Y_2^*(\hx_a) \ Y_a^*(\hat{K}) \, Y_a(\hx_a) \\ 
\nonumber & \qquad Y_b(\hat{K}) \, Y_b^*(\hx_b) \ Y_3^*(\hk_3) \, Y_3(\hx_b) \ Y_4(\hk_4) \, Y_4^*(\hx_b) \\
& \qquad \times T_\mr{gal}(K,k_{1234},z_{1234})\\
\nonumber &= \frac{\left(\frac{2}{\pi}\right)^5}{(2\ell+1)(2\ell'+1)} \int \dd V_{1234} \; k^2_{1234} \dd k_{1234} \; K^2\dd K \; \dd^3\xx_{ab} \\
\nonumber & \qquad j_\ell(k_1 r_1) \, j_\ell(k_1 x_a) \; j_\ell(k_2 r_2) \, j_\ell(k_2 x_a) \\ 
\nonumber & \qquad j_{\ell'}(k_3 r_3) \, j_{\ell'}(k_3 x_b) \;  j_{\ell'}(k_4 r_4) \, j_{\ell'}(k_4 x_b) \\
\nonumber & \qquad \sum_{m,m',a} j_{\ell_a}(K x_a) \, j_{\ell_a}(K x_b) \ Y_{\ell m}(\hx_a) \, Y_{\ell m}^*(\hx_a) \, Y_a(\hx_a) \\
& \qquad Y_a^*(\hx_b) \, Y_{\ell' m'}(\hx_b) \, Y_{\ell' m'}^*(\hx_b) \times T_\mr{gal}(K,k_{1234},z_{1234})\\
\nonumber &= \frac{T_{\ell\ell\ell'\ell'}^{\ell_\mr{diag}=0}}{4\pi}
\ea
where the angular trispectrum is
\ba\label{Eq:angtrisp-sqzdiag}
\nonumber T_{\ell\ell\ell'\ell'}^{\ell_\mr{diag}=0} &= T_{\ell,\ell,\ell',\ell',1+2}^{\ell,\ell,0,0,\ell',\ell'}\\
\nonumber &= \left(\frac{2}{\pi}\right)^5 \int \dd V_{1234} \; k^2_{1234} \dd k_{1234} \; K^2\dd K \; \dd V_{ab} \; j_0(K x_a) \, j_0(K x_b) \\
\nonumber & \qquad j_\ell(k_1 r_1) \, j_\ell(k_1 x_a) \; j_\ell(k_2 r_2) \, j_\ell(k_2 x_a) \\ 
\nonumber & \qquad j_{\ell'}(k_3 r_3) \, j_{\ell'}(k_3 x_b) \;  j_{\ell'}(k_4 r_4) \, j_{\ell'}(k_4 x_b) \\
& \qquad \times T_\mr{gal}(K,k_{1234},z_{1234})
\ea
with Limber's approximation on $k_{1234}$ (but not on $K$, since it aliases into the monopole) simplifying it to
\ba
\nonumber T_{\ell\ell\ell'\ell'}^{\ell_\mr{diag}=0} = \delta_{i_z,j_z} \ \delta_{k_z,l_z} \ \frac{2}{\pi} \int K^2\,\dd K \ \dd V_{ab} \ j_0(K x_a) \, j_0(K x_b) \\
\times T_\mr{gal}(K,k_{\ell},k_{\ell},k_{\ell'},k_{\ell'},z_{aabb})
\ea
where $x_{a/b}=r(z_{a/b})$, $z_a\in i_z$, $z_b\in k_z$, and
\ba
k_\ell = \frac{\ell+1/2}{x_a} \quad \mr{and} \quad k_{\ell'} = \frac{\ell'+1/2}{x_b}
\ea

\subsection{Trispectrum depending on one of the other diagonals}\label{App:2Dproj-trisp-altdiag}
In this case the trispectrum depends on $K=k_{1+3}$ (or $k_{1+4}$, which is a symmetric case), additionally to $k_{1234}$. The case proceeds similarly to section \ref{App:2Dproj-trisp-sqzdiag}, except that the diagonal will now produce a mixing between $\ell$ and $\ell'$
\ba
\nonumber \mathcal{C}_{\ell,\ell'} &= \frac{(-1)^{\ell+\ell'}}{(2\ell+1)(2\ell'+1)} \int \dd V_{1234} \frac{\dd^3\kk_{1234}}{(2\pi^2)^4} \frac{\dd^3\mathbf{K}}{(2\pi)^3} j_\ell(k_1 r_1) \,  j_\ell(k_2 r_2) \\ 
\nonumber & \qquad j_{\ell'}(k_3 r_3) \,  j_{\ell'}(k_4 r_4) \sum_{m,m'} Y_{\ell m}(\hk_1) Y_{\ell m}^*(\hk_2) Y_{\ell' m'}(\hk_3) Y_{\ell' m'}^*(\hk_4)\\
& \qquad \times \int \dd^3\xx_{ab} \ \mre^{\ii \xx_a \cdot (\kk_1+\kk_3-\mathbf{K})} \ \mre^{\ii \xx_b \cdot (\mathbf{K}+\kk_2+\kk_4)} \ T_\mr{gal}(K,k_{1234},z_{1234}) \\
\nonumber &= \frac{\left(\frac{2}{\pi}\right)^5 (-1)^{\ell+\ell'}}{(2\ell+1)(2\ell'+1)} \int \dd V_{1234} \ \dd^3\kk_{1234} \ \dd^3\mathbf{K} \ \dd^3\xx_{ab} \\ 
\nonumber & \qquad j_\ell(k_1 r_1) \,  j_\ell(k_2 r_2) \ j_{\ell'}(k_3 r_3) \,  j_{\ell'}(k_4 r_4) \\ 
\nonumber & \qquad  \sum_{m,m'} Y_{\ell m}(\hk_1) Y_{\ell m}^*(\hk_2) Y_{\ell' m'}(\hk_3) Y_{\ell' m'}^*(\hk_4)\\
\nonumber & \qquad \sum_{1,2,a,3,4,b} \ii^{\lu+\ld-\ell_a+\ell_b+\lt+\lq} j_{\ell_1}(k_1 x_a) j_{\ell_3}(k_3 x_a) j_{\ell_a}(K x_a) \\
\nonumber & \qquad j_{\ell_b}(K x_b) j_{\ell_2}(k_2 x_b) j_{\ell_4}(k_4 x_b)\\
\nonumber & \qquad Y_1^*(\hk_1) \, Y_1(\hx_a) \ Y_3^*(\hk_3) \, Y_3(\hx_a) \ Y_a^*(\hat{K}) \, Y_a(\hx_a) \\ 
\nonumber & \qquad Y_b(\hat{K}) \, Y_b^*(\hx_b) \ Y_2(\hk_2) \, Y_2^*(\hx_b) \ Y_4(\hk_4) \, Y_4^*(\hx_b) \\
& \qquad \times T_\mr{gal}(K,k_{1234},z_{1234}) \\
\nonumber &= \frac{\left(\frac{2}{\pi}\right)^5}{(2\ell+1)(2\ell'+1)} \int \dd V_{1234} \ k^2_{1234} \, \dd k_{1234} \ K^2 \, \dd K \ \dd^3\xx_{ab} \\
\nonumber & \qquad j_\ell(k_1 r_1) \, j_\ell(k_1 x_a) \; j_\ell(k_2 r_2) \, j_\ell(k_2 x_b) \\ 
\nonumber & \qquad j_{\ell'}(k_3 r_3) \, j_{\ell'}(k_3 x_a) \;  j_{\ell'}(k_4 r_4) \, j_{\ell'}(k_4 x_b) \\
\nonumber & \qquad \sum_{m,m',a} j_{\ell_a}(K x_a) j_{\ell_a}(K x_b) \ Y_{\ell m}(\hx_a) Y_{\ell m}^*(\hx_b) Y_a(\hx_a)  \\
& \qquad Y_a^*(\hx_b) Y_{\ell' m'}(\hx_a) Y_{\ell' m'}^*(\hx_b) \times T_\mr{gal}(K,k_{1234},z_{1234}) \\
\nonumber &= \left(\frac{2}{\pi}\right)^5 \sum_{\ell_a} \frac{2\ell_a+1}{4\pi} \int \dd V_{1234} \ k^2_{1234} \, \dd k_{1234} \ K^2 \, \dd K \ \dd V_{ab} \ \dd \mu \\
\nonumber & \qquad j_\ell(k_1 r_1) \, j_\ell(k_1 x_a) \; j_\ell(k_2 r_2) \, j_\ell(k_2 x_b) \\ 
\nonumber & \qquad j_{\ell'}(k_3 r_3) \, j_{\ell'}(k_3 x_a) \;  j_{\ell'}(k_4 r_4) \, j_{\ell'}(k_4 x_b) \\
\nonumber & \qquad j_{\ell_a}(K x_a) j_{\ell_a}(K x_b) \ \frac{1}{2} P_\ell(\mu) P_{\ell'}(\mu) P_{\ell_a}(\mu) \\
& \qquad \times T_\mr{gal}(K,k_{1234},z_{1234}) \\
&= \sum_{\ell_a} \frac{2\ell_a+1}{4\pi} \threeJz{\ell}{\ell'}{\ell_a}^2 T_{\ell\ell\ell'\ell'}^{\ell_a}
\ea
where the sum runs over multipoles following the triangular inequality $|\ell-\ell'|\leq\ell_a\leq\ell+\ell'$ and the parity condition $\ell+\ell'+\ell_a$ even, and the angular trispectrum is
\ba\label{Eq:angtrisp-altdiag}
\nonumber T_{\ell\ell\ell'\ell'}^{\ell_a} =& T_{\ell,\ell,\ell',\ell',1+3}^{\ell,\ell,\ell_a,\ell_a,\ell',\ell'}\\
\nonumber =& \left(\frac{2}{\pi}\right)^5 \int \dd V_{1234} \ k^2_{1234}\,\dd k_{1234} \ K^2\,\dd K \ \dd V_{ab} \ j_\ell(k_1 r_1) \, j_\ell(k_1 x_a) \\
\nonumber & j_\ell(k_2 r_2) \, j_\ell(k_2 x_b) \ j_{\ell_a}(K x_a) \, j_{\ell_a}(K x_b) \ j_{\ell'}(k_3 r_3) \, j_{\ell'}(k_3 x_a) \\ 
&  j_{\ell'}(k_4 r_4) \, j_{\ell'}(k_4 x_b) \times T_\mr{gal}(K,k_{1234},z_{1234})
\ea
Using Limber's approximation on $k_{1234}$, the angular trispectrum simplifies to
\ba
\nonumber T_{\ell\ell\ell'\ell'}^{\ell_a} = \delta_{i_z,k_z} \ \delta_{j_z,l_z} \ \frac{2}{\pi} \int K^2\,\dd K \ \dd V_{ab} \ j_{\ell_a}(K x_a) \, j_{\ell_a}(K x_b) \\
\times T_\mr{gal}(K,k_{\ell_{1234}},z_{abab})
\ea
where $x_{a/b}=r(z_{a/b})$ and
\ba
k_{\ell_1} = \frac{\ell+1/2}{x_a} \quad k_{\ell_2} = \frac{\ell+1/2}{x_b} \quad k_{\ell_3} = \frac{\ell'+1/2}{x_a} \quad k_{\ell_4} = \frac{\ell'+1/2}{x_b}
\ea
If Limber's approximation is also applied on $\ell_a$, one gets
\ba
T_{\ell\ell\ell'\ell'}^{\ell_a} = \delta_{i_z,j_z,k_z,l_z} \ \int \dd V \ T_\mr{gal}(K_{\ell_a},k_{\ell_{1234}},z)
\ea
where $z_a=z_b=z$ and $K_{\ell_a}=(\ell_a+1/2)/r(z)$

\section{Projection of the 3D trispectrum to the covariance of the angular power spectrum II : angle-dependent case}\label{App:2Dproj-trisp-angdep}

The three-halo and four-halo terms of the covariance involve respectively the halo bispectrum and trispectrum. For both these polyspectra, the terms coming from the tidal tensor and from perturbation theory contain kernels depending on angles between some wavevectors. Using Al-Kashi's theorem (/law of cosines) one could express all these angles in term of just wavenumbers $k_i$ and $k_{ij}$, and thus reduce to the cases studied in the previous section Appendix~\ref{App:2Dproj-trisp-angindep}. This is however not adequate as it yields non-separable expressions, and will be incorrect when using the Limber's approximation, due to angles being potentially fast-varying function of wavenumbers, particularly in the squeezed limit. The next subsection give the correct equations for projecting these angle-dependent trispectra into angular covariances.

\subsection{Trispectrum depending on one angle}\label{App:2Dproj-trisp-angdep-1angle}

The following properties of Gaunt coefficients will be useful
\ba
Y_{\ell,m}(\hx) \ Y_{\ell_1,m_1}(\hx)  = \sum_{\ell_a,m_a} G_{\ell,\ell_1,\ell_a}^{m,m_1,m_a} \ Y_{\ell_a,m_a}^*(\hx)
\ea
and
\ba
\sum_{m_1,m_2} G_{\lu\ld\ell}^{m_1 m_2 m} \ G_{\lu\ld\ell'}^{m_1 m_2 m'} = \frac{(2\ell+1)_{12}}{4\pi} \threeJz{\lu}{\ld}{\ell}^2 \ \delta_{\ell \ell'} \ \delta_{m m'}
\ea
implying
\ba
\sum_{m_1,m_2,m_3} G_{\lu\ld\lt}^{m_1 m_2 m_3} \ G_{\lu\ld\lt}^{m_1 m_2 m_3} = \frac{(2\ell+1)_{123}}{4\pi} \threeJz{\lu}{\ld}{\lt}^2
\ea

\subsubsection{Angle between base wavevectors $\hk_i$}\label{App:2Dproj-trisp-angdep-1angle-base}

A first useful remark is that a trispectrum dependence on an angle $\hk_i\cdot\hk_j$ never appears together with a dependence on one diagonal ($k_{1+2}$, $k_{1+3}$ or $k_{1+4}$). So, accounting for symmetries, the two case I have to consider are
\ba
T_\mr{gal}(\kk_{1234},z_{1234}) = P_n(\hk_1\cdot\hk_2) \ f(k_1,k_2,k_3,k_4)
\ea
and
\ba
T_\mr{gal}(\kk_{1234},z_{1234}) = P_n(\hk_1\cdot\hk_3) \ f(k_1,k_2,k_3,k_4)
\ea
where $P_n$ is the $n$-th order Legendre polynomial.

\myparagraph{The case of $\hk_1\cdot\hk_2$ }

In the case involving $\hk_1\cdot\hk_2$, the covariance reads
\ba
\nonumber \mathcal{C}_{\ell,\ell'} =& \frac{(-1)^{\ell+\ell'}}{(2\ell+1)(2\ell'+1)} \int \dd V_{1234} \frac{\dd^3\kk_{1234}}{(2\pi^2)^{4}} j_\ell(k_1 r_1) \,  j_\ell(k_2 r_2) \\ 
\nonumber & j_{\ell'}(k_3 r_3) \,  j_{\ell'}(k_4 r_4) \sum_{m,m'} Y_{\ell m}(\hk_1) Y_{\ell m}^*(\hk_2) Y_{\ell' m'}(\hk_3) Y_{\ell' m'}^*(\hk_4)\\
& \times \int \dd^3\xx \ \mre^{\ii \xx \cdot (\kk_1+\kk_2+\kk_3+\kk_4)} \ P_n(\hk_1\cdot\hk_2) \ f(k_{1234}) \\
\nonumber =& \frac{\left(\frac{2}{\pi}\right)^4 4\pi \, (-1)^{\ell+\ell'}}{(2\ell+1)(2\ell'+1)(2n+1)}  \int \dd V_{1234} \, \dd^3\kk_{1234} \, \dd^3\xx \\
\nonumber & j_\ell(k_1 r_1) \,  j_\ell(k_2 r_2) j_{\ell'}(k_3 r_3) \,  j_{\ell'}(k_4 r_4)  \\
\nonumber & \sum_{m,m',m_n} Y_{\ell m}(\hk_1) Y_{\ell m}^*(\hk_2) Y_{\ell' m'}(\hk_3) Y_{\ell' m'}^*(\hk_4) Y_{n m_n}(\hk_1) Y_{n m_n}^*(\hk_2)\\
\nonumber & \times \sum_{1,2,3,4} \ii^{\lu+\ld+\lt+\lq} \left[j_{\ell_i}(k_i x)\right]_{i=1234} Y_1(\hk_1) \, Y_1^*(\hx) \ Y_2^*(\hk_2) Y_2(\hx) \\ 
& \times Y_3^*(\hk_3) \, Y_3(\hx) \ Y_4(\hk_4) \, Y_4^*(\hx) \ f(k_{1234})
\ea
\ba
\nonumber \mathcal{C}_{\ell,\ell'}=& \frac{\left(\frac{2}{\pi}\right)^4 4\pi \, (-1)^{\ell}}{(2\ell+1)(2\ell'+1)(2n+1)}  \int \dd V_{1234} \, k_{1234}^2 \dd k_{1234} \, \dd^3\xx \\
\nonumber & j_\ell(k_1 r_1) \,  j_\ell(k_2 r_2) j_{\ell'}(k_3 r_3) \,  j_{\ell'}(k_4 r_4)  \\
\nonumber & \sum_{m,m',m_n,1,2} \ii^{\lu+\ld} \ G_{\ell,\ell_1,n}^{m,m_1,m_n} \ G_{\ell,\ell_2,n}^{m,m_2,m_n} \ j_{\lu}(k_1 x) \ j_{\ld}(k_2 x) \\
& \times j_{\ell'}(k_3 x) j_{\ell'}(k_4 x)  \ Y_1^*(\hx) \, Y_2(\hx) \ Y_{\ell' m'}(\hx) \, Y_{\ell' m'}^*(\hx) \ f(k_{1234}) \\
\nonumber =& \frac{\left(\frac{2}{\pi}\right)^4 (-1)^{\ell}}{(2\ell+1)(2n+1)}  \int \dd V_{1234} \, k_{1234}^2 \dd k_{1234} \, x^2\dd x \\
\nonumber & j_\ell(k_1 r_1) \,  j_\ell(k_2 r_2) j_{\ell'}(k_3 r_3) \,  j_{\ell'}(k_4 r_4) \, j_{\ell'}(k_3 x) j_{\ell'}(k_4 x) \\
\nonumber & \sum_{m,m_n,1,2} \ii^{\lu+\ld} \ G_{\ell,\ell_1,n}^{m,m_1,m_n} \ G_{\ell,\ell_2,n}^{m,m_2,m_n} \ \left[j_{\ell_i}(k_i x)\right]_{i=12} \\
& \times \delta_{\ell_1,\ell_2} \, \delta_{m_1,m_2} \ f(k_{1234}) \\
\nonumber =&  \left(\frac{2}{\pi}\right)^4  \int \dd V_{1234} \ k_{1234}^2  \, \dd k_{1234} \ x^2 \, \dd x \\
\nonumber & j_\ell(k_1 r_1) \,  j_\ell(k_2 r_2) j_{\ell'}(k_3 r_3) \,  j_{\ell'}(k_4 r_4) \, j_{\ell'}(k_3 x) j_{\ell'}(k_4 x) \\
& \sum_{\ell_a} \frac{2\ell_a+1}{4\pi} \ j_{\ell_a}(k_1 x) \ j_{\ell_a}(k_2 x) \threeJz{\ell}{\ell_a}{n}^2 \ (-1)^{\ell+\ell_a} f(k_{1234}) \\
=& \sum_{\ell_a} \frac{2\ell_a+1}{4\pi} \threeJz{\ell}{\ell_a}{n}^2 (-1)^n \ T_{\ell\ell\ell'\ell'}^{\ell_a}
\ea
where the sum runs over multipoles following the triangular inequality $|\ell-n|\leq\ell_a\leq\ell+n$ and the parity condition $\ell+n+\ell_a$ even, and the angular trispectrum is
\ba
\nonumber T_{\ell\ell\ell'\ell'}^{\ell_a} =& T_{\ell,\ell,\ell',\ell',\ci}^{\ell_a,\ell_a,\ell',\ell'} \\
\nonumber =& \left(\frac{2}{\pi}\right)^4 \int \dd V_{1234} \ k_{1234}^2  \, \dd k_{1234} \ x^2 \, \dd x \ j_\ell(k_1 r_1) \, j_{\ell_a}(k_1 x) \\
& j_\ell(k_2 r_2) \, j_{\ell_a}(k_2 x) \ j_{\ell'}(k_3 r_3) \, j_{\ell'}(k_3 x) \ j_{\ell'}(k_4 r_4)  j_{\ell'}(k_4 x) \ f(k_{1234})
\ea
Limber's approximation, if valid, can be used on $k_3$ and $k_4$, but not on $k_1$ and $k_2$, because a closed form expression for the integral of two distinct Bessel function\footnote{That is, for two identical multipoles one has the identity $\int k^2\dd k \ j_\ell(kr) \ j_\ell(kx) = \frac{\pi}{2 r^2} \delta(r-x)$ ; but I do not know any simplification of $\int k^2\dd k \ j_\ell(kr) \ j_{\ell'}(kx)$ for $\ell\neq\ell'$.} is unknown to the author. One finds
\ba
\nonumber T_{\ell\ell\ell'\ell'}^{\ell_a} =& \left(\frac{2}{\pi}\right)^2 \int \dd V_{12} \ k_{12}^2  \, \dd k_{12} \ x^2 \, \dd x \ j_\ell(k_1 r_1) \, j_{\ell_a}(k_1 x) \\
& \times j_\ell(k_2 r_2) \, j_{\ell_a}(k_2 x) \ f(k_1,k_2,k_{\ell'},k_{\ell'})
\ea
with $k_{\ell'}=(\ell'+1/2)/x$

\myparagraph{The case of $\hk_1\cdot\hk_3$ }
In the case involving $\hk_1\cdot\hk_3$, the covariance reads
\ba
\nonumber \mathcal{C}_{\ell,\ell'} =& \frac{(-1)^{\ell+\ell'}}{(2\ell+1)(2\ell'+1)} \int \dd V_{1234} \frac{\dd^3\kk_{1234}}{(2\pi^2)^{4}} j_\ell(k_1 r_1) \,  j_\ell(k_2 r_2) \\ 
\nonumber & j_{\ell'}(k_3 r_3) \,  j_{\ell'}(k_4 r_4) \sum_{m,m'} Y_{\ell m}(\hk_1) Y_{\ell m}^*(\hk_2) Y_{\ell' m'}^*(\hk_3) Y_{\ell' m'}(\hk_4)\\
& \times \int \dd^3\xx \ \mre^{\ii \xx \cdot (\kk_1+\kk_2+\kk_3+\kk_4)} \ P_n(\hk_1\cdot\hk_3) \ f(k_{1234}) \\
\nonumber =& \frac{\left(\frac{2}{\pi}\right)^4 4\pi \, (-1)^{\ell+\ell'}}{(2\ell+1)(2\ell'+1)(2n+1)}  \int \dd V_{1234} \, \dd^3\kk_{1234} \, \dd^3\xx \\
\nonumber & j_\ell(k_1 r_1) \,  j_\ell(k_2 r_2) j_{\ell'}(k_3 r_3) \,  j_{\ell'}(k_4 r_4)  \\
\nonumber & \sum_{m,m',m_n} Y_{\ell m}(\hk_1) Y_{\ell m}^*(\hk_2) Y_{\ell' m'}^*(\hk_3) Y_{\ell' m'}(\hk_4) Y_{n m_n}(\hk_1) Y_{n m_n}^*(\hk_3)\\
\nonumber & \times \sum_{1,2,3,4} \ii^{\lu+\ld+\lt+\lq} \ \left[j_{\ell_i}(k_i x)\right]_{i=1234} Y_1(\hk_1) \, Y_1^*(\hx) \\ 
& \times Y_2(\hk_2) \, Y_2^*(\hx) \ Y_3^*(\hk_3) \, Y_3(\hx) \ Y_4^*(\hk_4) \, Y_4(\hx) \ f(k_{1234}) \\
\nonumber =& \frac{\left(\frac{2}{\pi}\right)^4 4\pi \, (-1)^{\ell+\ell'}}{(2\ell+1)(2\ell'+1)(2n+1)}  \int \dd V_{1234} \, k_{1234}^2\dd k_{1234} \, \dd^3\xx \\
\nonumber & j_\ell(k_1 r_1) \,  j_\ell(k_2 r_2) \, j_{\ell'}(k_3 r_3) \,  j_{\ell'}(k_4 r_4) \, j_{\ell}(k_2 x) \, j_{\ell'}(k_4 x) \\
\nonumber & \sum_{m,m',m_n,1,3} \ii^{\lu+\ell+\lt+\ell'} \ G_{\ell,\ell_1,n}^{m,m_1,m_n} \ G_{\ell',\ell_3,n}^{m',m_3,m_n} \ \left[j_{\ell_i}(k_i x)\right]_{i=13} \\
& \times Y_1^*(\hx) \  Y_{\ell m}^*(\hx) \ Y_3(\hx) \ Y_{\ell' m'}(\hx) \ f(k_{1234}) 
\ea
\ba
\nonumber \mathcal{C}_{\ell,\ell'}=& \frac{\left(\frac{2}{\pi}\right)^4 4\pi \, (-1)^{\ell+\ell'}}{(2\ell+1)(2\ell'+1)(2n+1)}  \int \dd V_{1234} \, k_{1234}^2\dd k_{1234} \, \dd^3\xx \\
\nonumber & j_\ell(k_1 r_1) \,  j_\ell(k_2 r_2) \, j_{\ell'}(k_3 r_3) \,  j_{\ell'}(k_4 r_4) \, j_{\ell}(k_2 x) \, j_{\ell'}(k_4 x) \\
\nonumber & \sum_{m,m',m_n,1,3} \ii^{\lu+\ell+\lt+\ell'} \ G_{\ell,\ell_1,n}^{m,m_1,m_n} \ G_{\ell',\ell_3,n}^{m',m_3,m_n} \ \left[j_{\ell_i}(k_i x)\right]_{i=13} \\
& \times \sum_{a,b} G_{\ell,\ell_1,\ell_a}^{m,m_1,m_a} \ Y_{\ell_a,m_a}(\hx) \ G_{\ell',\ell_3,\ell_b}^{m',m_3,m_b} \ Y_{\ell_b,m_b}^*(\hx) \ f(k_{1234}) \\
\nonumber =& \frac{\left(\frac{2}{\pi}\right)^4 4\pi \, (-1)^{\ell+\ell'}}{(2\ell+1)(2\ell'+1)(2n+1)}  \int \dd V_{1234} \, k_{1234}^2\dd k_{1234} \, x^2\dd x \\
\nonumber & j_\ell(k_1 r_1) \,  j_\ell(k_2 r_2) \, j_{\ell'}(k_3 r_3) \,  j_{\ell'}(k_4 r_4) \, j_{\ell}(k_2 x) \, j_{\ell'}(k_4 x) \\
\nonumber & \sum_{m,m',m_n,1,3} \ii^{\lu+\ell+\lt+\ell'} \ G_{\ell,\ell_1,n}^{m,m_1,m_n} \ G_{\ell',\ell_3,n}^{m',m_3,m_n} \ \left[j_{\ell_i}(k_i x)\right]_{i=13} \\
& \times \sum_{a,b} G_{\ell,\ell_1,\ell_a}^{m,m_1,m_a} \ Y_{\ell_a,m_a}(\hx) \ G_{\ell',\ell_3,\ell_b}^{m',m_3,m_b} \ Y_{\ell_b,m_b}^*(\hx) \ f(k_{1234}) 
\ea
\ba
\nonumber \mathcal{C}_{\ell,\ell'}=& (-1)^{\ell+\ell'} (2n+1) \left(\frac{2}{\pi}\right)^4 \int \dd V_{1234} \ k_{1234}^2 \, \dd k_{1234} \ x^2 \, \dd x \\
\nonumber & j_\ell(k_1 r_1) \,  j_\ell(k_2 r_2) \, j_{\ell'}(k_3 r_3) \,  j_{\ell'}(k_4 r_4) \, j_{\ell}(k_2 x) \, j_{\ell'}(k_4 x) \\
\nonumber & \sum_{m_n,\ell_1,\ell_3,a} \ii^{\lu+\ell+\lt+\ell'} \ \frac{(2\ell+1)_{13}}{4\pi} \threeJz{\ell}{\ell_1}{n}^2 \\
& \times \threeJz{\ell'}{\ell_3}{n}^2 \delta_{n,\ell_a} \ \delta_{m_n,m_a} \ \left[j_{\ell_i}(k_i x)\right]_{i=13} \ f(k_{1234}) \\
\nonumber =& (-1)^{\ell+\ell'} (2n+1) \sum_{\ell_1,\ell_3} \ii^{\lu+\ell+\lt+\ell'} \frac{(2\ell+1)_{13}}{4\pi} \threeJz{\ell}{\ell_1}{n}^2 \\
& \times \threeJz{\ell'}{\ell_3}{n}^2 T^{\ell_1,\ell_3}_{\ell,\ell,\ell',\ell'}
\ea
where the sum runs over multipoles following the triangular inequalities $|\ell-n|\leq\ell_1\leq\ell+n$ and $|\ell'-n|\leq\ell_3\leq\ell'+n$, and the parity conditions $\ell+n+\ell_1$ and $\ell'+n+\ell_3$ even\footnote{This ensures that the covariance equation is real, since $\lu+\ell+\lt+\ell' = (\lu+\ell+n)+(\lt+\ell'+n)-2n$ is even.}, and the angular trispectrum is
\ba
\nonumber T^{\ell_1,\ell_3}_{\ell,\ell,\ell',\ell'} =& T_{\ell,\ell,\ell',\ell',\ci}^{\lu,\ell,\lt,\ell'} \\
\nonumber =& \left(\frac{2}{\pi}\right)^4 \int \dd V_{1234} \ k_{1234}^2 \, \dd k_{1234} \ x^2 \, \dd x \ f(k_{1234}) \ j_\ell(k_1 r_1) \, j_{\ell_1}(k_1 x) \\
& \times  j_\ell(k_2 r_2) \, j_{\ell}(k_2 x) \ j_{\ell'}(k_3 r_3) \, j_{\ell_3}(k_3 x) \ j_{\ell'}(k_4 r_4) \, j_{\ell'}(k_4 x)
\ea
Limber's approximation, if valid, can be used on $k_2$ and $k_4$, but not on $k_1$ and $k_3$. This yields:
\ba
\nonumber T^{\ell_1,\ell_3}_{\ell,\ell,\ell',\ell'} =& \left(\frac{2}{\pi}\right)^2  \int \dd V_{13} \ k_{13}^2  \, \dd k_{13} \ x^2 \, \dd x \ j_\ell(k_1 r_1) \, j_{\ell_1}(k_1 x) \\
& \times j_{\ell'}(k_3 r_3) \, j_{\ell_3}(k_3 x) \ f(k_1,k_{\ell},k_3,k_{\ell'})
\ea
with $k_{\ell}=(\ell+1/2)/x$ and $k_{\ell'}=(\ell'+1/2)/x$

\subsubsection{Angle with a diagonal $\hk_{1+2},\hk_{1+3}$ or $\hk_{1+4}$}\label{App:2Dproj-trisp-angdep-1angle-diag}

A useful remark is that there is never a dependence on an angle between two diagonals, for example, $\hk_{1+2}\cdot\hk_{1+3}$, the angles involved will always be between a diagonal and a base wavevector $\hk_i$. Moreover, using that for example, $\kk_{1+2}=-\kk_{3+4}$, the case can always reduced to that of an angle $\hk_{i+j}\cdot\hk_l$ where $l\neq i,j$. \\
Finally, a dependence on an angle with one diagonal never appears together with a dependence on a wavenumber of another diagonal. For instance if there is a dependence on $\hk_{1+2}\cdot\hk_{3}$, there can be a dependence on $k_{1+2}$ (and $k_1,k_2,k_3,k_4$) but there will be no dependence on $k_{1+3}$ or $k_{1+4}$.\\
Armed with these considerations and accounting for symmetries, there are two cases I have to consider:
\ba
T_\mr{gal}(\kk_{1234},z_{1234}) = P_n(\hk_{1+2}\cdot\hk_3) \ f(k_{1+2},k_1,k_2,k_3,k_4)
\ea
and
\ba
T_\mr{gal}(\kk_{1234},z_{1234}) = P_n(\hk_{1+3}\cdot\hk_2) \ f(k_{1+3},k_1,k_2,k_3,k_4)
\ea
where $P_n$ is the $n$-th order Legendre polynomial.

\myparagraph{Angle with the squeezed diagonal}

In this case,
\ba
T_\mr{gal}(\kk_{1234},z_{1234}) = P_n(\hk_{1+2}\cdot\hk_3) \ f(k_{1+2},k_1,k_2,k_3,k_4)
\ea
Then the covariance reads
\ba
\nonumber \mathcal{C}_{\ell,\ell'} =& \frac{(-1)^{\ell+\ell'}}{(2\ell+1)(2\ell'+1)} \int \dd V_{1234} \frac{\dd^3\kk_{1234}}{(2\pi^2)^{4}} \frac{\dd^3\mathbf{K}}{(2\pi)^3} j_\ell(k_1 r_1) \,  j_\ell(k_2 r_2) \\ 
\nonumber & j_{\ell'}(k_3 r_3) \,  j_{\ell'}(k_4 r_4) \sum_{m,m'} Y_{\ell m}(\hk_1) Y_{\ell m}^*(\hk_2) Y_{\ell' m'}^*(\hk_3) Y_{\ell' m'}(\hk_4)\\
\nonumber & \int \dd^3\xx_{ab} \ \mre^{\ii \xx_a \cdot (\kk_1+\kk_2-\mathbf{K})} \ \mre^{\ii \xx_b \cdot (\mathbf{K}+\kk_3+\kk_4)} \ P_n(\hat{K}\cdot\hk_3) \\
& \times f(K,k_1,k_2,k_3,k_4) \\
\nonumber =& \frac{\left(\frac{2}{\pi}\right)^5 4\pi \, (-1)^{\ell+\ell'}}{(2\ell+1)(2\ell'+1)(2n+1)} \int \dd V_{1234} \ \dd^3\kk_{1234} \ \dd^3\mathbf{K} \ \dd^3\xx_{ab} \\ 
\nonumber & j_\ell(k_1 r_1) \,  j_\ell(k_2 r_2) \, j_{\ell'}(k_3 r_3) \,  j_{\ell'}(k_4 r_4) \sum_{m,m'} Y_{\ell m}(\hk_1) Y_{\ell m}^*(\hk_2) \\
\nonumber & Y_{\ell' m'}^*(\hk_3) Y_{\ell' m'}(\hk_4) \sum_{1,2,a,b,3,4} \ii^{\lu+\ld-\ell_a+\ell_b+\lt+\lq} j_{\ell_1}(k_1 x_a) j_{\ell_2}(k_2 x_a) \\
\nonumber & j_{\ell_a}(K x_a) j_{\ell_b}(K x_b) j_{\ell_3}(k_3 x_b) j_{\ell_4}(k_4 x_b)\\
\nonumber & Y_1^*(\hk_1) \, Y_1(\hx_a) \ Y_2(\hk_2) \, Y_2^*(\hx_a) \ Y_a(\hat{K}) \, Y_a^*(\hx_a) \\ 
\nonumber & Y_b(\hat{K}) \, Y_b^*(\hx_b) \ Y_3^*(\hk_3) \, Y_3(\hx_b) \ Y_4^*(\hk_4) \, Y_4(\hx_b) \\
& \times \sum_{m_n} Y_{n m_n}(\hat{K}) \ Y_{n m_n}^*(\hk_3) \ f(K,k_1,k_2,k_3,k_4)\\
\nonumber =& \frac{\left(\frac{2}{\pi}\right)^5 4\pi \, (-1)^{\ell'}}{(2\ell+1)(2\ell'+1)(2n+1)} \int \dd V_{1234} k^2_{1234}\dd k_{1234} K^2\dd K \dd^3\xx_{ab} \\ 
\nonumber & j_\ell(k_1 r_1) \,  j_\ell(k_2 r_2) \, j_{\ell'}(k_3 r_3) \,  j_{\ell'}(k_4 r_4) \\
\nonumber & \sum_{m,m',m_n,a,3,b} \ii^{-\ell_a+\ell_b+\lt+\ell'} \ G_{\ell_a,\ell_b,n}^{m_a,m_b,m_n} \ G_{\ell',\ell_3,n}^{m',m_3,m_n} \\
\nonumber & j_{\ell}(k_1 x_a) j_{\ell}(k_2 x_a) j_{\ell_a}(K x_a) j_{\ell_b}(K x_b) j_{\ell_3}(k_3 x_b) j_{\ell'}(k_4 x_b)\\
\nonumber & Y_{\ell m}(\hx_a) \ Y_{\ell m}^*(\hx_a) \ Y_a^*(\hx_a) Y_b^*(\hx_b) \ Y_3(\hx_b) \ Y_{\ell' m'}(\hx_b) \\ 
& \times f(K,k_1,k_2,k_3,k_4)
\ea
\ba
\nonumber \mathcal{C}_{\ell,\ell'} =& \frac{\left(\frac{2}{\pi}\right)^5 (-1)^{\ell'}}{(2\ell'+1)(2n+1)} \int \dd V_{1234} k^2_{1234}\dd k_{1234} K^2\dd K \dd^3\xx_{ab} \\ 
\nonumber & j_\ell(k_1 r_1) \,  j_\ell(k_2 r_2) \, j_{\ell'}(k_3 r_3) \,  j_{\ell'}(k_4 r_4) \\
\nonumber & \sum_{m',m_n,3,b} \ii^{\ell_b+\lt+\ell'} \ G_{0,\ell_b,n}^{0,m_b,m_n} \ G_{\ell',\ell_3,n}^{m',m_3,m_n} \\
\nonumber & j_{\ell}(k_1 x_a) j_{\ell}(k_2 x_a) j_{0}(K x_a) j_{\ell_b}(K x_b) j_{\ell_3}(k_3 x_b) j_{\ell'}(k_4 x_b)\\
\nonumber & Y_{00}^*(\hx_a) Y_b^*(\hx_b) \ Y_3(\hx_b) \ Y_{\ell' m'}(\hx_b) \\ 
& \times f(K,k_1,k_2,k_3,k_4)\\
\nonumber =& \frac{\left(\frac{2}{\pi}\right)^5 (-1)^{\ell'}}{(2\ell'+1)(2n+1)} \int \dd V_{1234} k^2_{1234}\dd k_{1234} K^2\dd K x_{ab}^2\dd x_{ab} \\ 
\nonumber & j_\ell(k_1 r_1) \,  j_\ell(k_2 r_2) \, j_{\ell'}(k_3 r_3) \,  j_{\ell'}(k_4 r_4) \\
\nonumber & \sum_{m',m_n,3} \ii^{n+\lt+\ell'} \ G_{\ell',\ell_3,n}^{m',m_3,m_n} \ G_{\ell',\ell_3,n}^{m',m_3,m_n} \\
\nonumber & j_{\ell}(k_1 x_a) j_{\ell}(k_2 x_a) j_{0}(K x_a) j_{n}(K x_b) j_{\ell_3}(k_3 x_b) j_{\ell'}(k_4 x_b) \\ 
& \times f(K,k_1,k_2,k_3,k_4)\\
\nonumber =& (-1)^{\ell'} \sum_{\lt} \frac{2\lt+1}{4\pi} \threeJz{\ell'}{n}{\lt}^2 \ii^{\ell'+n+\lt} T_{\ell\ell\ell'\ell'}^{\lt,n} 
\ea
where the sum runs over multipoles following the triangular inequality $|\ell'-n|\leq\ell_3\leq\ell'+n$, and the parity condition $\ell'+n+\ell_3$ even, and the angular trispectrum is
\ba
\nonumber T_{\ell\ell\ell'\ell'}^{\lt,n} =& T_{\ell\ell\ell'\ell',1+2}^{\ell,\ell,0,n,\lt,\ell'} \\
\nonumber =& \left(\frac{2}{\pi}\right)^5 \int \dd V_{1234} \ k^2_{1234}\,\dd k_{1234} \ K^2\,\dd K \ x_{ab}^2\,\dd x_{ab} \\
\nonumber & j_\ell(k_1 r_1) \, j_{\ell}(k_1 x_a) \ j_\ell(k_2 r_2) \, j_{\ell}(k_2 x_a) \ j_{\ell'}(k_3 r_3) \,  j_{\ell_3}(k_3 x_b)  \\
& j_{\ell'}(k_4 r_4) \, j_{\ell'}(k_4 x_b) \ j_{0}(K x_a) \, j_{n}(K x_b)   \times f(K,k_1,k_2,k_3,k_4)
\ea
Limber's approximation, if valid, can be used on $k_1$, $k_2$ and $k_4$, but not on $k_3$ and $K$. It yields
\ba
\nonumber T_{\ell\ell\ell'\ell'}^{\lt,n} = \delta_{i_z,j_z} \left(\frac{2}{\pi}\right)^2 \int \dd V_{134} \ k^2_{3}\,\dd k_{3} \ K^2\,\dd K \ j_{\ell'}(k_3 r_3) \,  j_{\ell_3}(k_3 r_4) \\
\times j_{0}(K r_1) \, j_{n}(K r_4) \ f(K,k_\ell,k_\ell,k_3,k_{\ell'})
\ea
with $k_\ell=(\ell+1/2)/r_1$ and $k_{\ell'}=(\ell'+1/2)/r_4$.

\myparagraph{Angle with an alternate diagonal}

In this case,
\ba
T_\mr{gal}(\kk_{1234},z_{1234}) = P_n(\hk_{1+3}\cdot\hk_2) \ f(k_{1+3},k_1,k_2,k_3,k_4)
\ea
Then the covariance reads
\ba
\nonumber \mathcal{C}_{\ell,\ell'} =& \frac{(-1)^{\ell+\ell'}}{(2\ell+1)(2\ell'+1)} \int \dd V_{1234} \frac{\dd^3\kk_{1234}}{(2\pi^2)^{4}} \frac{\dd^3\mathbf{K}}{(2\pi)^3} j_\ell(k_1 r_1) \,  j_\ell(k_2 r_2) \\ 
\nonumber & j_{\ell'}(k_3 r_3) \,  j_{\ell'}(k_4 r_4) \sum_{m,m'} Y_{\ell m}(\hk_1) Y_{\ell m}^*(\hk_2) Y_{\ell' m'}^*(\hk_3) Y_{\ell' m'}(\hk_4)\\
\nonumber & \int \dd^3\xx_{ab} \ \mre^{\ii \xx_a \cdot (\kk_1+\kk_3-\mathbf{K})} \ \mre^{\ii \xx_b \cdot (\mathbf{K}+\mathbf{K}+\kk_2+\kk_4)} \ P_n(\hat{K}\cdot\hk_2) \\
& \times f(K,k_1,k_2,k_3,k_4) \\
\nonumber =& \frac{\left(\frac{2}{\pi}\right)^5 4\pi \, (-1)^{\ell+\ell'}}{(2\ell+1)(2\ell'+1)(2n+1)} \int \dd V_{1234} \ \dd^3\kk_{1234} \ \dd^3\mathbf{K} \ \dd^3\xx_{ab} \\ 
\nonumber & j_\ell(k_1 r_1) \,  j_\ell(k_2 r_2) \, j_{\ell'}(k_3 r_3) \,  j_{\ell'}(k_4 r_4) \sum_{m,m'} Y_{\ell m}(\hk_1) Y_{\ell m}^*(\hk_2) \\
\nonumber & Y_{\ell' m'}^*(\hk_3) Y_{\ell' m'}(\hk_4) \sum_{1,3,a,b,2,4} \ii^{\lu+\lt-\ell_a+\ell_b+\ld+\lq} j_{\ell_1}(k_1 x_a) j_{\ell_3}(k_3 x_a) \\
\nonumber & j_{\ell_a}(K x_a) j_{\ell_b}(K x_b) j_{\ell_2}(k_2 x_b) j_{\ell_4}(k_4 x_b)\\
\nonumber & Y_1^*(\hk_1) \, Y_1(\hx_a) \ Y_3(\hk_3) \, Y_3^*(\hx_a) \ Y_a(\hat{K}) \, Y_a^*(\hx_a) \\ 
\nonumber & Y_b(\hat{K}) \, Y_b^*(\hx_b) \ Y_2^*(\hk_2) \, Y_2(\hx_b) \ Y_4^*(\hk_4) \, Y_4(\hx_b) \\
& \times \sum_{m_n} Y_{n m_n}(\hat{K}) \ Y_{n m_n}^*(\hk_2) \ f(K,k_1,k_2,k_3,k_4)\\
\nonumber =& \frac{\left(\frac{2}{\pi}\right)^5 4\pi \, (-1)^{\ell}}{(2\ell+1)(2\ell'+1)(2n+1)} \int \dd V_{1234} k^2_{1234}\dd k_{1234} K^2\dd K \dd^3\xx_{ab} \\ 
\nonumber & j_\ell(k_1 r_1) \,  j_\ell(k_2 r_2) \, j_{\ell'}(k_3 r_3) \,  j_{\ell'}(k_4 r_4) \\
\nonumber & \sum_{m,m',m_n,a,b,2} \ii^{\ell-\ell_a+\ell_b+\ld} \ (-1)^{m+m_2+m_n}\ G_{\ell,\ell_2,n}^{-m,-m_2,-m_n} \ G_{\ell_a,\ell_b,n}^{m_a,m_b,m_n} \\
\nonumber & j_{\ell}(k_1 x_a) j_{\ell'}(k_3 x_a) j_{\ell_a}(K x_a) j_{\ell_b}(K x_b) j_{\ell_2}(k_2 x_b) j_{\ell'}(k_4 x_b)\\
\nonumber & Y_{\ell m}(\hx_a) \ Y_{\ell' m'}^*(\hx_a) \ Y_a^*(\hx_a) Y_b^*(\hx_b) \ Y_2(\hx_b) \ Y_{\ell' m'}(\hx_b) \\ 
& \times f(K,k_1,k_2,k_3,k_4)\\
\nonumber =& \frac{\left(\frac{2}{\pi}\right)^5 4\pi \, (-1)^{\ell}}{(2\ell+1)(2\ell'+1)(2n+1)} \int \dd V_{1234} k^2_{1234}\dd k_{1234} K^2\dd K x^2_{ab}\dd x_{ab} \\ 
\nonumber & j_\ell(k_1 r_1) \,  j_\ell(k_2 r_2) \, j_{\ell'}(k_3 r_3) \,  j_{\ell'}(k_4 r_4) \!\!\!\!\sum_{m,m',m_n,a,b,2} \!\!\!\! \ii^{\ell-\ell_a+\ell_b+\ld} \\
\nonumber & j_{\ell}(k_1 x_a) j_{\ell'}(k_3 x_a) j_{\ell_a}(K x_a) j_{\ell_b}(K x_b) j_{\ell_2}(k_2 x_b) j_{\ell'}(k_4 x_b)\\
\nonumber & (-1)^{m+m_2+m_n+m'+m_a+m_b} \ G_{\ell,\ell_2,n}^{-m,-m_2,-m_n} \ G_{\ell_a,\ell_b,n}^{m_a,m_b,m_n} \ G_{\ell,\ell',\ell_a}^{m,-m',-m_a} \\
& \times G_{\ell_b,\ell_2,\ell'}^{-m_b,m_2,m'} \ f(K,k_1,k_2,k_3,k_4)\\
\nonumber =& (-1)^{\ell} \sum_{\ell_a,\ell_b,\ell_2} \ii^{\ell-\ell_a+\ell_b+\ld} \frac{(2\ld+1)(2\ell_a+1)(2\ell_b+1)}{4\pi} \\
& \times H_{\ell_2,\ell,\ell'}^{\ell_a,\ell_b,n} \ T_{\ell,\ell'}^{\ell_a,\ell_b,\ell_2} 
\ea
where I defined the geometric coefficient
\ba
\nonumber H_{\ell_2,\ell,\ell'}^{\ell_a,\ell_b,n} =& \frac{(4\pi)^2}{(2\ell+1)(2\ell'+1)(2n+1)(2\ld+1)(2\ell_a+1)(2\ell_b+1)} \\
\nonumber & \sum_{m,m',m_n,m_a,m_b,m_2} \!\!\!\! (-1)^{m_a+m_b+m_n+m_2+m+m'} \ G_{\ell_a,\ell_b,n}^{m_a,m_b,m_n} \ G_{\ell_a,\ell,\ell'}^{-m_a,m,-m'}  \\
& \times G_{\ell_2,\ell_b,\ell'}^{m_2,-m_b,m'} \ G_{\ell_2,\ell,n}^{-m_2,-m,-m_n} 
\ea
where the parity conditions respected by the Gaunt coefficients ensure that $\mathcal{C}_{\ell,\ell'}$ is real\footnote{e.g. $\ell-\ell_a+\ell_b+\ld=(\ell+\ld+n)+(\ell_a+\ell_b+n)-2\ell_a-2n$ is even.}. This coefficient will be shown to be,  in fact, a 6J symbol in Appendix \ref{App:H-and-6J}.\\
The angular trispectrum is given by
\ba
\nonumber T_{\ell,\ell'}^{\ell_a,\ell_b,\ell_2} =& T_{\ell,\ell,\ell',\ell',1+3}^{\ell,\ld,\ell_a,\ell_b,\ell',\ell'} \\
\nonumber =& \left(\frac{2}{\pi}\right)^5 \int \dd V_{1234} \ k^2_{1234}\,\dd k_{1234} \ K^2\,\dd K \ x^2_{ab}\,\dd x_{ab} \\
\nonumber & j_\ell(k_1 r_1) \,  j_{\ell}(k_1 x_a) \ j_\ell(k_2 r_2) \, j_{\ell_2}(k_2 x_b) \ j_{\ell'}(k_3 r_3) \, j_{\ell'}(k_3 x_a) \\
& j_{\ell'}(k_4 r_4) \, j_{\ell'}(k_4 x_b) \ j_{\ell_a}(K x_a) j_{\ell_b}(K x_b) \times f(K,k_1,k_2,k_3,k_4)
\ea
Limber's approximation, if valid, can be used on $k_1$, $k_3$ and $k_4$, but not on $k_2$ and $K$. It yields
\ba
\nonumber T_{\ell,\ell'}^{\ell_a,\ell_b,\ell_2} =& \left(\frac{2}{\pi}\right)^2 \int \dd V_{2} \ k^2_{2}\,\dd k_{2} \ K^2\,\dd K \ x_{ab}^2\,\dd x_{ab} \ j_{\ell}(k_2 r_2) \,  j_{\ell_2}(k_2 x_b) \\
&  j_{\ell_a}(K x_a) \, j_{\ell_b}(K x_b) \times f(K,k_\ell,k_2,k_{\ell'}(x_a),k_{\ell'}(x_b))
\ea
with $k_\ell=(\ell+1/2)/x_a$ and $k_{\ell'}(x_{a/b})=(\ell'+1/2)/x_{a/b}$.

\subsection{Trispectrum depending on two angles}\label{App:2Dproj-trisp-angdep-2angles}

From Appendix \ref{App:3Dhalotrispec}, one sees that the halo trispectrum brings terms of the form
\ba\label{Eq:trisp_legendre_2X2_k1+3}
T^{X\times Y}_{n,n'} = P_n(-\hk_{1+3}\cdot\hk_1) \; P_{n'}(\hk_{1+3}\cdot\hk_2) \ f(k_1,k_2,k_3,k_4,k_{1+3})
\ea
for some function of the wavevector moduli.\\
Furthermore the 3PT halo trispectrum term brings subterms of the form
\ba
T^{3PT}_{n,n'} = P_n(\hk_{1}\cdot\hk_3) \; P_{n'}(\hk_{1+3}\cdot\hk_2) \ f(k_1,k_2,k_3,k_4)
\ea
for another function of the moduli. There will also be similar terms in $k_{2+3}=k_{1+4}$ which will be deducible from the $k_{1+3}$ case by symmetry, and terms in $k_{1+2}=k_{3+4}$ (squeezed diagonal).
In the following I will first tackle the two cases needed for the $X\times Y$ terms, then the cases needed for the 3PT term.

\subsubsection{$X\times Y$ terms}\label{App:2Dproj-trisp-angdep-2angles-XY}

\myparagraph{Squeezed diagonal case}

In this case,
\ba
T_\mr{gal}(\kk_{1234},z_{1234}) = P_n(-\hk_{1+2}\cdot\hk_1) \; P_{n'}(\hk_{1+2}\cdot\hk_3) \ f(k_{1+2},k_1,k_2,k_3,k_4)
\ea
Then the covariance reads
\ba
\nonumber \mathcal{C}_{\ell,\ell'} =& \frac{(-1)^{\ell+\ell'}}{(2\ell+1)(2\ell'+1)} \int \dd V_{1234} \frac{\dd^3\kk_{1234}}{(2\pi^2)^4} \frac{\dd^3\mathbf{K}}{(2\pi)^3} j_\ell(k_1 r_1) \,  j_\ell(k_2 r_2) \\ 
\nonumber & j_{\ell'}(k_3 r_3) \,  j_{\ell'}(k_4 r_4) \sum_{m,m'} Y_{\ell m}(\hk_1) Y_{\ell m}^*(\hk_2) Y_{\ell' m'}(\hk_3) Y_{\ell' m'}^*(\hk_4)\\
\nonumber & \int \dd^3\xx_{ab} \ \mre^{\ii \xx_a \cdot (\kk_1+\kk_2-\mathbf{K})} \ \mre^{\ii \xx_b \cdot (\mathbf{K}+\kk_3+\kk_4)} \ P_n(-\hat{K}\cdot\hk_1) \ P_{n'}(\hat{K}\cdot\hk_3)\\
& \times f(K,k_1,k_2,k_3,k_4) \\
\nonumber =& \frac{\left(\frac{2}{\pi}\right)^5 \ (4\pi)^2 \ (-1)^{\ell+\ell'+n}}{(2\ell+1)(2\ell'+1)(2n+1)(2n'+1)} \int \dd V_{1234} \ \dd^3\kk_{1234} \ \dd^3\mathbf{K} \\ 
\nonumber & \dd^3\xx_{ab} \ j_\ell(k_1 r_1) \,  j_\ell(k_2 r_2) \, j_{\ell'}(k_3 r_3) \,  j_{\ell'}(k_4 r_4) \\
\nonumber &  \sum_{m,m'} Y_{\ell m}(\hk_1) Y_{\ell m}^*(\hk_2) Y_{\ell' m'}(\hk_3) Y_{\ell' m'}^*(\hk_4) \!\! \sum_{1,2,a,b,3,4} \!\! \ii^{\lu+\ld-\ell_a+\ell_b+\lt+\lq} \\
\nonumber & j_{\ell_1}(k_1 x_a) \, j_{\ell_2}(k_2 x_a) \, j_{\ell_a}(K x_a) \, j_{\ell_b}(K x_b) \, j_{\ell_3}(k_3 x_b) \, j_{\ell_4}(k_4 x_b)\\
\nonumber & Y_1(\hk_1) \ Y_1^*(\hx_a) \ Y_2(\hk_2) \ Y_2^*(\hx_a) \ Y_a(\hat{K}) \ Y_a^*(\hx_a) \\ 
\nonumber & Y_b(\hat{K}) \ Y_b^*(\hx_b) \ Y_3(\hk_3) \ Y_3^*(\hx_b) \ Y_4(\hk_4) \ Y_4^*(\hx_b) \\
\nonumber & \sum_{m_n,m_{n'}} Y_{n m_n}^*(\hat{K}) \ Y_{n m_n}(\hk_1) \ Y_{n' m_{n'}}^*(\hat{K}) \ Y_{n' m_{n'}}(\hk_3) \\ 
& \times f(K,k_1,k_2,k_3,k_4)\\
\nonumber =& \frac{\left(\frac{2}{\pi}\right)^5 \ (4\pi)^2 \ (-1)^{\ell+\ell'+n}}{(2\ell+1)(2\ell'+1)(2n+1)(2n'+1)} \int \dd V_{1234} \ k^2_{1234}\,\dd k_{1234} \\
\nonumber & K^2\dd K \ \dd^3\xx_{ab} \ j_\ell(k_1 r_1) \,  j_\ell(k_2 r_2) \, j_{\ell'}(k_3 r_3) \,  j_{\ell'}(k_4 r_4) \sum_{\substack{m,m',1,3,c \\ a,b,m_n,m_{n'}}} \\
\nonumber & j_{\ell_1}(k_1 x_a) \, j_{\ell}(k_2 x_a) \, j_{\ell_a}(K x_a) \, j_{\ell_b}(K x_b) \, j_{\ell_3}(k_3 x_b) \, j_{\ell'}(k_4 x_b)\\
\nonumber & \ii^{\lu+\ell-\ell_a+\ell_b+\lt+\ell'} \ G_{\ell,\ell_1,n}^{m,m_1,m_n} \ G_{\ell',\ell_3,n'}^{m',m_3,m_{n'}} \ (-1)^{m_n+m_{n'}+m_c} \ G_{\ell_a,n,\ell_c}^{m_a,-m_n,m_c} \\
\nonumber & G_{\ell_b,n',\ell_c}^{m_b,-m_{n'},-m_c} \ Y_1^*(\hx_a) \ Y_{\ell m}^*(\hx_a) \ Y_a^*(\hx_a) \ Y_b^*(\hx_b) \ Y_3^*(\hx_b) \ Y_{\ell' m'}^*(\hx_b) \\ 
& \times f(K,k_1,k_2,k_3,k_4) 
\ea
\ba
\nonumber \mathcal{C}_{\ell,\ell'} =& \frac{\left(\frac{2}{\pi}\right)^5 \ (4\pi)^2 \ (-1)^{\ell+\ell'+n}}{(2\ell+1)(2\ell'+1)(2n+1)(2n'+1)} \int \dd V_{1234} \ k^2_{1234}\,\dd k_{1234} \\ 
\nonumber & K^2\,\dd K \ x_{ab}^2\,\dd x_{ab} \ j_\ell(k_1 r_1) \,  j_\ell(k_2 r_2) \, j_{\ell'}(k_3 r_3) \,  j_{\ell'}(k_4 r_4) \\
\nonumber & \!\!\!\! \sum_{\substack{m,m',1,3,c \\ a,b,m_n,m_{n'}}} \!\!\!\! j_{\ell_1}(k_1 x_a) \, j_{\ell}(k_2 x_a) \, j_{\ell_a}(K x_a) \, j_{\ell_b}(K x_b) \, j_{\ell_3}(k_3 x_b) \, j_{\ell'}(k_4 x_b)\\
\nonumber & \ii^{\lu+\ell-\ell_a+\ell_b+\lt+\ell'} \ G_{\ell,\ell_1,n}^{m,m_1,m_n} \ G_{\ell',\ell_3,n'}^{m',m_3,m_{n'}} \ G_{\ell,\ell_1,\ell_a}^{m,m_1,m_a} \ G_{\ell',\ell_3,\ell_b}^{m',m_3,m_{b}} \\
& (-1)^{m_n+m_{n'}+m_c} \ G_{\ell_a,n,\ell_c}^{m_a,-m_n,m_c} \ G_{\ell_b,n',\ell_c}^{m_b,-m_{n'},-m_c} \times f(K,k_1,k_2,k_3,k_4) \\ 
\nonumber =& \frac{\left(\frac{2}{\pi}\right)^5 \ (4\pi)^2 \ (-1)^{\ell+\ell'}}{(2n+1)(2n'+1)} \int \dd V_{1234} \ k^2_{1234}\,\dd k_{1234} \ K^2\,\dd K \ x_{ab}^2\,\dd x_{ab}\\ 
\nonumber & j_\ell(k_1 r_1) \, j_\ell(k_2 r_2) \, j_{\ell'}(k_3 r_3) \,  j_{\ell'}(k_4 r_4) \sum_{\substack{\lu,\lt,c \\ m_n,m_{n'}}} j_{\ell_1}(k_1 x_a) \, j_{\ell}(k_2 x_a) \\
\nonumber & j_{n}(K x_a) \, j_{n'}(K x_b) \, j_{\ell_3}(k_3 x_b) \, j_{\ell'}(k_4 x_b) \ \ii^{\lu+\ell+n+n'+\lt+\ell'} \\
\nonumber & \frac{(2\ell+1)_{13}}{(4\pi)^2} \ \threeJz{\ell}{\lu}{n}^2 \ \threeJz{\ell'}{\lt}{n'}^2 \ (-1)^{m_n+m_{n'}+m_c} \\
& G_{n,n,\ell_c}^{m_n,-m_n,m_c} \ G_{n',n',\ell_c}^{m_{n'},-m_{n'},-m_c} \times f(K,k_1,k_2,k_3,k_4) \\ 
\nonumber =& (-1)^{\ell+\ell'} \sum_{\lu,\lt} \ii^{\lu+\ell+n+n'+\lt+\ell'} \ \frac{(2\ell+1)_{13}}{4\pi} \threeJz{\ell}{\lu}{n}^2 \\ 
& \times \threeJz{\ell'}{\lt}{n'}^2 T_{\ell,\ell,\ell',\ell'}^{\lu,\lt,n,n'}
\ea
where the sum runs over multipoles following the triangular inequalities $|\ell-n|\leq\ell_1\leq\ell+n$, $|\ell'-n'|\leq\ell_3\leq\ell'+n'$ and $|n-n'|\leq\ell_c\leq n+n'$, and the parity conditions $\ell+n+\ell_1$, $\ell'+n'+\ell_3$ and $n+n'+\ell_c$ even (which ensure that $\mathcal{C}_{\ell,\ell'}$ is real). The angular trispectrum is
\ba
\nonumber T_{\ell,\ell,\ell',\ell'}^{\lu,\lt,n,n'} =& T_{\ell,\ell,\ell',\ell',1+2}^{\lu,\ell,n,n',\lt,\ell'}\\
\nonumber =& \left(\frac{2}{\pi}\right)^5 \int \dd V_{1234} \ k^2_{1234}\,\dd k_{1234} \ K^2\,\dd K \ x_{ab}^2\,\dd x_{ab} \\
\nonumber & j_\ell(k_1 r_1) \, j_{\lu}(k_1 x_a) \ j_\ell(k_2 r_2) \, j_{\ell}(k_2 x_a) \ j_{\ell'}(k_3 r_3) \,  j_{\ell_3}(k_3 x_b)  \\
& j_{\ell'}(k_4 r_4) \, j_{\ell'}(k_4 x_b) \ j_{n}(K x_a) \, j_{n'}(K x_b)   \times f(K,k_1,k_2,k_3,k_4)
\ea
Limber's approximation, if valid, can be used on $k_2$ and $k_4$, but not on $k_1$, $k_3$ and $K$. It yields:
\ba
\nonumber T_{\ell,\ell,\ell',\ell'}^{\lu,\lt,n,n'} =& \left(\frac{2}{\pi}\right)^3 \int \dd V_{13} \ k^2_{13}\,\dd k_{13} \ K^2\,\dd K \ x_{ab}^2\,\dd x_{ab} \ j_\ell(k_1 r_1) \, j_{\lu}(k_1 x_a) \\
&  j_{\ell'}(k_3 r_3) \,  j_{\lt}(k_3 x_b) \ j_{n}(K x_a) \, j_{n'}(K x_b) \times f(K,k_1,k_\ell,k_3,k_{\ell'})
\ea
with $k_\ell=(\ell+1/2)/x_a$ and $k_{\ell'}=(\ell'+1/2)/x_b$.\\

\myparagraph{Alternate diagonal case}

In this case,
\ba
\nonumber T_\mr{gal}(\kk_{1234},z_{1234}) = P_n(-\hk_{1+3}\cdot\hk_1) \; P_{n'}(\hk_{1+3}\cdot\hk_2) \\
\times f(k_{1+3},k_1,k_2,k_3,k_4)
\ea
Then the covariance reads
\ba
\nonumber \mathcal{C}_{\ell,\ell'} =& \frac{(-1)^{\ell+\ell'}}{(2\ell+1)(2\ell'+1)} \int \dd V_{1234} \frac{\dd^3\kk_{1234}}{(2\pi^2)^4} \frac{\dd^3\mathbf{K}}{(2\pi)^3} j_\ell(k_1 r_1) \,  j_\ell(k_2 r_2) \\ 
\nonumber & j_{\ell'}(k_3 r_3) \,  j_{\ell'}(k_4 r_4) \sum_{m,m'} Y_{\ell m}(\hk_1) Y_{\ell m}^*(\hk_2) Y_{\ell' m'}(\hk_3) Y_{\ell' m'}^*(\hk_4)\\
& \times \int \dd^3\xx_{ab} \ \mre^{\ii \xx_a \cdot (\kk_1+\kk_3-\mathbf{K})} \ \mre^{\ii \xx_b \cdot (\mathbf{K}+\kk_2+\kk_4)} \ T^{X\times Y}_{n,n'} \\
\nonumber =& \frac{\left(\frac{2}{\pi}\right)^5 \ (4\pi)^2 \ (-1)^{\ell+\ell'+n}}{(2\ell+1)(2\ell'+1)(2n+1)(2n'+1)} \int \dd V_{1234} \ \dd^3\kk_{1234} \\ 
\nonumber & \dd^3\mathbf{K} \ \dd^3\xx_{ab} \ j_\ell(k_1 r_1) \,  j_\ell(k_2 r_2) \ j_{\ell'}(k_3 r_3) \,  j_{\ell'}(k_4 r_4) \\ 
\nonumber & \sum_{m,m'} Y_{\ell m}(\hk_1) \, Y_{\ell m}^*(\hk_2) \, Y_{\ell' m'}(\hk_3) \, Y_{\ell' m'}^*(\hk_4) \!\!\! \sum_{1,3,a,b,2,4} \!\!\! \ii^{\lu+\lt-\ell_a+\ell_b+\ld+\lq} \\
\nonumber & j_{\ell_1}(k_1 x_a) \, j_{\ell_3}(k_3 x_a) \, j_{\ell_a}(K x_a) \, j_{\ell_b}(K x_b) \, j_{\ell_2}(k_2 x_b) \, j_{\ell_4}(k_4 x_b) \\
\nonumber & Y_1(\hk_1) \, Y_1^*(\hx_a) \ Y_3^*(\hk_3) \, Y_3(\hx_a) \ Y_a^*(\hat{K}) \, Y_a(\hx_a) \\ 
\nonumber & Y_b(\hat{K}) \, Y_b^*(\hx_b) \ Y_2^*(\hk_2) \, Y_2(\hx_b) \ Y_4(\hk_4) \, Y_4^*(\hx_b) \\
\nonumber & \sum_{m_n,m_{n'}} Y_{n,m_n}^*(\hat{K}) \, Y_{n,m_n}(\hk_1)  \ Y_{n',m_{n'}}(\hat{K}) \, Y^*_{n',m_{n'}}(\hk_2) \\
& \times f(K,k_1,k_2,k_3,k_4) \\
\nonumber =& \frac{\left(\frac{2}{\pi}\right)^5 \ (4\pi)^2 \ (-1)^{\ell+n}}{(2\ell+1)(2\ell'+1)(2n+1)(2n'+1)} \int \dd V_{1234} \ k^2_{1234}\dd k_{1234} \\ 
\nonumber & K^2\dd K \ \dd^3\xx_{ab} \ j_\ell(k_1 r_1) \,  j_\ell(k_2 r_2) \, j_{\ell'}(k_3 r_3) \,  j_{\ell'}(k_4 r_4)  \sum_{\substack{m,m',1,2,c \\ a,b,m_n,m_{n'}}} \\
\nonumber & j_{\ell_1}(k_1 x_a) \, j_{\ell'}(k_3 x_a) \, j_{\ell_a}(K x_a) \, j_{\ell_b}(K x_b) \, j_{\ell_2}(k_2 x_b) \, j_{\ell'}(k_4 x_b)\\
\nonumber &  \ii^{\lu-\ell_a+\ell_b+\ld} \ (-1)^{m+m_2+m_{n'}+m_a+m_n} \ G_{\ell,\ell_1,n}^{m,m_1,m_n} \ G_{\ell,\ell_2,n'}^{-m,-m_2,-m_{n'}} \\
\nonumber & G_{\ell_a,n',\ell_c}^{-m_a,m_{n'},m_c} \ G_{\ell_b,n,\ell_c}^{m_b,-m_{n},-m_c} \ Y_1^*(\hx_a) \ Y_{\ell' m'}(\hx_a) \ Y_a(\hx_a) \\ 
& \times Y_b^*(\hx_b) \ Y_2(\hx_b) \ Y_{\ell' m'}^*(\hx_b) \ f(K,k_1,k_2,k_3,k_4)\\
\nonumber =& (-1)^{\ell+n} \sum_{\ell_1,\ell_2,\ell_a,\ell_b,\ell_c} \ii^{\lu-\ell_a+\ell_b+\ld} \ \frac{(2\ell+1)_{12abc}}{4\pi} \ J^{\ld,\ell,n';\ell_b,n,\ell_c}_{\ell',\lu,\ell_a} \\
& \times T_{\ell,\ell',\lu,\ld}^{\ell_a,\ell_b}
\ea
where I had to define the geometric coefficient\footnote{The parity conditions of the last two Gaunt coefficients ensure that $\mathcal{C}_{\ell,\ell'}$ is real, since $\lu-\ell_a+\ell_b+\ld=(\lu+\ell'+\ell_a)+(\ell_b+\ld+\ell')-2\ell'-2\ell_a$ is even.}
\ba
\nonumber J^{\ld,\ell,n';\ell_b,n,\ell_c}_{\ell',\lu,\ell_a} =& \frac{(4\pi)^3}{(2\ell+1)(2\ell'+1) \ (2n+1)(2n'+1) \ (2\ell+1)_{12abc}} \\
\nonumber & \sum_{\substack{m,m',m_1,m_2,m_c \\ m_a,m_b,m_n,m_{n'}}} (-1)^{m+m'+m_n+m_{n'}+m_1+m_2+m_a+m_b+m_c} \\
\nonumber &  G_{\ell_2,\ell,n'}^{-m_2,-m,-m_{n'}} \ G_{\ell_b,n,\ell_c}^{m_b,-m_{n},-m_c} \ G_{\ell',\ell_1,\ell_a}^{m',-m_1,m_a} \\
& G_{\ell_b,\ell_2,\ell'}^{-m_b,m_2,-m'} \ G_{\ell,n,\ell_1}^{m,m_n,m_1} \ G_{n',\ell_c,\ell_a}^{m_{n'},m_c,-m_a}
\ea
which will be shown in Appendix \ref{App:J-and-9J} to be, in fact, a 9J symbol.\\
The angular trispectrum is
\ba
\nonumber T_{\ell,\ell',\lu,\ld}^{\ell_a,\ell_b} =& T_{\ell,\ell,\ell',\ell',1+3}^{\lu,\ld,\ell_a,\ell_b,\ell',\ell'}\\
\nonumber =& \left(\frac{2}{\pi}\right)^5 \int \dd V_{1234} \ k^2_{1234}\,\dd k_{1234} \ K^2\,\dd K \ x_{ab}^2\,\dd x_{ab} \\
\nonumber & j_\ell(k_1 r_1) \, j_{\lu}(k_1 x_a) \ j_\ell(k_2 r_2) \, j_{\ld}(k_2 x_b) \ j_{\ell'}(k_3 r_3) \,  j_{\ell'}(k_3 x_a)  \\
& j_{\ell'}(k_4 r_4) \, j_{\ell'}(k_4 x_b) \ j_{\ell_a}(K x_a) \, j_{\ell_b}(K x_b)   \times f(K,k_1,k_2,k_3,k_4)
\ea
Limber's approximation, if valid, can be used on $k_3$ and $k_4$, but not on $k_1$, $k_2$ and $K$. It yields
\ba
\nonumber T_{\ell,\ell',\lu,\ld}^{\ell_a,\ell_b} =& \left(\frac{2}{\pi}\right)^3 \int \dd V_{1234} \ k^2_{12}\,\dd k_{12} \ K^2\,\dd K \ j_\ell(k_1 r_1) \, j_{\lu}(k_1 r_3) \\
\nonumber & j_{\ell}(k_2 r_2) \,  j_{\ld}(k_2 r_4) \ j_{\ell_a}(K r_3) \, j_{\ell_b}(K r_4) \\
& \times f(K,k_1,k_2,k_{\ell'}(r_3),k_{\ell'}(r_4))
\ea
with $k_{\ell'}(r_i)=(\ell'+1/2)/r_i$.

\subsubsection{3PT terms}\label{App:2Dproj-trisp-angdep-2angles-3PT}

Following Eq.~\ref{Eq:F_3^s-Lacasa} for the 3PT kernel, and accounting for symmetries, the two cases I have to consider are
\ba
T_\mr{gal}(\kk_{1234},z_{1234}) = P_n(\hk_1\cdot\hk_2) \; P_{n'}(\hk_{1+2}\cdot\hk_3) \ f(k_1,k_2,k_3,k_4)
\ea
and
\ba
T_\mr{gal}(\kk_{1234},z_{1234}) = P_n(\hk_1\cdot\hk_3) \; P_{n'}(\hk_{1+3}\cdot\hk_2) \ f(k_1,k_2,k_3,k_4)
\ea

\myparagraph{Squeezed diagonal case}

In this case,
\ba
T_\mr{gal}(\kk_{1234},z_{1234}) = P_n(\hk_1\cdot\hk_2) \; P_{n'}(\hk_{1+2}\cdot\hk_3) \ f(k_1,k_2,k_3,k_4)
\ea
Then the covariance reads
\ba
\nonumber \mathcal{C}_{\ell,\ell'} =& \frac{(-1)^{\ell+\ell'}}{(2\ell+1)(2\ell'+1)} \int \dd V_{1234} \frac{\dd^3\kk_{1234}}{(2\pi^2)^4} \frac{\dd^3\mathbf{K}}{(2\pi)^3} j_\ell(k_1 r_1) \,  j_\ell(k_2 r_2) \\ 
\nonumber & j_{\ell'}(k_3 r_3) \,  j_{\ell'}(k_4 r_4) \sum_{m,m'} Y_{\ell m}(\hk_1) Y_{\ell m}^*(\hk_2) Y_{\ell' m'}(\hk_3) Y_{\ell' m'}^*(\hk_4)\\
& \times \int \dd^3\xx_{ab} \ \mre^{\ii \xx_a \cdot (\kk_1+\kk_2-\mathbf{K})} \ \mre^{\ii \xx_b \cdot (\mathbf{K}+\kk_3+\kk_4)} \ T_\mr{gal}(\kk_{1234},z_{1234}) \\
\nonumber =& \frac{\left(\frac{2}{\pi}\right)^5 \ (4\pi)^2 \ (-1)^{\ell+\ell'}}{(2\ell+1)(2\ell'+1)(2n+1)(2n'+1)} \int \dd V_{1234} \ \dd^3\kk_{1234} \ \dd^3\mathbf{K} \\ 
\nonumber & \dd^3\xx_{ab} \ j_\ell(k_1 r_1) \,  j_\ell(k_2 r_2) \, j_{\ell'}(k_3 r_3) \,  j_{\ell'}(k_4 r_4) \\
\nonumber &  \sum_{m,m'} Y_{\ell m}(\hk_1) Y_{\ell m}^*(\hk_2) Y_{\ell' m'}(\hk_3) Y_{\ell' m'}^*(\hk_4) \!\! \sum_{1,2,a,b,3,4} \!\! \ii^{\lu+\ld-\ell_a+\ell_b+\lt+\lq} \\
\nonumber & j_{\ell_1}(k_1 x_a) \, j_{\ell_2}(k_2 x_a) \, j_{\ell_a}(K x_a) \, j_{\ell_b}(K x_b) \, j_{\ell_3}(k_3 x_b) \, j_{\ell_4}(k_4 x_b)\\
\nonumber & Y_1(\hk_1) \ Y_1^*(\hx_a) \ Y_2^*(\hk_2) \ Y_2(\hx_a) \ Y_a(\hat{K}) \ Y_a^*(\hx_a) \\ 
\nonumber & Y_b(\hat{K}) \ Y_b^*(\hx_b) \ Y_3(\hk_3) \ Y_3^*(\hx_b) \ Y_4(\hk_4) \ Y_4^*(\hx_b) \\
\nonumber & \sum_{m_n,m_{n'}} Y_{n m_n}(\hk_1) \ Y_{n m_n}^*(\hk_2) \ Y_{n' m_{n'}}^*(\hat{K}) \ Y_{n' m_{n'}}(\hk_3) \\ 
& \times f(k_1,k_2,k_3,k_4)\\
\nonumber =& \frac{\left(\frac{2}{\pi}\right)^5 \ (4\pi)^2\ (-1)^{\ell+\ell'}}{(2\ell+1)(2\ell'+1)(2n+1)(2n'+1)} \int \dd V_{1234} \ k_{1234}^2\dd k_{1234} \\ 
\nonumber & K^2\dd K \ \dd^3\xx_{ab} \ j_\ell(k_1 r_1) \,  j_\ell(k_2 r_2) \, j_{\ell'}(k_3 r_3) \,  j_{\ell'}(k_4 r_4) \sum_{\substack{m,m',1,2,3 \\ a,b,m_n,m_{n'}}} \\
\nonumber & j_{\ell_1}(k_1 x_a) \, j_{\ell_2}(k_2 x_a) \, j_{\ell_a}(K x_a) \, j_{\ell_b}(K x_b) \, j_{\ell_3}(k_3 x_b) \, j_{\ell'}(k_4 x_b)\\
\nonumber & \ii^{\lu+\ld-\ell_a+\ell_b+\lt+\ell'} \ G_{\ell,\ell_1,n}^{m,m_1,m_{n}} \ G_{\ell,\ell_2,n}^{m,m_2,m_{n}} \ G_{\ell',\ell_3,n'}^{m',m_3,m_{n'}}  \ (-1)^{m_{n'}} \\
\nonumber & G_{\ell_a,\ell_b,n'}^{m_a,m_b,-m_{n'}} \ Y_1^*(\hx_a) \ Y_2(\hx_a) \ Y_a^*(\hx_a) \ Y_b^*(\hx_b) \ Y_3^*(\hx_b) \ Y_{\ell' m'}^*(\hx_b) \\ 
& \times f(k_1,k_2,k_3,k_4) 
\ea
\ba
\nonumber \mathcal{C}_{\ell,\ell'} =& \frac{\left(\frac{2}{\pi}\right)^5 \ 4\pi \ (-1)^{\ell+\ell'}}{(2\ell'+1)(2n'+1)} \int \dd V_{1234} \ k_{1234}^2\,\dd k_{1234} \ K^2\,\dd K \ x^2_{ab}\,\dd x_{ab} \\ 
\nonumber & j_\ell(k_1 r_1) \,  j_\ell(k_2 r_2) \, j_{\ell'}(k_3 r_3) \,  j_{\ell'}(k_4 r_4) \\
\nonumber & \sum_{\substack{m',1,2,3 \\ a,b,m_{n'}}} \ii^{\lu+\ld-\ell_a+\ell_b+\lt+\ell'} \ \threeJz{\ell}{\lu}{n}^2 \delta_{1,2} \ G_{\ell',\ell_3,n'}^{m',m_3,m_{n'}} \\
\nonumber & (-1)^{m_{n'}} G_{\ell_a,\ell_b,n'}^{m_a,m_b,-m_{n'}} \ (-1)^{m_2} \ G_{\ell_1,\ell_2,\ell_a}^{m_1,-m_2,m_{a}} \ G_{\ell_b,\ell_3,\ell'}^{m_b,m_3,m'} \\
\nonumber & j_{\ell_1}(k_1 x_a) \, j_{\ell_2}(k_2 x_a) \, j_{\ell_a}(K x_a) \, j_{\ell_b}(K x_b) \, j_{\ell_3}(k_3 x_b) \, j_{\ell'}(k_4 x_b)\\
& \times f(k_1,k_2,k_3,k_4)\\
\nonumber =& \frac{\left(\frac{2}{\pi}\right)^5 \ (-1)^{\ell+\ell'}}{(2n'+1)} \int \dd V_{1234} \ k_{1234}^2\,\dd k_{1234} \ K^2\,\dd K \ x^2_{ab}\,\dd x_{ab} \\ 
\nonumber & j_\ell(k_1 r_1) \,  j_\ell(k_2 r_2) \, j_{\ell'}(k_3 r_3) \,  j_{\ell'}(k_4 r_4) \\
\nonumber & \sum_{\substack{\lu,\lt \\ a,b,m_{n'}}} \ii^{2\lu+n'+\lt+\ell'} \ \threeJz{\ell}{\lu}{n}^2 \ (2\lt+1) \threeJz{\ell'}{\lt}{n'}^2 \\
\nonumber & \delta_{n',\ell_b} \, \delta_{m_{n'},m_b} (-1)^{m_{n'}} G_{\ell_a,\ell_b,n'}^{m_a,m_b,-m_{n'}} \ \frac{2\lu+1}{4\pi} \sqrt{4\pi} \ \delta_{\ell_a,0} \, \delta_{m_a,0} \\
\nonumber & j_{\ell_1}(k_1 x_a) \, j_{\ell_1}(k_2 x_a) \, j_{\ell_a}(K x_a) \, j_{\ell_b}(K x_b) \, j_{\ell_3}(k_3 x_b) \, j_{\ell'}(k_4 x_b)\\
& \times f(k_1,k_2,k_3,k_4)\\
\nonumber =& (-1)^{\ell+\ell'} \sum_{\lu\lt} \ii^{2\lu+n'+\lt+\ell'} \ \frac{(2\lu+1)(2\lt+1)}{4\pi} \threeJz{\ell}{\lu}{n}^2 \\
&\times \threeJz{\ell'}{\lt}{n'}^2 T_{\ell,\ell,\ell',\ell'}^{\lu,\lt,n'}
\ea
where the sum runs over multipoles following the triangular inequalities $|\ell-n|\leq\ell_1\leq\ell+n$ and $|\ell'-n'|\leq\ell_3\leq\ell'+n'$, and the parity conditions $\ell+n+\ell_1$ and $\ell'+n'+\ell_3$ even (the latter ensuring that $\mathcal{C}_{\ell,\ell'}$ is real). The angular trispectrum is
\ba
\nonumber T_{\ell,\ell,\ell',\ell'}^{\lu,\lt,n'} =& T_{\ell,\ell,\ell',\ell',1+2}^{\lu,\lu,0,n',\lt,\ell'} \\
\nonumber =& \left(\frac{2}{\pi}\right)^5 \int \dd V_{1234} \ k^2_{1234}\,\dd k_{1234} \ K^2\,\dd K \ x_{ab}^2\,\dd x_{ab} \\
\nonumber & j_\ell(k_1 r_1) \, j_{\lu}(k_1 x_a) \ j_\ell(k_2 r_2) \, j_{\lu}(k_2 x_a) \ j_{\ell'}(k_3 r_3) \,  j_{\ell_3}(k_3 x_b) \\
& j_{\ell'}(k_4 r_4) \, j_{\ell'}(k_4 x_b) \ j_{0}(K x_a) \, j_{n'}(K x_b)   \times f(k_1,k_2,k_3,k_4)
\ea
Limber's approximation, if valid, can be used on $k_4$, but not on $k_1$, $k_2$, $k_3$ and $K$. It yields:
\ba
\nonumber T_{\ell,\ell,\ell',\ell'}^{\lu,\lt,n'} =& \left(\frac{2}{\pi}\right)^4 \int \dd V_{1234} \ k^2_{1234}\,\dd k_{1234} \ K^2\,\dd K \ x_{a}^2\,\dd x_{a} \\
\nonumber & j_\ell(k_1 r_1) \, j_{\lu}(k_1 x_a) \ j_\ell(k_2 r_2) \, j_{\lu}(k_2 x_a) \ j_{\ell'}(k_3 r_3) \,  j_{\ell_3}(k_3 r_4) \\
& j_{0}(K x_a) \, j_{n'}(K r_4) \times f(k_1,k_2,k_3,k_{\ell'})
\ea
with $k_{\ell'}=(\ell'+1/2)/r_4$.\\

\myparagraph{Alternate diagonal case}

In this case,
\ba
T_\mr{gal}(\kk_{1234},z_{1234}) = P_n(\hk_1\cdot\hk_3) \; P_{n'}(\hk_{1+3}\cdot\hk_2) \ f(k_1,k_2,k_3,k_4)
\ea
Then the covariance reads
\ba
\nonumber \mathcal{C}_{\ell,\ell'} =& \frac{(-1)^{\ell+\ell'}}{(2\ell+1)(2\ell'+1)} \int \dd V_{1234} \frac{\dd^3\kk_{1234}}{(2\pi^2)^4} \frac{\dd^3\mathbf{K}}{(2\pi)^3} j_\ell(k_1 r_1) \,  j_\ell(k_2 r_2) \\ 
\nonumber & j_{\ell'}(k_3 r_3) \,  j_{\ell'}(k_4 r_4) \sum_{m,m'} Y_{\ell m}(\hk_1) Y_{\ell m}^*(\hk_2) Y_{\ell' m'}(\hk_3) Y_{\ell' m'}^*(\hk_4)\\
& \times \int \dd^3\xx_{ab} \ \mre^{\ii \xx_a \cdot (\kk_1+\kk_3-\mathbf{K})} \ \mre^{\ii \xx_b \cdot (\mathbf{K}+\kk_2+\kk_4)} \ T_\mr{gal}(\kk_{1234},z_{1234}) \\
\nonumber =& \frac{\left(\frac{2}{\pi}\right)^5 \ (4\pi)^2 (-1)^{\ell+\ell'}}{(2\ell+1)(2\ell'+1)(2n+1)(2n'+1)} \int \dd V_{1234} \ \dd^3\kk_{1234} \ \dd^3\mathbf{K} \\ 
\nonumber & \dd^3\xx_{ab} \ j_\ell(k_1 r_1) \,  j_\ell(k_2 r_2) \, j_{\ell'}(k_3 r_3) \,  j_{\ell'}(k_4 r_4) \sum_{m,m'} \\
\nonumber & Y_{\ell m}(\hk_1) \, Y_{\ell m}^*(\hk_2) \, Y_{\ell' m'}^*(\hk_3) \, Y_{\ell' m'}(\hk_4) \!\! \sum_{1,3,a,b,2,4} \!\! \ii^{\lu+\lt-\ell_a+\ell_b+\ld+\lq} \\
\nonumber & j_{\ell_1}(k_1 x_a) \, j_{\ell_2}(k_2 x_b) \, j_{\ell_a}(K x_a) \, j_{\ell_b}(K x_b) \, j_{\ell_3}(k_3 x_a) \, j_{\ell_4}(k_4 x_b)\\
\nonumber & Y_1(\hk_1) \ Y_1^*(\hx_a) \ Y_3^*(\hk_3) \ Y_3(\hx_a) \ Y_a(\hat{K}) \ Y_a^*(\hx_a) \\ 
\nonumber & Y_b(\hat{K}) \ Y_b^*(\hx_b) \ Y_2^*(\hk_2) \ Y_2(\hx_b) \ Y_4^*(\hk_4) \ Y_4(\hx_b) \\
\nonumber & \sum_{m_n,m_{n'}} Y_{n m_n}(\hk_1) \ Y_{n m_n}^*(\hk_3) \ Y_{n' m_{n'}}(\hat{K}) \ Y_{n' m_{n'}}^*(\hk_2) \\ 
& \times f(k_1,k_2,k_3,k_4)\\
\nonumber =& \frac{\left(\frac{2}{\pi}\right)^5 \ (4\pi)^2 (-1)^{\ell+\ell'}}{(2\ell+1)(2\ell'+1)(2n+1)(2n'+1)} \int \dd V_{1234} \ k_{1234}^2\dd k_{1234} \\ 
\nonumber & K^2\dd K \ \dd^3\xx_{ab} \ j_\ell(k_1 r_1) \,  j_\ell(k_2 r_2) \, j_{\ell'}(k_3 r_3) \,  j_{\ell'}(k_4 r_4) \sum_{\substack{m,m',1,2,3 \\ a,b,m_n,m_{n'}}} \\
\nonumber & j_{\ell_1}(k_1 x_a) \, j_{\ell_2}(k_2 x_b) \, j_{\ell_a}(K x_a) \, j_{\ell_b}(K x_b) \, j_{\ell_3}(k_3 x_a) \, j_{\ell'}(k_4 x_b)\\
\nonumber & \ii^{\lu+\lt-\ell_a+\ell_b+\ld+\ell'} (-1)^{m+m_2+m_{n'}+m'+m_3+m_{n}} \ G_{\ell,\ell_1,n}^{m,m_1,m_{n}} \ G_{\ell,\ell_2,n'}^{-m,-m_2,-m_{n'}}  \\
\nonumber & G_{\ell',\ell_3,n}^{-m',-m_3,-m_{n}} \ G_{\ell_a,\ell_b,n'}^{m_a,m_b,m_{n'}} \ Y_1^*(\hx_a) \ Y_3(\hx_a) \ Y_a^*(\hx_a) \\ 
& \times Y_b^*(\hx_b) \ Y_2(\hx_b) \ Y_{\ell' m'}(\hx_b) \ f(k_1,k_2,k_3,k_4)\\
\nonumber =& \frac{\left(\frac{2}{\pi}\right)^5 \ (4\pi)^2 (-1)^{\ell+\ell'}}{(2\ell+1)(2\ell'+1)(2n+1)(2n'+1)} \int \dd V_{1234} \ k_{1234}^2\dd k_{1234} \\ 
\nonumber & K^2\dd K \ x^2_{ab}\,\dd x_{ab} \ j_\ell(k_1 r_1) \,  j_\ell(k_2 r_2) \, j_{\ell'}(k_3 r_3) \,  j_{\ell'}(k_4 r_4) \sum_{\substack{m,m',1,2,3 \\ a,b,m_n,m_{n'}}} \\
\nonumber & \ii^{\lu+\lt-\ell_a+\ell_b+\ld+\ell'} (-1)^{m+m_2+m_{n'}+m'+m_3+m_n+m_1+m_a+m_b} \ G_{\ell,\ell_1,n}^{m,m_1,m_{n}} \\
\nonumber &  G_{\ell,\ell_2,n'}^{-m,-m_2,-m_{n'}} \ G_{\ell',\ell_3,n}^{m',m_3,m_{n}} \ G_{\ell_a,\ell_b,n'}^{m_a,m_b,m_{n'}} \ G_{\lu,\lt,\ell_a}^{-m_1,m_3,-m_{a}} \ G_{\ell_b,\ld,\ell'}^{-m_b,m_2,m'} \\
\nonumber & j_{\ell_1}(k_1 x_a) \, j_{\ell_2}(k_2 x_b) \, j_{\ell_a}(K x_a) \, j_{\ell_b}(K x_b) \, j_{\ell_3}(k_3 x_a) \, j_{\ell'}(k_4 x_b)\\ 
& \times f(k_1,k_2,k_3,k_4)\\
\nonumber =& (-1)^{\ell+\ell'} \sum_{\lu,\ld,\lt,\ell_a,\ell_b} \ii^{\lu+\lt-\ell_a+\ell_b+\ld+\ell'} \ \frac{(2\ell+1)_{123ab}}{4\pi} \ K^{\lu,\lt,n;\ell,\ell_a,\ell'}_{n',\ell_b,\ld} \\
& \times T_{\ell,\ell,\ell',\ell'}^{\lu,\ld,\ell_a,\ell_b,\lt}
\ea
where I had to define the geometric coefficient\footnote{The parity condition of the Gaunt coefficients ensure that $\mathcal{C}_{\ell,\ell'}$ is real.}
\ba
\nonumber K^{\lu,\lt,n;\ell,\ell_a,\ell'}_{n',\ell_b,\ld} =& \frac{(4\pi)^3}{(2\ell+1)(2\ell'+1) \ (2n+1)(2n'+1) \ (2\ell+1)_{123ab}}   \\
\nonumber & \sum_{\substack{m,m',m_{123} \\ m_{ab},m_n,m_{n'}}} (-1)^{m+m'+m_1+m_2+m_3+m_a+m_b+m_n+m_{n'}} \\
\nonumber & G_{\ell,n,\ell_1}^{m,m_{n},m_1} \ G_{\ell_a,\lu,\lt}^{-m_a,-m_1,m_3} \ G_{\ell',\ell_3,n}^{m',m_3,m_{n}} \\
& \times G_{\ell,\ell_2,n'}^{-m,-m_2,-m_{n'}}  \ G_{\ell_a,n',\ell_b}^{m_a,m_{n'},m_b}  \ G_{\ell',\ell_b,\ld}^{m',-m_b,m_2}
\ea
and the angular trispectrum is
\ba
\nonumber T_{\ell,\ell,\ell',\ell'}^{\lu,\ld,\ell_a,\ell_b,\lt} =& T_{\ell,\ell,\ell',\ell',1+3}^{\lu,\ld,\ell_a,\ell_b,\lt,\ell'}\\
\nonumber =& \left(\frac{2}{\pi}\right)^5 \int \dd V_{1234} \ k_{1234}^2\,\dd k_{1234} \ K^2\,\dd K \ x^2_{ab}\,\dd x_{ab} \\
\nonumber & j_\ell(k_1 r_1) \, j_{\lu}(k_1 x_a) \ j_\ell(k_2 r_2) \, j_{\ld}(k_2 x_b) \ j_{\ell_a}(K x_a) \, j_{\ell_b}(K x_b) \\
& j_{\ell'}(k_3 r_3) \, j_{\ell_3}(k_3 x_a) \  j_{\ell'}(k_4 r_4) \, j_{\ell'}(k_4 x_b) \times f(k_1,k_2,k_3,k_4)
\ea
Limber's approximation, if valid, can be used on $k_4$, but not on $k_1$, $k_2$, $k_3$ and $K$. It yields:
\ba
\nonumber T_{\ell,\ell,\ell',\ell'}^{\lu,\ld,\ell_a,\ell_b,\lt} =& \left(\frac{2}{\pi}\right)^4 \int \dd V_{1234} \ k^2_{1234}\,\dd k_{1234} \ K^2\,\dd K \ x_{a}^2\,\dd x_{a} \\
\nonumber & j_\ell(k_1 r_1) \, j_{\lu}(k_1 x_a) \ j_\ell(k_2 r_2) \, j_{\ld}(k_2 r_4) \ j_{\ell_a}(K x_a) \, j_{\ell_b}(K r_4) \\
& j_{\ell'}(k_3 r_3) \,  j_{\ell_3}(k_3 x_a) \times f(k_1,k_2,k_3,k_{\ell'})
\ea
with $k_{\ell'}=(\ell'+1/2)/r_4$.\\

\section{Reduction of covariance cases}\label{App:reductions}

In this section, I show that the various complex trispectrum cases computed in Appendices \ref{App:2Dproj-trisp-angindep} and \ref{App:2Dproj-trisp-angdep}, reduce to simpler trispectrum cases when appropriate. This is a consistency (sanity) check of the derivations presented above.

\subsection{Diagonal independence}\label{App:reduc-diagonal}
In this subsection, I show that the two trispectrum cases with a diagonal dependence, presented in Appendix \ref{App:2Dproj-trisp-angindep}, do reduce to the diagonal independent case when appropriate.\\
The following properties will be useful here:
\ba\label{Eq:reduc-prop-spherbess}
\int K^2\,\dd K \ j_\ell(K x_a) j_\ell(K x_b) = \frac{\pi}{2 \ x_a^2} \delta(x_a - x_b)
\ea
\ba\label{Eq:reduc-prop-3J-sum}
\sum_{\lt} \frac{2\lt+1}{4\pi} \threeJz{\lu}{\ld}{\lt}^2 = \frac{1}{4\pi}
\ea

\subsubsection{Squeezed diagonal}\label{App:reduc-diagindep-sqz}
In this case, the 3D trispectrum is
\ba
T_\mr{gal}(\kk_{1234},z_{1234}) = T_\mr{gal}(k_{1+2},k_{1234},z_{1234})
\ea
and the resulting covariance, derived in Sect. \ref{App:2Dproj-trisp-sqzdiag}, is
\ba
\mathcal{C}_{\ell,\ell'} = \frac{T_{\ell,\ell,\ell',\ell',1+2}^{\ell,\ell,0,0,\ell',\ell'}}{4\pi}
\ea
where the angular trispectrum is
\ba
\nonumber T_{\ell,\ell,\ell',\ell',1+2}^{\ell,\ell,0,0,\ell',\ell'} =& \left(\frac{2}{\pi}\right)^5 \int \dd V_{1234} \ k^2_{1234}\,\dd k_{1234} \ K^2\,\dd K \ \dd V_{ab} \\
\nonumber & j_\ell(k_1 r_1) \, j_\ell(k_1 x_a) \ j_\ell(k_2 r_2) \, j_\ell(k_2 x_a) \ j_0(K x_a)  \\ 
\nonumber & j_0(K x_b) \ j_{\ell'}(k_3 r_3) \, j_{\ell'}(k_3 x_b) \ j_{\ell'}(k_4 r_4) \, j_{\ell'}(k_4 x_b) \\ 
& \times T_\mr{gal}(K,k_{1234},z_{1234})
\ea
If in fact $T_\mr{gal}(k_{1+2},k_{1234},z_{1234})$ does not depend on $k_{1+2}$, using Eq.~\ref{Eq:reduc-prop-spherbess}, one sees that the angular trispectrum reduces to the diagonal-independent case:
\ba
\nonumber T_{\ell,\ell,\ell',\ell',1+2}^{\ell,\ell,0,0,\ell',\ell'} =& \left(\frac{2}{\pi}\right)^4 \int \dd V_{1234} \ k^2_{1234}\,\dd k_{1234} \ K^2\,\dd K \ \dd V \\
\nonumber & j_\ell(k_1 r_1) \, j_\ell(k_1 x) \ j_\ell(k_2 r_2) \, j_\ell(k_2 x) \ j_{\ell'}(k_3 r_3) \, j_{\ell'}(k_3 x) \\
& \times j_{\ell'}(k_4 r_4) \, j_{\ell'}(k_4 x) \ T_\mr{gal}(k_{1234},z_{1234})
\ea
And the covariance equation is already equal to the diagonal-independent case.

\subsubsection{Alternate diagonal}\label{App:reduc-diagindep-alt}
In this case, the 3D trispectrum is
\ba
T_\mr{gal}(\kk_{1234},z_{1234}) = T_\mr{gal}(k_{1+3},k_{1234},z_{1234})
\ea
and the resulting covariance, derived in Sect. \ref{App:2Dproj-trisp-altdiag}, is
\ba
\mathcal{C}_{\ell,\ell'} = \sum_{\ell_a} \frac{2\ell_a+1}{4\pi} \threeJz{\ell}{\ell'}{\ell_a}^2 T_{\ell,\ell,\ell',\ell',1+3}^{\ell,\ell,\ell_a,\ell_a,\ell',\ell'}
\ea
where the angular trispectrum is
\ba
\nonumber T_{\ell,\ell,\ell',\ell',1+3}^{\ell,\ell,\ell_a,\ell_a,\ell',\ell'} =& \left(\frac{2}{\pi}\right)^5 \int \dd V_{1234} \ k^2_{1234}\,\dd k_{1234} \ K^2\,\dd K \ \dd V_{ab} \\
\nonumber & j_\ell(k_1 r_1) \, j_\ell(k_1 x_a) \ j_\ell(k_2 r_2) \, j_\ell(k_2 x_b) \ j_{\ell_a}(K x_a) \\ 
\nonumber &  j_{\ell_a}(K x_b) \ j_{\ell'}(k_3 r_3) \, j_{\ell'}(k_3 x_a) \ j_{\ell'}(k_4 r_4) \, j_{\ell'}(k_4 x_b)\\
& \times T_\mr{gal}(K,k_{1234},z_{1234})
\ea
If in fact $T_\mr{gal}(k_{1+3},k_{1234},z_{1234})$ does not depend on $k_{1+3}$, as in previous subsection \ref{App:reduc-diagindep-sqz}, one sees that the angular trispectrum reduces to the diagonal-independent case, becoming independent of $\ell_a$. Then using Eq.~\ref{Eq:reduc-prop-3J-sum}, one sees that the covariance equation also reduces to the diagonal-independent case.

\subsection{Angle independence}\label{App:reduc-angle}
In this subsection, I show that the trispectrum cases with an angle dependence, presented in Appendix \ref{App:2Dproj-trisp-angdep-1angle}, do reduce to the angle independent cases of Appendix \ref{App:2Dproj-trisp-angindep} when appropriate. I then show that the most complex cases of a trispectrum depending on two angles, presented in Appendix \ref{App:2Dproj-trisp-angdep-2angles}, do reduce to the cases depending on a single angle when appropriate. This ends the reduction tree from the most complex to the most simple cases.\\
All these reductions will correspond to take the case $n=0$ for some Legendre polynomial. The following properties will thus be useful:
\ba\label{Eq:reduc-prop-Gaunt}
G_{\lu,\ld,n}^{m_1,m_2,m_n} = \frac{(-1)^{m_1}}{\sqrt{4\pi}} \ \delta_{\lu,\ld} \ \delta_{m_1,-m_2} \qquad \mr{if} \qquad n=0
\ea
\ba\label{Eq:reduc-prop-3J-n0}
\frac{2\ell_a+1}{4\pi} \threeJz{\ell}{\ell_a}{n}^2 = \frac{\delta_{\ell,\ell_a}}{4\pi} \qquad \mr{if} \qquad n=0
\ea

As a side note, Eq. \ref{Eq:reduc-prop-Gaunt} can be interpreted diagrammatically with the diagrams introduced in Appendix \ref{App:3nJ-symbols}: it corresponds to cutting a  line, and on each extremity vertex letting the two concurring lines reconnect into a single one.

\subsubsection{One angle}\label{App:reduc-angle-1}

\myparagraph{$\hk_1\cdot\hk_2$ case}
In this case, the 3D trispectrum is
\ba
T_\mr{gal}(\kk_{1234},z_{1234}) = P_n(\hk_1\cdot\hk_2) \ f(k_1,k_2,k_3,k_4)
\ea
and the resulting covariance, derived in Appendix \ref{App:2Dproj-trisp-angdep-1angle-base}, is
\ba
\mathcal{C}_{\ell,\ell'} = \sum_{\ell_a} \frac{2\ell_a+1}{4\pi} \threeJz{\ell}{\ell_a}{n}^2 (-1)^n \ T_{\ell,\ell,\ell',\ell',\ci}^{\ell_a,\ell_a,\ell',\ell'}
\ea
where the angular trispectrum is
\ba
\nonumber T_{\ell,\ell,\ell',\ell',\ci}^{\ell_a,\ell_a,\ell',\ell'} =& \left(\frac{2}{\pi}\right)^4 \int \dd V_{1234} \ k_{1234}^2  \, \dd k_{1234} \ x^2 \, \dd x \ j_\ell(k_1 r_1) \, j_{\ell_a}(k_1 x) \\
\nonumber & j_\ell(k_2 r_2) \, j_{\ell_a}(k_2 x) \ j_{\ell'}(k_3 r_3) \, j_{\ell'}(k_3 x) \ j_{\ell'}(k_4 r_4)  j_{\ell'}(k_4 x) \\
& \times f(k_1,k_2,k_3,k_4).
\ea
For $n=0$, Eq.~\ref{Eq:reduc-prop-3J-n0} shows that the covariance equation reduces to the simplest trispectrum case (angle and diagonal independent) and forces $\ell_a=\ell$, implying that the angular trispectrum also reduces to the standard case.

\myparagraph{The case of $\hk_1\cdot\hk_3$ }
In this case, the 3D trispectrum is
\ba
T_\mr{gal}(\kk_{1234},z_{1234}) = P_n(\hk_1\cdot\hk_3) \ f(k_1,k_2,k_3,k_4)
\ea
and the resulting covariance, derived in Appendix \ref{App:2Dproj-trisp-angdep-1angle-base}, is
\ba
\nonumber \mathcal{C}_{\ell,\ell'} =& (-1)^{\ell+\ell'} (2n+1) \sum_{\ell_1,\ell_3} \ii^{\lu+\ell+\lt+\ell'} \frac{(2\ell+1)_{13}}{4\pi} \threeJz{\ell}{\ell_1}{n}^2 \\
& \times \threeJz{\ell'}{\ell_3}{n}^2 T_{\ell,\ell,\ell',\ell',\ci}^{\lu,\ell,\lt,\ell'}
\ea
where the angular trispectrum is
\ba
\nonumber T_{\ell,\ell,\ell',\ell',\ci}^{\lu,\ell,\lt,\ell'} =& \left(\frac{2}{\pi}\right)^4 \int \dd V_{1234} \ k_{1234}^2 \, \dd k_{1234} \ x^2 \, \dd x \ j_\ell(k_1 r_1) \, j_{\ell_1}(k_1 x) \\
\nonumber & j_\ell(k_2 r_2) \, j_{\ell}(k_2 x) \ j_{\ell'}(k_3 r_3) \, j_{\ell_3}(k_3 x) \ j_{\ell'}(k_4 r_4) \, j_{\ell'}(k_4 x) \\
& \times f(k_1,k_2,k_3,k_4)
\ea
For $n=0$, Eq.~\ref{Eq:reduc-prop-3J-n0} shows that the covariance equation reduces to the simplest trispectrum case (angle and diagonal independent) and forces $\ell_1=\ell$ and $\ell_3=\ell'$, implying that the angular trispectrum also reduces to the standard case.

\myparagraph{The case of $\hk_{1+2}\cdot\hk_3$ }
In this case, the 3D trispectrum is
\ba
T_\mr{gal}(\kk_{1234},z_{1234}) = P_n(\hk_{1+2}\cdot\hk_3) \ f(k_{1+2},k_1,k_2,k_3,k_4)
\ea
and the resulting covariance, derived in Appendix \ref{App:2Dproj-trisp-angdep-1angle-diag}, is
\ba
\nonumber \mathcal{C}_{\ell,\ell'} =& (-1)^{\ell'} \sum_{\lt} \ii^{\ell'+n+\lt} \frac{2\lt+1}{4\pi} \threeJz{\ell'}{n}{\lt}^2 T_{\ell\ell\ell'\ell',1+2}^{\ell,\ell,0,n,\lt,\ell'} 
\ea
where the angular trispectrum is
\ba
\nonumber T_{\ell\ell\ell'\ell',1+2}^{\ell,\ell,0,n,\lt,\ell'} =& \left(\frac{2}{\pi}\right)^5 \int \dd V_{1234} \ k^2_{1234}\,\dd k_{1234} \ K^2\,\dd K \ x_{ab}^2\,\dd x_{ab} \\
\nonumber & j_\ell(k_1 r_1) \, j_{\ell}(k_1 x_a) \ j_\ell(k_2 r_2) \, j_{\ell}(k_2 x_a) \ j_{0}(K x_a)   \\
\nonumber & j_{n}(K x_b) \ j_{\ell'}(k_3 r_3) \,  j_{\ell_3}(k_3 x_b) \ j_{\ell'}(k_4 r_4) \, j_{\ell'}(k_4 x_b) \\
& \times f(K,k_1,k_2,k_3,k_4).
\ea
For $n=0$, Eq.~\ref{Eq:reduc-prop-3J-n0} shows that the covariance equation reduces to the angle independent trispectrum case (depending on the squeezed diagonal) and forces $\ell_3=\ell'$, implying that the angular trispectrum also reduces to the appropriate case.

\myparagraph{The case of $\hk_{1+3}\cdot\hk_2$ }
In this case, the 3D trispectrum is
\ba
T_\mr{gal}(\kk_{1234},z_{1234}) = P_n(\hk_{1+3}\cdot\hk_2) \ f(k_{1+3},k_1,k_2,k_3,k_4)
\ea
and the resulting covariance, derived in Appendix \ref{App:2Dproj-trisp-angdep-1angle-diag}, is
\ba
\nonumber \mathcal{C}_{\ell,\ell'}=& (-1)^{\ell} \sum_{\ell_a,\ell_b,\ell_2} \ii^{\ell-\ell_a+\ell_b+\ld} \frac{(2\ld+1)(2\ell_a+1)(2\ell_b+1)}{4\pi} \\
& \times H_{\ell_2,\ell,\ell'}^{\ell_a,\ell_b,n} \ T_{\ell,\ell,\ell',\ell',1+3}^{\ell,\ld,\ell_a,\ell_b,\ell',\ell'} 
\ea
where the geometric coefficient is
\ba
\nonumber H_{\ell_2,\ell,\ell'}^{\ell_a,\ell_b,n} =& \frac{(4\pi)^2}{(2\ell+1)(2\ell'+1)(2n+1)(2\ld+1)(2\ell_a+1)(2\ell_b+1)} \\
\nonumber & \sum_{m,m',m_n,m_a,m_b,m_2} \!\!\!\! (-1)^{m_a+m_b+m_n+m_2+m+m'} \ G_{\ell_a,\ell_b,n}^{m_a,m_b,m_n} \ G_{\ell_a,\ell,\ell'}^{-m_a,m,-m'}  \\
& \times G_{\ell_2,\ell_b,\ell'}^{m_2,-m_b,m'} \ G_{\ell_2,\ell,n}^{-m_2,-m,-m_n} 
\ea
and the angular trispectrum is given by
\ba
\nonumber T_{\ell,\ell,\ell',\ell',1+3}^{\ell,\ld,\ell_a,\ell_b,\ell',\ell'} =& \left(\frac{2}{\pi}\right)^5 \int \dd V_{1234} \ k^2_{1234}\,\dd k_{1234} \ K^2\,\dd K \ x^2_{ab}\,\dd x_{ab} \\
\nonumber & j_\ell(k_1 r_1) \,  j_{\ell}(k_1 x_a) \ j_\ell(k_2 r_2) \, j_{\ell_2}(k_2 x_b) \ j_{\ell_a}(K x_a) \\
\nonumber & j_{\ell_b}(K x_b) \ j_{\ell'}(k_3 r_3) \, j_{\ell'}(k_3 x_a) \ j_{\ell'}(k_4 r_4) \, j_{\ell'}(k_4 x_b) \\
& \times f(K,k_1,k_2,k_3,k_4).
\ea
For $n=0$, Eq.~\ref{Eq:reduc-prop-Gaunt} reduces the $H$ coefficient to
\ba
\nonumber H_{\ell_2,\ell,\ell'}^{\ell_a,\ell_b,0} =& \frac{4\pi \ \delta_{\ell,\ld} \ \delta_{\ell_a,\ell_b}}{(2\ell+1)^2 (2\ell'+1) (2\ell_a+1)^2} \\
\nonumber & \sum_{m,m',m_a} \!\!\!\! (-1)^{m+m'+m_a} \ G_{\ell_a,\ell,\ell'}^{-m_a,m,-m'} \ G_{\ell,\ell_a,\ell'}^{-m,m_a,m'} \\
&= \frac{\delta_{\ell,\ld} \ \delta_{\ell_a,\ell_b}}{(2\ell+1) (2\ell_a+1)} \threeJz{\ell}{\ell'}{\ell_a}^2.
\ea
Diagrammatically, this reduction can be seen as the transformation of the tetrahedron of Fig.~\ref{Fig:6J-tetrahedron} into an oat grain diagram (Fig.~\ref{Fig:oatgrain}) when the side $n$ is cut. Inputting this to the covariance equation, one sees that the angular trispectrum reduces to the appropriate angle independent case (depending on the alternate diagonal), and the covariance is
\ba
\nonumber \mathcal{C}_{\ell,\ell'} = \sum_{\ell_a} \frac{(2\ell_a+1)}{4\pi} \threeJz{\ell}{\ell'}{\ell_a}^2 T_{\ell,\ell,\ell',\ell',1+3}^{\ell,\ell,\ell_a,\ell_a,\ell',\ell'} 
\ea
as required.

\subsubsection{Two angles}\label{App:reduc-angle-2}

\myparagraph{X$\times$Y squeezed diagonal case}
In this case, the 3D trispectrum is
\ba
T_\mr{gal}(\kk_{1234},z_{1234}) = P_n(-\hk_{1+2}\cdot\hk_1) \; P_{n'}(\hk_{1+2}\cdot\hk_3) \ f(k_{1+2},k_1,k_2,k_3,k_4)
\ea
and the resulting covariance, derived in Appendix \ref{App:2Dproj-trisp-angdep-2angles-XY}, is
\ba
\nonumber \mathcal{C}_{\ell,\ell'} =& (-1)^{\ell+\ell'} \sum_{\lu,\lt} \ii^{\lu+\ell+n+n'+\lt+\ell'} \ \frac{(2\ell+1)_{13}}{4\pi} \threeJz{\ell}{\lu}{n}^2 \\ 
& \times \threeJz{\ell'}{\lt}{n'}^2 T_{\ell,\ell,\ell',\ell',1+2}^{\lu,\ell,n,n',\lt,\ell'}
\ea
where the angular trispectrum is
\ba
\nonumber T_{\ell,\ell,\ell',\ell',1+2}^{\lu,\ell,n,n',\lt,\ell'} =& \left(\frac{2}{\pi}\right)^5 \int \dd V_{1234} \ k^2_{1234}\,\dd k_{1234} \ K^2\,\dd K \ x_{ab}^2\,\dd x_{ab} \\
\nonumber & j_\ell(k_1 r_1) \, j_{\lu}(k_1 x_a) \ j_\ell(k_2 r_2) \, j_{\ell}(k_2 x_a) \ j_{n}(K x_a) \\
\nonumber & j_{n'}(K x_b) \ j_{\ell'}(k_3 r_3) \,  j_{\ell_3}(k_3 x_b) \ j_{\ell'}(k_4 r_4) \, j_{\ell'}(k_4 x_b) \\
& \times f(K,k_1,k_2,k_3,k_4).
\ea
For $n=0$, Eq.~\ref{Eq:reduc-prop-3J-n0} shows that the covariance equation reduces to the $\hk_{1+2}\cdot\hk_3$ trispectrum case, and forces $\ell_1=\ell$, implying that the angular trispectrum also reduces to the appropriate case.\\
The manifest symmetry between $n$ and $n'$ means that one does not have to check the case $n'=0$.

\myparagraph{X$\times$Y alternate diagonal case}
In this case, the 3D trispectrum is
\ba
\nonumber T_\mr{gal}(\kk_{1234},z_{1234}) = P_n(-\hk_{1+3}\cdot\hk_1) \; P_{n'}(\hk_{1+3}\cdot\hk_2) \\
\times f(k_{1+3},k_1,k_2,k_3,k_4)
\ea
and the resulting covariance, derived in Appendix \ref{App:2Dproj-trisp-angdep-2angles-XY}, is
\ba
\nonumber \mathcal{C}_{\ell,\ell'}=& (-1)^{\ell+n} \sum_{\ell_1,\ell_2,\ell_a,\ell_b,\ell_c} \ii^{\lu-\ell_a+\ell_b+\ld} \ \frac{(2\ell+1)_{12abc}}{4\pi} \\
& \times J^{\ld,\ell,n';\ell_b,n,\ell_c}_{\ell',\lu,\ell_a} \ T_{\ell,\ell,\ell',\ell',1+3}^{\lu,\ld,\ell_a,\ell_b,\ell',\ell'}
\ea
where the geometric coefficient is
\ba
\nonumber J^{\ld,\ell,n';\ell_b,n,\ell_c}_{\ell',\lu,\ell_a} =& \frac{(4\pi)^3}{(2\ell+1)(2\ell'+1) \ (2n+1)(2n'+1) \ (2\ell+1)_{12abc}} \\
\nonumber & \sum_{\substack{m,m',m_1,m_2,m_c \\ m_a,m_b,m_n,m_{n'}}} (-1)^{m+m'+m_n+m_{n'}+m_1+m_2+m_a+m_b+m_c} \\
\nonumber &  G_{\ell_2,\ell,n'}^{-m_2,-m,-m_{n'}} \ G_{\ell_b,n,\ell_c}^{m_b,-m_{n},-m_c} \ G_{\ell',\ell_1,\ell_a}^{m',-m_1,m_a} \\
& G_{\ell_b,\ell_2,\ell'}^{-m_b,m_2,-m'} \ G_{\ell,n,\ell_1}^{m,m_n,m_1} \ G_{n',\ell_c,\ell_a}^{m_{n'},m_c,-m_a}
\ea
and the angular trispectrum is
\ba
\nonumber T_{\ell,\ell,\ell',\ell',1+3}^{\lu,\ld,\ell_a,\ell_b,\ell',\ell'} =& \left(\frac{2}{\pi}\right)^5 \int \dd V_{1234} \ k^2_{1234}\,\dd k_{1234} \ K^2\,\dd K \ x_{ab}^2\,\dd x_{ab} \\
\nonumber & j_\ell(k_1 r_1) \, j_{\lu}(k_1 x_a) \ j_\ell(k_2 r_2) \, j_{\ld}(k_2 x_b) \ j_{\ell_a}(K x_a) \\
\nonumber & j_{\ell_b}(K x_b) \ j_{\ell'}(k_3 r_3) \,  j_{\ell'}(k_3 x_a) \ j_{\ell'}(k_4 r_4) \, j_{\ell'}(k_4 x_b) \\
& \times f(K,k_1,k_2,k_3,k_4)
\ea
For $n=0$, Eq.~\ref{Eq:reduc-prop-Gaunt} reduces the $J$ coefficient to
\ba
\nonumber J^{\ld,\ell,n';\ell_b,0,\ell_c}_{\ell',\lu,\ell_a} =& \frac{(4\pi)^2 \ \delta_{\ell,\lu} \ \delta_{\ell_b,\ell_c} }{(2\ell+1)^2(2\ell'+1) \ (2n+1)(2n'+1) \ (2\ell+1)_{2abb}} \\
\nonumber & \sum_{\substack{m,m',m_2 \\ m_a,m_b,m_n,m_{n'}}} (-1)^{m+m'+m_n+m_{n'}+m_2+m_a+m_b} \\
\nonumber &  G_{\ell_2,\ell,n'}^{-m_2,-m,-m_{n'}} \ G_{\ell',\ell,\ell_a}^{m',m,m_a} \ G_{\ell_b,\ell_2,\ell'}^{-m_b,m_2,-m'} \ G_{n',\ell_b,\ell_a}^{m_{n'},m_b,-m_a} \\
=& \frac{\delta_{\ell,\lu} \ \delta_{\ell_b,\ell_c}}{(2\ell+1)(2\ell_b+1)} \ H_{\ld,\ell,\ell'}^{\ell_a,\ell_b,n'}
\ea
Diagrammatically, this reduction can be seen as the transformation of the hexagon of Fig.~\ref{Fig:9J-hexagon} into a tetrahedron diagram (Fig.~\ref{Fig:6J-tetrahedron}) when the side $n$ is cut.\\
The covariance then becomes
\ba
\nonumber \mathcal{C}_{\ell,\ell'}=& (-1)^{\ell} \sum_{\ell_2,\ell_a,\ell_b} \ii^{\ell-\ell_a+\ell_b+\ld} \ \frac{(2\ell+1)_{2ab}}{4\pi} \ H_{\ld,\ell,\ell'}^{\ell_a,\ell_b,n'} \ T_{\ell,\ell,\ell',\ell',1+3}^{\ell,\ld,\ell_a,\ell_b,\ell',\ell'}
\ea
i.e. reducing to the $\hk_{1+3}\cdot\hk_2$ case as required.\\
The manifest symmetry between $n$ and $n'$ means that one does not have to check the case $n'=0$.

\myparagraph{3PT squeezed diagonal case}
In this case, the 3D trispectrum is
\ba
T_\mr{gal}(\kk_{1234},z_{1234}) = P_n(\hk_1\cdot\hk_2) \; P_{n'}(\hk_{1+2}\cdot\hk_3) \ f(k_1,k_2,k_3,k_4)
\ea
and the resulting covariance, derived in Appendix \ref{App:2Dproj-trisp-angdep-2angles-3PT}, is
\ba
\nonumber \mathcal{C}_{\ell,\ell'}=& (-1)^{\ell+\ell'} \sum_{\lu\lt} \ii^{2\lu+n'+\lt+\ell'} \ \frac{(2\lu+1)(2\lt+1)}{4\pi} \threeJz{\ell}{\lu}{n}^2 \\
&\times \threeJz{\ell'}{\lt}{n'}^2 T_{\ell,\ell,\ell',\ell',1+2}^{\lu,\ell,0,n',\lt,\ell'}
\ea
where the angular trispectrum is
\ba
\nonumber T_{\ell,\ell,\ell',\ell',1+2}^{\lu,\ell,0,n',\lt,\ell'} =& \left(\frac{2}{\pi}\right)^5 \int \dd V_{1234} \ k^2_{1234}\,\dd k_{1234} \ K^2\,\dd K \ x_{ab}^2\,\dd x_{ab} \\
\nonumber & j_\ell(k_1 r_1) \, j_{\lu}(k_1 x_a) \ j_\ell(k_2 r_2) \, j_{\lu}(k_2 x_a) \ j_{0}(K x_a) \\
\nonumber & j_{n'}(K x_b) \ j_{\ell'}(k_3 r_3) \,  j_{\ell_3}(k_3 x_b) \ j_{\ell'}(k_4 r_4) \, j_{\ell'}(k_4 x_b) \\
& \times f(k_1,k_2,k_3,k_4).
\ea
For $n=0$, Eq.~\ref{Eq:reduc-prop-3J-n0} shows that the covariance equation reduces to the $\hk_{1+2}\cdot\hk_3$ trispectrum case, and forces $\ell_1=\ell$, implying that the angular trispectrum also reduces to the appropriate case.\\
For $n'=0$, Eq.~\ref{Eq:reduc-prop-3J-n0}, and using that $\ell+\lu+n$ is even, shows that the covariance equation reduces to the $\hk_{1}\cdot\hk_2$ trispectrum case. Furthermore it forces $\ell_3=\ell'$, together with Eq.~\ref{Eq:reduc-prop-spherbess}, this implies that the angular trispectrum also reduces to the appropriate case.

\myparagraph{3PT alternate diagonal case}
In this case, the 3D trispectrum is
\ba
T_\mr{gal}(\kk_{1234},z_{1234}) = P_n(\hk_1\cdot\hk_3) \; P_{n'}(\hk_{1+3}\cdot\hk_2) \ f(k_1,k_2,k_3,k_4)
\ea
and the resulting covariance, derived in Appendix \ref{App:2Dproj-trisp-angdep-2angles-3PT}, is
\ba
\nonumber \mathcal{C}_{\ell,\ell'}=& (-1)^{\ell+\ell'} \sum_{\lu,\ld,\lt,\ell_a,\ell_b} \ii^{\lu+\lt-\ell_a+\ell_b+\ld+\ell'} \ \frac{(2\ell+1)_{123ab}}{4\pi} \\
& \times K^{\lu,\ell,n;n',\ell_b,\ell_a}_{\ell',\ell_b,\ld} \ T_{\ell,\ell,\ell',\ell',1+3}^{\lu,\ld,\ell_a,\ell_b,\lt,\ell'}
\ea
where the geometric coefficient is
\ba
\nonumber K^{\lu,\lt,n;\ell,\ell_a,\ell'}_{n',\ell_b,\ld} =& \frac{(4\pi)^3}{(2\ell+1)(2\ell'+1) \ (2n+1)(2n'+1) \ (2\ell+1)_{123ab}}   \\
\nonumber & \sum_{\substack{m,m',m_{123} \\ m_{ab},m_n,m_{n'}}} (-1)^{m+m'+m_1+m_2+m_3+m_a+m_b+m_n+m_{n'}} \\
\nonumber & G_{\ell,n,\ell_1}^{m,m_{n},m_1} \ G_{\ell_a,\lu,\lt}^{-m_a,-m_1,m_3} \ G_{\ell',\ell_3,n}^{m',m_3,m_{n}} \\
& \times G_{\ell,\ell_2,n'}^{-m,-m_2,-m_{n'}}  \ G_{\ell_a,n',\ell_b}^{m_a,m_{n'},m_b}  \ G_{\ell',\ell_b,\ld}^{m',-m_b,m_2}
\ea
and the angular trispectrum is
\ba
\nonumber T_{\ell,\ell,\ell',\ell',1+3}^{\lu,\ld,\ell_a,\ell_b,\lt,\ell'} =& \left(\frac{2}{\pi}\right)^5 \int \dd V_{1234} \ k_{1234}^2\,\dd k_{1234} \ K^2\,\dd K \ x^2_{ab}\,\dd x_{ab} \\
\nonumber & j_\ell(k_1 r_1) \, j_{\lu}(k_1 x_a) \ j_\ell(k_2 r_2) \, j_{\ld}(k_2 x_b) \ j_{\ell_a}(K x_a) \\
\nonumber & j_{\ell_b}(K x_b) \ j_{\ell'}(k_3 r_3) \, j_{\ell_3}(k_3 x_a) \  j_{\ell'}(k_4 r_4) \, j_{\ell'}(k_4 x_b) \\
& \times f(k_1,k_2,k_3,k_4)
\ea

For $n=0$, Eq.~\ref{Eq:reduc-prop-Gaunt} reduces the $K$ coefficient to
\ba
\nonumber K^{\lu,\lt,0;\ell,\ell_a,\ell'}_{n',\ell_b,\ld} =& \frac{(4\pi)^2 \ \delta_{\ell,\lu} \ \delta_{\ell',\lt}}{(2\ell+1)^2 (2\ell'+1)^2 \ (2n'+1) \ (2\ell+1)_{2ab}}   \\
\nonumber & \sum_{\substack{m,m',m_{2} \\ m_{ab},m_n,m_{n'}}} (-1)^{m+m'+m_2+m_a+m_b+m_n+m_{n'}} \\
\nonumber & G_{\ell_a,\ell,\ell'}^{-m_a,m,-m'} \ G_{\ell,\ell_2,n'}^{-m,-m_2,-m_{n'}}  \ G_{\ell_a,n',\ell_b}^{m_a,m_{n'},m_b}  \ G_{\ell',\ell_b,\ld}^{m',-m_b,m_2} \\
=& \frac{\delta_{\ell,\lu} \ \delta_{\ell',\lt}}{(2\ell+1) (2\ell'+1)} \ H_{\ld,\ell,\ell'}^{\ell_a,\ell_b,n'}.
\ea
Diagrammatically, this reduction can be seen as the transformation of the prism of Fig.~\ref{Fig:K-prism} into a tetrahedron diagram (Fig.~\ref{Fig:6J-tetrahedron}) when the side $n$ is cut. Hence the covariance is
\ba
\nonumber \mathcal{C}_{\ell,\ell'}=& (-1)^{\ell} \sum_{\ell_2,\ell_a,\ell_b} \ii^{\ell-\ell_a+\ell_b+\ld} \ \frac{(2\ell+1)_{2ab}}{4\pi} \ H_{\ld,\ell,\ell'}^{\ell_a,\ell_b,n'} \ T_{\ell,\ell,\ell',\ell',1+3}^{\ell,\ld,\ell_a,\ell_b,\ell',\ell'}
\ea
i.e. reducing to the $\hk_{1+3}\cdot\hk_2$ case as required.

For $n'=0$, Eq.~\ref{Eq:reduc-prop-Gaunt} reduces the $K$ coefficient to
\ba
\nonumber K^{\lu,\lt,n;\ell,\ell_a,\ell'}_{0,\ell_b,\ld} =& \frac{(4\pi)^2 \ \delta_{\ell,\ld} \ \delta_{\ell_a,\ell_b}}{(2\ell+1)^2 (2\ell'+1) \ (2n+1) \ (2\ell+1)_{13aa}} \\
\nonumber & \sum_{\substack{m,m',m_{13} \\ m_{a},m_n,m_{n'}}} (-1)^{m+m'+m_1+m_3+m_a+m_n+m_{n'}} \\
\nonumber & G_{\ell,n,\ell_1}^{m,m_{n},m_1} \ G_{\ell_a,\lu,\lt}^{-m_a,-m_1,m_3} \ G_{\ell',\ell_3,n}^{m',m_3,m_{n}} \ G_{\ell',\ell_a,\ell}^{m',m_a,-m} \\
=& \frac{\delta_{\ell,\ld} \ \delta_{\ell_a,\ell_b}}{(2\ell+1) (2\ell_a+1)} \ H_{\ell_3,\ell_a,\ell'}^{\ell,n,\lu}.
\ea
Diagrammatically, this reduction can be seen as the transformation of the prism of Fig.~\ref{Fig:K-prism} into a tetrahedron diagram (Fig.~\ref{Fig:6J-tetrahedron}) when the side $n$ is cut.\\
Then further reductions occur : the Kronecker symbols reduce the angular trispectrum to $T_{\ell,\ell,\ell',\ell',1+3}^{\lu,\ell,\ell_a,\ell_a,\lt,\ell'}$. But given that the 3D trispectrum does not depend on $k_{1+3}$, I can employ Eq.~\ref{Eq:reduc-prop-spherbess} to show that the angular trispectrum further reduces to $T_{\ell,\ell,\ell',\ell',\ci}^{\lu,\ell,\lt,\ell'}$, independent of $\ell_a$ and corresponding to the angular trispectrum required for the $\hk_1\cdot\hk_3$ case. Now I can perform the sum over $\ell_a$ and use the identity
\ba
\nonumber \sum_{\ell_a,m_a} G_{\ell_a,\lu,\lt}^{-m_a,-m_1,m_3} \ G_{\ell',\ell_a,\ell}^{m',m_a,-m} =& \int \dd^2\hn \ Y_1^*(\hn) \ Y_3(\hn) \ Y_{\ell,m}^*(\hn) \ Y_{\ell',m'}(\hn) \\
=& \sum_{\ell_c,m_c} G_{\ell,\ell_c,\lu}^{-m,-m_c,-m_1} \ G_{\ell',\ell_3,\ell_c}^{m',m_3,m_c}.
\ea
So that, when performing the partial sum over $(\ell_a,\ell_b)$ in the covariance equation, one finds 
\ba
\nonumber \mr{Partial\,sum} \equiv & \sum_{\ell_a,\ell_b} i^{-\ell_a+\ell_b} (2\ell+1)_{ab} \ K^{\lu,\lt,n;\ell,\ell_a,\ell'}_{0,\ell_b,\ld} \\
\nonumber =& \frac{(4\pi)^2 \ \delta_{\ell,\ld}}{(2\ell+1)^2 (2\ell'+1) \ (2n+1) \ (2\ell+1)_{13}} \\
\nonumber & \sum_{\substack{m,m',m_{13} \\ c,m_n,m_{n'}}} (-1)^{m+m'+m_1+m_3+m_a+m_n+m_{n'}} \\
\nonumber & G_{\ell,n,\ell_1}^{m,m_{n},m_1} \ G_{\ell,\ell_c,\lu}^{-m,-m_c,-m_1} \ G_{\ell',\ell_3,n}^{m',m_3,m_{n}} \ G_{\ell',\ell_3,\ell_c}^{m',m_3,m_c} \\
\nonumber =& \frac{ (2n+1) \ \delta_{\ell,\ld}}{(2\ell+1)} \threeJz{\ell}{n}{\lu}^2 \threeJz{\ell'}{\lt}{n}^2.
\ea
Hence the covariance equation becomes
\ba
\nonumber \mathcal{C}_{\ell,\ell'}=& (-1)^{\ell+\ell'} (2n+1) \sum_{\lu,\lt} \ii^{\lu+\lt+\ell+\ell'} \ \frac{(2\ell+1)_{13}}{4\pi} \threeJz{\ell}{n}{\lu}^2 \\
& \times \threeJz{\ell'}{\lt}{n}^2 \ T_{\ell,\ell,\ell',\ell',\ci}^{\lu,\ell,\lt,\ell'}
\ea
in other words, finally, reducing to the $\hk_{1}\cdot\hk_3$ case as required.

With this final reduction, I have completed the reduction tree, that is, proved that all equations for the complex bispectrum case reduce to the simpler case when appropriate. Hence the covariance derivations presented in all the preceding appendices have successfully passed this sanity/consistency check.

\section{Relation of geometric coefficients with 3n-J symbols}\label{App:3nJ-symbols}

The following relation between Gaunt coefficients and 3J symbols will be useful in this section:
\ba\label{Eq:3J-to-Gaunt}
G_{\lu,\ld,\lt}^{m_1,m_2,m_3} = \sqrt{\frac{(2\ell+1)_{123}}{4\pi}} \threeJz{\lu}{\ld}{\lt} \threeJm{\lu}{\ld}{\lt}{m_1}{m_2}{m_3} 
\ea

\subsection{Diagrammatic of 3J symbols}\label{App:3J-diagrammatic}
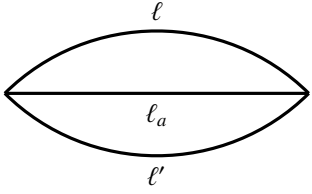
\begin{figure}[htbp]
\begin{center}
\begin{tikzpicture}
\draw [-, very thick] (-2,0) -- node[below] {$\ell_a$} (2,0);
\draw [-, very thick] (-2,0) to [bend left=45] node[above] {$\ell$} (2,0);
\draw [-, very thick] (-2,0) to [bend right=45] node[below] {$\ell'$} (2,0);
\end{tikzpicture}
\caption{Oat grain diagram}
\label{Fig:oatgrain}
\end{center}
\end{figure}

Geometric coefficients found in the covariance derivation of the previous appendices can be represented diagrammatically, by representing a 3J symbol as a vertex from which  start 3 lines labelled with the corresponding multipoles. This representation was devised by \cite{Yutsis1962}, and in the following I am going to call these Yutsis diagrams.\\
For example the identity
\ba
\sum_{m,m',m_a} G_{\ell,\ell',\ell_a}^{m,m',m_a} \ G_{\ell,\ell',\ell_a}^{-m,-m',-m_a}  = \threeJz{\ell}{\ell'}{\ell_a}^2
\ea
equivalent to
\ba
\sum_{m,m',m_a} \threeJm{\ell}{\ell'}{\ell_a}{m}{m'}{m_a} \ \threeJm{\ell}{\ell'}{\ell_a}{-m}{-m'}{-m_a}  = 1
\ea
appearing for example, in the covariance case derived in Appendix \ref{App:2Dproj-trisp-altdiag}, can be represented with the oat grain diagram shown in Fig.~\ref{Fig:oatgrain}

\subsection{$H_{\ell_2,\ell,\ell'}^{\ell_a,\ell_b,n}$ and 6J symbols}\label{App:H-and-6J}

6J symbols are defined as a sum over products of four 3J symbols (\url{https://en.wikipedia.org/wiki/6-j_symbol}, see also Eq. 34.4.1 of \cite{NIST:DLMF} for an equivalent definition)
\ba
\nonumber \sixJ{j_1}{j_2}{j_3}{j_4}{j_5}{j_6} = (-1)^{\sum_i j_i} \sum_{m_1,\dots,m_6} (-1)^{\sum_i m_i} \threeJm{j_1}{j_2}{j_3}{-m_1}{-m_2}{-m_3} \\
\times \threeJm{j_1}{j_5}{j_6}{m_1}{-m_5}{m_6} \threeJm{j_4}{j_2}{j_6}{m_4}{m_2}{-m_6} \threeJm{j_4}{j_5}{j_3}{-m_4}{m_5}{m_3}.
\ea

Recall the definition of $H_{\ell_2,\ell,\ell'}^{\ell_a,\ell_b,n}$
\ba
\nonumber H_{\ell_2,\ell,\ell'}^{\ell_a,\ell_b,n} =& \frac{(4\pi)^2}{(2\ell+1)(2\ell'+1)(2n+1)(2\ld+1)(2\ell_a+1)(2\ell_b+1)} \\
\nonumber & \sum_{m,m',m_n,m_a,m_b,m_2} \!\!\!\! (-1)^{m_a+m_b+m_n+m_2+m+m'} \ G_{\ell_a,\ell_b,n}^{m_a,m_b,m_n} \ G_{\ell_a,\ell,\ell'}^{-m_a,m,-m'}  \\
& \times G_{\ell_2,\ell_b,\ell'}^{m_2,-m_b,m'} \ G_{\ell_2,\ell,n}^{-m_2,-m,-m_n}. 
\ea
If one uses Eq. \ref{Eq:3J-to-Gaunt} to relate Gaunt coefficients to 3J symbols, up to change of silent variables $m\rightarrow -m$, one sees that $H$ can be put in the form of a 6J symbol, namely
\ba
H_{\ell_2,\ell,\ell'}^{\ell_a,\ell_b,n} = (-1)^{\ell_a+\ell_b+n+\ell_2+\ell+\ell'} \sixJ{\ell_a}{\ell_b}{n}{\ell_2}{\ell}{\ell'}.
\ea

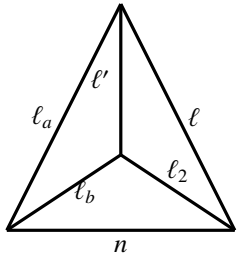
\begin{figure}[htbp]
\begin{center}
\begin{tikzpicture}
\draw [-, very thick] (1.5,-1) -- node[below] {$n$} (-1.5,-1);
\draw [-, very thick] (-1.5,-1) -- node[left] {$\ell_a$} (0,2);
\draw [-, very thick] (0,2) -- node[right] {$\ell$} (1.5,-1);
\draw [-, very thick] (0,0) -- node[left] {$\ell'$} (0,2);
\draw [-, very thick] (0,0) -- node[above] {$\ell_2$} (1.5,-1);
\draw [-, very thick] (0,0) -- node[right] {$\ell_b$} (-1.5,-1);
\end{tikzpicture}
\caption{Tetrahedron corresponding to the 6J symbol involved in this article.}
\label{Fig:6J-tetrahedron}
\end{center}
\end{figure}

The Yutsis diagram of this symbol is a tetrahedron, visible in Fig.~\ref{Fig:6J-tetrahedron} where each vertex represents a 3J symbol.

\subsection{$J^{\ld,\ell,n';\ell_b,n,\ell_c}_{\ell',\lu,\ell_a}$ and 9J symbols}\label{App:J-and-9J}

9J symbols are defined as a sum over products of six 3J symbols (see Eq. 34.6.1 of \cite{NIST:DLMF}, \url{http://dlmf.nist.gov/34.6})
\ba
\nonumber \nineJ{j_1}{j_2}{j_3}{j_4}{j_5}{j_6}{j_7}{j_8}{j_9} =& \sum_{m_1,\cdots,m_9} \\
\nonumber & \threeJm{j_1}{j_2}{j_3}{m_1}{m_2}{m_3} \threeJm{j_4}{j_5}{j_6}{m_4}{m_5}{m_6} \threeJm{j_7}{j_8}{j_9}{m_7}{m_8}{m_9} \\
\times & \threeJm{j_1}{j_4}{j_7}{m_1}{m_4}{m_7} \threeJm{j_2}{j_5}{j_8}{m_2}{m_5}{m_8} \threeJm{j_3}{j_6}{j_9}{m_3}{m_6}{m_9}.
\ea

Recall the definition of $J^{\ld,\ell,n';\ell_b,n,\ell_c}_{\ell',\lu,\ell_a}$
\ba
\nonumber J^{\ld,\ell,n';\ell_b,n,\ell_c}_{\ell',\lu,\ell_a} =& \frac{(4\pi)^3}{(2\ell+1)(2\ell'+1) \ (2n+1)(2n'+1) \ (2\ell+1)_{12abc}} \\
\nonumber & \sum_{\substack{m,m',m_1,m_2,m_c \\ m_a,m_b,m_n,m_{n'}}} (-1)^{m+m'+m_n+m_{n'}+m_1+m_2+m_a+m_b+m_c} \\
\nonumber &  G_{\ell_2,\ell,n'}^{-m_2,-m,-m_{n'}} \ G_{\ell_b,n,\ell_c}^{m_b,-m_{n},-m_c} \ G_{\ell',\ell_1,\ell_a}^{m',-m_1,m_a} \\
& G_{\ell_b,\ell_2,\ell'}^{-m_b,m_2,-m'} \ G_{\ell,n,\ell_1}^{m,m_n,m_1} \ G_{n',\ell_c,\ell_a}^{m_{n'},m_c,-m_a}.
\ea
If one uses Eq. \ref{Eq:3J-to-Gaunt} to relate Gaunt coefficients to 3J symbols, up to change of silent variables $m\rightarrow -m$ and using the parity condition on azimuthal parameters (e.g. $m_1+m_2+m_3=0$), one sees that $J$ is in fact a 9J symbol, namely
\ba
J^{\ld,\ell,n';\ell_b,n,\ell_c}_{\ell',\lu,\ell_a} = \nineJ{\ell_2}{\ell}{n'}{\ell_b}{n}{\ell_c}{\ell'}{\ell_1}{\ell_a}.
\ea

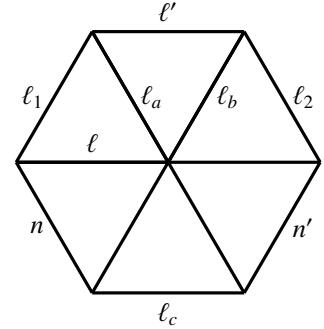
\begin{figure}[htbp]
\begin{center}
\begin{tikzpicture}
\draw [-, very thick] (-2,0) -- (2,0);
\draw [-, very thick] (-1,-1.7321) -- (1,1.7321);
\draw [-, very thick] (-1,1.7321) -- (1,-1.7321);
\draw [-, very thick] (-2,0) -- node[above] {$\ell$} (0,0);
\draw [-, very thick] (1,1.7321) -- node[right] {$\ell_b$} (0,0);
\draw [-, very thick] (-1,1.7321) -- node[right] {$\ell_a$} (0,0);
\draw [-, very thick] (-2,0) -- node[left] {$\ell_1$} (-1,1.7321);
\draw [-, very thick] (-1,1.7321) -- node[above] {$\ell'$} (1,1.7321);
\draw [-, very thick] (1,1.7321) -- node[right] {$\ell_2$} (2,0);
\draw [-, very thick] (2,0) -- node[right] {$n'$} (1,-1.7321);
\draw [-, very thick] (1,-1.7321) -- node[below] {$\ell_c$} (-1,-1.7321);
\draw [-, very thick] (-1,-1.7321) -- node[left] {$n$} (-2,0);
\end{tikzpicture}
\caption{Hexagon corresponding to the 9J symbol involved in this article.}
\label{Fig:9J-hexagon}
\end{center}
\end{figure}

The Yutsis diagram of this symbol is an hexagon, visible in Fig.~\ref{Fig:9J-hexagon}.

\subsection{$K^{\lu,\lt,n;\ell,\ell_a,\ell'}_{n',\ell_b,\ld}$ and 12J symbols}\label{App:K-and-12J}

Recall the definition of $K^{\lu,\lt,n;\ell,\ell_a,\ell'}_{n',\ell_b,\ld}$
\ba
\nonumber K^{\lu,\lt,n;\ell,\ell_a,\ell'}_{n',\ell_b,\ld} =& \frac{(4\pi)^3}{(2\ell+1)(2\ell'+1) \ (2n+1)(2n'+1) \ (2\ell+1)_{123ab}}   \\
\nonumber & \sum_{\substack{m,m',m_{123} \\ m_{ab},m_n,m_{n'}}} (-1)^{m+m'+m_1+m_2+m_3+m_a+m_b+m_n+m_{n'}} \\
\nonumber & G_{\ell,n,\ell_1}^{m,m_{n},m_1} \ G_{\ell_a,\lu,\lt}^{-m_a,-m_1,m_3} \ G_{\ell',\ell_3,n}^{m',m_3,m_{n}} \\
& \times G_{\ell,\ell_2,n'}^{-m,-m_2,-m_{n'}}  \ G_{\ell_a,n',\ell_b}^{m_a,m_{n'},m_b}  \ G_{\ell',\ell_b,\ld}^{m',-m_b,m_2}.
\ea
If one uses Eq. \ref{Eq:3J-to-Gaunt} to relate Gaunt coefficients to 3J symbols and the parity condition on azimuthal parameters, one sees that $K$ can be rewritten as a sum of six 3J symbol, namely
\ba
\nonumber K^{\lu,\lt,n;\ell,\ell_a,\ell'}_{n',\ell_b,\ld} \!=\!& \sum_{\substack{m,m',m_{123} \\ m_{ab},m_n,m_{n'}}} \\
\nonumber & \threeJm{\ell}{n}{\lu}{m}{m_n}{m_1} \threeJm{\ell_a}{\lu}{\lt}{\smallminus\! m_a}{\smallminus\! m_1}{m_3} \threeJm{\ell'}{\lt}{n}{m'}{m_3}{m_n} \\
\times & \threeJm{\ell}{\ld}{n'}{\smallminus\! m}{\smallminus\! m_2}{\smallminus\! m_{n'}} \threeJm{\ell_a}{n'}{\ell_b}{m_a}{m_{n'}}{m_b} \threeJm{\ell'}{\ell_b}{\ld}{m'}{\smallminus\! m_b}{m_2}.
\ea

\begin{figure}[htbp]
\begin{center}
\begin{tikzpicture}
\draw [-, very thick] (0,0) -- node[below] {$\ell_2$} (3,0);
\draw [-, very thick] (3,0) -- node[above] {$\ell_b$} (1,1);
\draw [-, very thick] (1,1) -- node[above] {$n'$} (0,0);
\draw [-, very thick] (0,3) -- node[below] {$n$} (3,3);
\draw [-, very thick] (3,3) -- node[above] {$\ell_3$} (1,4);
\draw [-, very thick] (1,4) -- node[above] {$\ell_1$} (0,3);
\draw [-, very thick] (0,0) -- node[left] {$\ell$} (0,3);
\draw [-, very thick] (3,0) -- node[right] {$\ell'$} (3,3);
\draw [dotted, very thick] (1,3) -- (1,4);
\draw [dotted, very thick] (1,1) -- node[right] {$\ell_a$} (1,3);
\end{tikzpicture}
\caption{Triangular prism corresponding to the $K$ symbol involved in this article.}
\label{Fig:K-prism}
\end{center}
\end{figure}
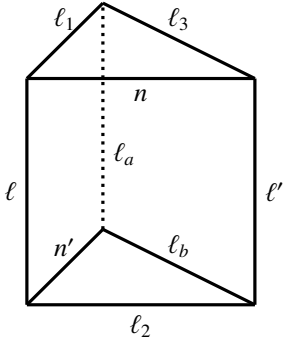

The corresponding diagram is a triangular prism visible in Fig.~\ref{Fig:K-prism}.

This coefficient can not be written in terms of a 9J symbol nor a simple sum of a few 6J symbols. It can however be written as reduced 12J symbol of the second kind (without braiding), that is the symbol whose Yutsis diagram is a cube, reduced to the case where one multipole is zero. See, for instance, \cite{Alisauskas2000} for a definition and a representation of this 12J symbol. Explicitly, one finds
\ba
K^{\lu,\lt,n;\ell,\ell_a,\ell'}_{n',\ell_b,\ld} = \begin{Bmatrix} n' & \lu & \ld & n \\ \ell_b & \ld & \lt & n \\ \ell_a & \ell' & \ell & 0 \end{Bmatrix}
\ea

\section{Shot noise}\label{App:shot-noise}

\subsection{Case of overlapping redshift bins}\label{App:shot-overlapping}

Equation~\ref{Eq:Clshot} for the shot-noise angular power spectrum assumed that the redshift bins were not overlapping. In the case of overlapping bins one would have instead
\ba
C_\ell^\mr{shot}(i_z,j_z) = N_\mr{gal}(i_z\cap j_z)
\ea
where $i_z\cap j_z$ denotes the range of redshift overlap.

\subsection{Poissonianity ?}\label{App:shotvsPoisson}

In Section \ref{Sect:power-spectrum}, the power spectrum shot-noise term is given by Eq.~\ref{Eq:Clshot}, restated here for convenience
\ba
C_\ell^\mr{shot}(i_z,j_z) = N_\mr{gal}(i_z) \ \delta_{i_z,j_z}.
\ea
This power spectrum is the same as the one of a random map with galaxies drawn independently with a Poisson distribution. Thus an underlying Poissonian assumption seems hidden in the modelling.

I clarify here that there is no such explicit assumption: the power spectrum shot-noise term simply results from its definition as the part of the correlation function with galaxy coincidence (third diagram in Fig.~\ref{Fig:diagrams-spectrum}).\\
In a map with N galaxies, the two-point correlation function hits twice the same galaxy exactly N times. Equations~\ref{Eq:Pshot} and \ref{Eq:Clshot} are simply a restatement of this fact, which has no underlying assumption.

In the modelling used in this article, the total power spectrum asymptotes to the shot-noise value as $\ell \rightarrow\infty$. In practice in a given survey analysis, the power spectrum may not asymptote to the shot-noise value due either to data analysis or physical effects, see for example, \cite{Paech2017}. One such physical effect is non-linear clustering, whose modelisation is absent from \cite{Paech2017}. In the present article, non-linear clustering is given by the one-halo term of the power spectrum. This term is constant on scales larger than the typical halo radius, so that a survey limited to large scale would indeed see a constant component in the power spectrum with value larger than the shot-noise value: `super-Poissonian noise'. Another such physical effect is halo and/or galaxy exclusion, which is not modeled in this article and would lead to an anti-clustering on halo and/or galaxy scales: `sub-Poissonian noise'. 

Hence in a survey the high frequency limit of the power spectrum may not be given by the shot-noise value. However I refrain from calling this a non-Poissonian shot-noise or similar, as this would just create confusion in my opinion. Instead I reserve the word shot noise for the discreteness effect given by Eqs.~\ref{Eq:Pshot} and \ref{Eq:Clshot}, and call other high frequency terms by the physical effect they originate from (non-linear clustering, halo and/or galaxy exclusion...). Finally, I note that as in Sect.~\ref{Sect:shot-subs}, the shot-noise value can be subtracted from the measured power spectrum, and thus the measurement can be used to constraint these high frequency effects and the underlying physics.

\end{document}